\documentclass[aos,preprint]{imsart}

\RequirePackage{amsthm,amsmath,amsfonts,amssymb,mathtools,mathrsfs}
\RequirePackage[authoryear]{natbib}
\RequirePackage{enumerate}
\RequirePackage[colorlinks,citecolor=blue,urlcolor=blue,linkcolor=blue]{hyperref}

\hypersetup{bookmarksdepth=2}
\startlocaldefs

\numberwithin{equation}{section}
\theoremstyle{plain}
\newtheorem{theorem}{Theorem}[section]
\newtheorem{lemma}[theorem]{Lemma}
\newtheorem{proposition}[theorem]{Proposition}
\newtheorem{corollary}[theorem]{Corollary}
\theoremstyle{definition}
\newtheorem{remark}[theorem]{Remark}
\newcommand{\R}{\mathbb R}

\endlocaldefs
\begin{document}

\begin{frontmatter}
\title{Convergence and global optimality of Wasserstein--Fisher--Rao descent for Gaussian mixtures}
\runtitle{Convergence and global optimality of WFR descent}
\runauthor{Wang et al.}

\begin{aug}
\author[A]{\fnms{Xingzhou}~\snm{Wang}\thanksref{authororder}%
  \ead[label=e1]{xwang268@ncsu.edu}}
\author[B]{\fnms{Ziyang}~\snm{Jiang}%
  \ead[label=e2]{zjiang29@ncsu.edu}}
\author[C]{\fnms{Zheng}~\snm{Li}%
  \ead[label=e3]{zli42@ncsu.edu}}
\author[A]{\fnms{Mingyan}~\snm{Liu}%
  \ead[label=e4]{mliu32@ncsu.edu}}
\author[D]{\fnms{Yiling}~\snm{Wang}%
  \ead[label=e5]{ywang327@scut.edu.cn}}
\author[B]{\fnms{Jingze}~\snm{Yang}%
  \ead[label=e6]{jyang76@ncsu.edu}}

\thankstext{authororder}{The first author led the research.
The remaining authors are listed alphabetically by surname.}

\address[A]{Department of Statistics, North Carolina State University%
  \printead[presep={,\ }]{e1,e4}}
\address[B]{Department of Mathematics, North Carolina State University%
  \printead[presep={,\ }]{e2,e6}}
\address[C]{Department of Agricultural and Resource Economics, North Carolina State University%
  \printead[presep={,\ }]{e3}}
\address[D]{School of Mathematics,
South China University of Technology%
  \printead[presep={,\ }]{e5}}
\end{aug}

\begin{abstract}
We study explicit Wasserstein--Fisher--Rao descent for fitting Gaussian
location mixtures by nonparametric maximum likelihood. The full sequence
converges weakly and has finite length in the bounded-Lipschitz metric
throughout the original step-size range. These conclusions hold for any
fixed data set in any dimension and any initial probability measure. By an earlier conditional optimality theorem, the limit maximizes the likelihood under full-support initialization.

Fix any particle step size in $(0,1]$. Initialize the particles by
resampling the observations. For an open set of triples of points in
the plane, the probability of convergence to a strictly suboptimal
fixed point tends to one exponentially as the particle count grows. Without reweighting, a centered Gaussian initialization can give a strictly suboptimal limit. Every
finite iterate has a smooth, strictly positive density. The limit assigns positive mass to a nonglobal local maximum of the associated Gaussian sum. We also give the exact pure-transport limit
for a one-dimensional initialization with full support and an atom.

An exact finite-dimensional representation preserves the truncated
likelihood moments and the retained region's contribution to the current
dissipation. This lets us transfer an analytic gradient inequality to the
measure trajectory.
\end{abstract}
\end{frontmatter}

\section{Introduction}
A nonparametric maximum likelihood estimator (NPMLE) for a Gaussian
location mixture maximizes the likelihood over all mixing distributions.
An algorithm can decrease the negative log-likelihood and still converge
to a suboptimal fixed point. We study convergence and optimality for the
explicit Wasserstein--Fisher--Rao iteration of \citet{YWR2024}. We also
analyze its particle implementation with initial locations drawn by
resampling the observations. We then study a pure-transport variant that
omits the reweighting step.

Let $X_1,\ldots,X_N\in\R^d$ be fixed observations, where $N,d\ge1$, and put
\begin{equation}\label{eq:kernel}
 \phi(x)=(2\pi)^{-d/2}e^{-|x|^2/2},\qquad \psi_i(x)=\phi(x-X_i).
\end{equation}
For a Borel probability measure $\rho\in\mathcal P(\R^d)$, define
\begin{equation}\label{eq:likelihood}
 a_i(\rho)=\int\psi_i\,d\rho,\qquad
 L_X(\rho)=-\frac1N\sum_{i=1}^N\log a_i(\rho),\qquad
 L_X^*=\inf_{\rho\in\mathcal P(\R^d)}L_X(\rho).
\end{equation}
An NPMLE is a minimizer of $L_X$. Every fitted value $a_i(\rho)$
is positive. We omit the subscript $X$ when the dependence on the
data is clear. Throughout the paper, we treat the observations as
fixed. We study convergence as the number of iterations tends to infinity. For the particle result, we also let the number of particles tend to infinity.

For $a\in(0,\infty)^N$, write
\begin{equation}\label{eq:score}
 g_a(x)=\frac1N\sum_{i=1}^N\frac{\psi_i(x)}{a_i},\qquad
 g_\rho=g_{a(\rho)}.
\end{equation}
The first variation of $L$ at $\rho$ is $-g_\rho$, and
$\int g_\rho\,d\rho=1$. The explicit WFR descent of \citet{YWR2024}
alternates between the following two updates.
\begin{align}
 T_n(x)&=x+\eta\nabla g_{\rho_n}(x),\qquad
 \widetilde\rho_n=(T_n)_\#\rho_n,\label{eq:transport}\\
 \rho_{n+1}(dx)&=h_n(x)\widetilde\rho_n(dx),\qquad
 h_n(x)=1-\eta+\eta g_{\widetilde\rho_n}(x).\label{eq:reaction}
\end{align}
These are the transport and reweighting steps in Section~3.2.1 of that paper, written in terms of $g_\rho$. The reaction is evaluated at the transported measure. This distinction is used in the convergence and particle proofs below.

For a sufficiently small fixed step size and an initial measure with
full support, \citet[Theorem~2]{YWR2024} show that the loss is
nonincreasing and the iterates are tight. Every weak subsequential limit satisfies the WFR stationarity conditions. If the full sequence converges weakly, its limit is an NPMLE.
The authors leave open whether this weak convergence holds
\citep[p.~1781]{YWR2024}.

Their Algorithm~1 draws the initial particle locations independently
from the empirical distribution and gives all particles equal weight. They suggest that it may be possible to guarantee convergence with enough particles \citep[p.~1782]{YWR2024}.

We prove that the full sequence of measures converges weakly.
This holds for any initial probability measure and every fixed step
size in the original small-step range. If the initial measure has full
support, the limit is an NPMLE by \citet[Theorem~2(iii)]{YWR2024}.

In our particle example, we resample the observations to obtain
the initial particle locations. The iterates can converge to a strictly suboptimal limit even with reweighting. In our pure-transport example, the initial measure has a smooth, strictly positive density. Yet the iterates converge to a strictly suboptimal limit.

Our convergence proof transfers a finite-dimensional analytic gradient
inequality to the measure trajectory. We first restrict the measure to a fixed ball without renormalizing. The resulting auxiliary loss is an upper bound on the original loss. We then construct a finite-dimensional representation that exactly preserves the truncated likelihood moments and the contribution of the retained mass to the
original dissipation.

We use the square roots of the weights as coordinates. We add a
separate coordinate for the square root of the discarded mass.
In these coordinates, we work with a single analytic function
defined on an open neighborhood of a fixed compact set.

The gradient inequality, the descent estimate, and the step-length
estimate together show that the trajectory has finite length in the
bounded-Lipschitz metric. This holds for any initial probability
measure, including singular measures and measures with infinite
moments. The proof allows nonunique minimizers and nonisolated
critical points.

Each resampled particle starts at one of the three observed sites.
Particles that start at the same site follow the same trajectory.
For each group, we track its common position and total weight.
As the number of particles grows, the initial total weight at each
site concentrates around one third.

For a particular equilateral data configuration, we compute the Hessian
of the loss in these coordinates. We use the Hessian to show that the
split update is a contraction near an explicit three-atom fixed point.
If we start with equal total weights at the three observed sites,
the iterates enter this neighborhood.

For nearby data, the fixed point persists and remains strictly
suboptimal. The split update is still a contraction near it. When the
initial weights are nearly equal, continuity of the finite-step update
ensures that the iterates still enter the contraction neighborhood.
Once there, they converge to the suboptimal fixed point. A concentration
bound for the resampling counts shows that the probability of nearly
equal initial weights tends to one exponentially fast as the number of
particles grows.

Theorem~7 of \citet{YWR2024} requires a positive reaction parameter
$\gamma$. On p.~1793, the authors ask whether the theorem extends
to pure transport, where $\gamma=0$. In our example, the three
observations form an equilateral triangle centered at the origin.
The initial measure is a centered isotropic Gaussian. Each iterate
is invariant under rotation by $2\pi/3$. This symmetry keeps the three
fitted values equal. The transport step therefore updates each point
by gradient ascent on the same fixed sum of Gaussian kernels.
The effective step size depends on the current measure.

We choose the triangle so that the origin is an attracting local
maximum. The sum takes a higher value at three points related by
the same rotation. The initial Gaussian assigns positive mass to
the basin of attraction of the origin. Under the step-size condition
of our pure-transport theorem, every pointwise trajectory converges.
We take some of the mass that the limiting measure assigns to the origin and divide it equally among the three points. This strictly increases the likelihood.
So the optimality conclusion of Theorem~7(iii) can fail at
$\gamma=0$ even when the iterates converge.

In a separate example with two observations, we compute the exact limit
from an initial measure with full support and a stationary atom.

\subsection{Related work and attribution}
\citet[p.~1781]{YWR2024} compare their conditional optimality result
(Theorem~2(iii)) with the gradient-flow results of \citet{ChizatBach2018}.
Under full-support initialization, their theorem gives global optimality
if the full sequence of WFR iterates converges weakly. We prove this
weak convergence.

Other methods for optimizing over measures have convergence guarantees
under different assumptions. \citet{Chizat2022Sparse} gives quantitative
convergence bounds for conic descent. The analysis assumes a unique
sparse minimizer, curvature and strict slackness conditions, and
quantitative control of the initialization. \citet{Chizat2022Rates}
studies Bregman methods for convex optimization over measures and
gives rates for the objective gap. \citet{Yao2024} analyze an implicit
KL proximal method for minimizing convex functionals over probability
measures.

The Gaussian NPMLE problem also has a mass-penalized formulation. For $m>0$
and $\rho\in\mathcal P(\R^d)$, the penalized objective
$J$ satisfies
\begin{equation}\label{eq:masspenalty}
 J(m\rho)=L(\rho)-\log m+m,\qquad \min_{m>0}J(m\rho)=L(\rho)+1.
\end{equation}
The minimum over $m$ is attained at $m=1$, so the two formulations
have the same minimizing measures. The convergence guarantees depend
on the update rule and its assumptions.

For univariate Gaussian mixtures, \citet[Section~3.2.1]{FanGuanShenWu2023}
study the continuous-time Fisher--Rao flow on mixing distributions
supported on a compact interval. They show that if the initial density
is strictly positive on that interval, the negative log-likelihood
converges to its minimum over those distributions. In our convergence
proof, we use the classical {\L}ojasiewicz gradient inequality
\citep{Bolte2007}. \citet{Hauer2019} discuss how gradient inequalities can establish
finite length for gradient-flow trajectories. \citet{Cosentino2020}
use moment-preserving reduction for discrete measures. For optimal
design on a finite set, \citet{Piazzon2022} writes the weights as
squares of coordinates and compresses the computed design while
preserving selected moments. \citet[Section~4.2]{Carrillo2026}
reduce a Fisher--Rao gradient-dominance inequality for
$f$-divergences to finite-dimensional inequalities.

\citet[Theorem~1.1]{Isobe2026} studies a continuous-time
Wasserstein-type flow for an $L^2$-regularized mean-field neural
network model. Under Assumptions~1.3--1.5 of that paper, the proof
uses a {\L}ojasiewicz--Simon inequality to show that the flow
converges to a critical point.

Our atomic representation matches the truncated fitted values and the
contribution of the retained mass to the original dissipation. We bound
the discarded mass separately. The resulting gradient bound holds
with the same constant for all such representations, even when some
weights are zero. We then use the transport and reaction estimates to
transfer the analytic gradient inequality to the split iteration.

\citet{CW2003} describe Duistermaat's example of a mixture of three
isotropic Gaussians centered at the vertices of an equilateral triangle.
For a suitable common variance, the mixture has four modes.
\citet[Lemma~2]{SGS2025} choose a scale at which all four modes are global
maxima. They use this geometry to construct distinct Gaussian NPMLEs
for the same data.

Fix any step size covered by our particle result. For data triples
in an open set, the three-group particle dynamics under the split
update \eqref{eq:transport}--\eqref{eq:reaction} have a strictly
suboptimal attracting fixed point. When we initialize $m$ particles
by resampling the observations, the probability of convergence to
this point is at least $1-Ce^{-cm}$, where $C,c>0$ are independent of $m$.

In our pure-transport example, every finite iterate from the Gaussian
initialization has a smooth, strictly positive density. Yet the weak
limit is strictly suboptimal and assigns positive mass to a nonglobal
local maximum of the Gaussian sum.

\citet{Jin2016} show that the population log-likelihood of equally
weighted Gaussian mixtures can have suboptimal local maxima. They
also show that population EM can remain strictly suboptimal at every
iteration under their random initialization. \citet[Theorem~3.1]{Zhou2026} study population gradient EM with random initialization. They give recovery guarantees for well-separated
Gaussian mixtures when the algorithm is sufficiently overparameterized
and the weight updates are near-optimal. For certain high-dimensional mixtures, \citet[Theorem~2]{ZhangGradientEM2026}
show that the population KL loss of a hybrid EM scheme can stay bounded
away from zero for a number of iterations that grows exponentially with
the dimension. This can occur even with overparameterization.

\citet{PolyanskiySellke2025} study certified computation and generic
structural properties of the univariate Gaussian NPMLE.
\citet{Wang2026Modes} proves that every finite Gaussian mixture whose
components share a covariance matrix has finitely many critical
points. As a consequence, every Gaussian location NPMLE in this
setting has finite support. \citet{Wang2026Unique} proves that the
Gaussian location NPMLE is unique for Lebesgue-almost every data set.
Finite Gaussian mixtures with unequal component covariances also
have finitely many critical points \citep{AmendolaRodriguez2026,Okuno2026}.

Our convergence and counterexample proofs are independent of these
results. We show directly in Lemma~\ref{sw:geometry} that the Gaussian
sum in the symmetric pure-transport example has finitely many critical
points. After the main proofs, we use Wang's theorem
\citep[Theorem~1]{Wang2026Modes} and stationarity of the limit to show
that the limit of the split iteration has finite support.

\citet{Gallouet2017} develop a JKO scheme for Kantorovich--Fisher--Rao
gradient flows with successive implicit Wasserstein and Fisher--Rao
steps. For sampling from a prescribed target, \citet{Crucinio2026}
analyze time discretizations that compose the exact Wasserstein and
Fisher--Rao subflows. \citet{CrucinioLogConcavity2026} study preservation
of strong log-concavity and convergence for the continuous-time WFR flow.
\citet[Theorem~3.5]{DuanHertrichSteidl2026} analyze an exact implicit
WFR--JKO scheme on finite positive measures. With a fixed step size
and a strictly positive target density, the extended
Kullback--Leibler divergence decays exponentially along the iterates.

Section~2 states the three main theorems and the one-dimensional
pure-transport result. Sections~\ref{sec:compression}
and~\ref{sec:finite-length} prove that the split iteration has finite
length and converges. Section~\ref{sec:particles} proves the particle
result. Section~\ref{sec:smooth-W} presents the pure-transport
counterexample. Its proof is in Section~6 of the Supplement.
Section~7 discusses the role of initialization. It shows that the
number of iterations needed to reach a sufficiently small likelihood
gap cannot be bounded uniformly over the class considered there.

The Supplement gives estimates for the transport and reaction steps
and a uniform admissible step-size bound for empirical initializations.
It provides detailed calculations for the Hessian, the Gaussian
geometry, and the concentration of the resampling counts. It also
contains the proofs of the Wasserstein convergence corollary, the
one-dimensional result, and the nonuniformity corollary.

\section{Main results}
Let
\begin{equation}\label{eq:kernelbounds}
 \begin{gathered}
 M=\phi(0),\qquad M_1=\sup_x|\nabla\phi(x)|,\qquad
 M_2=\sup_x\|\nabla^2\phi(x)\|_{\mathrm{op}},\\
 D_X=\operatorname{diam}\{X_1,\ldots,X_N\}.
 \end{gathered}
\end{equation}
Given $\rho_0\in\mathcal P(\R^d)$, define
\begin{equation}\label{eq:threshold}
 \begin{gathered}
 b=e^{-L(\rho_0)},\qquad r_0=\sqrt{2\log(2M/b)},\qquad
 c_0=\frac b2e^{-(r_0+D_X)^2/2},\\
 \eta_0=\frac{c_0}{M_2+M_1^2/c_0}>0.
 \end{gathered}
\end{equation}
The threshold $\eta_0$ is the Gaussian step-size bound from
\citet[Remark~2]{YWR2024}. The initial measure enters this bound
through $L(\rho_0)$. For $\mu,\nu\in\mathcal P(\R^d)$, define the
bounded-Lipschitz distance by
\begin{equation}\label{eq:bl}
 d_{\mathrm{BL}}(\mu,\nu)=\sup_{\|f\|_\infty\le1,\,\operatorname{Lip}(f)\le1}
 \left|\int f\,d\mu-\int f\,d\nu\right|.
\end{equation}
Convergence in $d_{\mathrm{BL}}$ is equivalent to weak convergence
on $\mathcal P(\R^d)$.

\begin{theorem}\label{thm:convergence}
Let $X_1,\ldots,X_N\in\R^d$ be fixed and let
$\rho_0\in\mathcal P(\R^d)$. Fix $0<\eta\le\eta_0$, where $\eta_0$
is defined in \eqref{eq:threshold}. The iterates
\eqref{eq:transport}--\eqref{eq:reaction} lie in $\mathcal P(\R^d)$
for every $n\ge0$ and satisfy
\begin{equation}\label{eq:finlength}
 \sum_{n=0}^{\infty}d_{\mathrm{BL}}(\rho_{n+1},\rho_n)<\infty.
\end{equation}
The full sequence converges weakly to a measure
$\widehat\rho\in\mathcal P(\R^d)$. The limit satisfies
\begin{equation}\label{eq:stationary}
 g_{\widehat\rho}=1,\qquad \nabla g_{\widehat\rho}=0
 \quad\widehat\rho\text{-almost everywhere}.
\end{equation}
If $\operatorname{supp}(\rho_0)=\R^d$, then
\[
 L(\widehat\rho)=\inf_{\rho\in\mathcal P(\R^d)}L(\rho).
\]
No moment, absolute-continuity, separation, uniqueness or nondegeneracy assumption is required.
\end{theorem}
\noindent The proof is given in Sections~\ref{sec:compression}
and~\ref{sec:finite-length}, with the substep estimates proved in
Appendix~A of the Supplement.

\begin{corollary}[Wasserstein convergence]\label{cor:wp}
Assume the hypotheses of Theorem~\ref{thm:convergence} and fix
$1\le p<\infty$. If $\int|x|^p\,d\rho_0(x)<\infty$, then
$W_p(\rho_n,\widehat\rho)\to0$. If $\rho_0$ also has full support,
then $\widehat\rho$ is an NPMLE.
\end{corollary}
\noindent The proof of Corollary~\ref{cor:wp} is given in the
Supplement.
\setcounter{equation}{5}\setcounter{theorem}{2}\setcounter{subsection}{0}

The next theorem shows that the particle iteration can converge to a
strictly suboptimal limit. Let $d=2$, $N=3$, and define
\begin{equation}\label{eq:triangle}
 \begin{gathered}
 e_1=(1,0),\quad e_2=(-1/2,\sqrt3/2),\quad e_3=(-1/2,-\sqrt3/2),\\
 r_\triangle=\sqrt{\tfrac53\log3},\quad X_i^\triangle=r_\triangle e_i,\quad
 r_*=2r_\triangle/5,\\
 \rho_*=\tfrac13\sum_{j=1}^3\delta_{r_*e_j},\quad
 \delta_\triangle=\tfrac{22}{15}\log3-\log5>0.
 \end{gathered}
\end{equation}
For a data triple $X=(X_1,X_2,X_3)$ and each particle count $m$,
we use the resampling initialization
\begin{equation}\label{eq:particleinit}
 J_1,\ldots,J_m\overset{\mathrm{iid}}\sim\operatorname{Unif}\{1,2,3\},\qquad
 \rho_0^{(m)}=\frac1m\sum_{\ell=1}^m\delta_{X_{J_\ell}}.
\end{equation}
All probabilities in the theorem are conditional on fixed observations.

\begin{theorem}[Suboptimal particle limits under resampling]\label{thm:particles}
For every fixed $\eta\in(0,1]$, there exist an open neighborhood $U_\eta\subset(\R^2)^3$ of $X^\triangle$, a constant $0<\varepsilon_\eta\le 1/12$, and smooth maps
\[
 \begin{gathered}
 p_j:U_\eta\to(0,1),\qquad z_j:U_\eta\to\R^2,\qquad j=1,2,3,\\
 \sum_{j=1}^3p_j(X)=1,\qquad z_j(X)\ne z_k(X)\ (j\ne k),\qquad
 \rho_{\mathrm{bad}}(X)=\sum_{j=1}^3p_j(X)\delta_{z_j(X)},
 \end{gathered}
\]
such that:
\begin{enumerate}[(i)]
\item $\rho_{\mathrm{bad}}(X^\triangle)=\rho_*$ and $\rho_{\mathrm{bad}}(X)$ is a fixed point of \eqref{eq:transport}--\eqref{eq:reaction};
\item $L_X(\rho_{\mathrm{bad}}(X))-L_X^*\ge\delta_\triangle/2$ for all $X\in U_\eta$;
\item for every $X\in U_\eta$ and $m\ge1$,
\begin{equation}\label{eq:particleprob}
 \begin{split}
 \mathbb P\bigl\{\rho_n^{(m)}\Rightarrow\rho_{\mathrm{bad}}(X),\quad
  \lim_{n\to\infty}L_X(\rho_n^{(m)})-L_X^*\ge\delta_\triangle/2\bigr\}
 \ge1-6e^{-2m\varepsilon_\eta^2}.
 \end{split}
\end{equation}
On the count event
\[
 \mathcal A_m=\left\{\max_{1\le j\le3}\left|
 \frac1m\sum_{\ell=1}^m\mathbf1_{\{J_\ell=j\}}-\frac13\right|
 \le\varepsilon_\eta\right\},
\]
let $z_{n,j}$ and $p_{n,j}$ be the position and total weight at
iteration $n$ of the group initialized at $X_j$. Then
\[
 \rho_n^{(m)}=\sum_{j=1}^3p_{n,j}\delta_{z_{n,j}},
 \qquad
 z_{0,j}=X_j,\qquad
 p_{0,j}=\frac1m\sum_{\ell=1}^m\mathbf1_{\{J_\ell=j\}}.
\]
There exist $K\in\mathbb N_0$, $C>0$, and $0<\theta<1$, which may
depend on $\eta$ but are uniform over $X\in U_\eta$, $m\ge1$, and all
realizations in $\mathcal A_m$, such that
\[
 \max_{1\le j\le3}
 \bigl\{|z_{n,j}-z_j(X)|+|p_{n,j}-p_j(X)|\bigr\}
 \le C\theta^{n-K}\qquad(n\ge K).
\]
\end{enumerate}
If the initialization for each $m$ uses the first $m$ terms of one infinite iid index sequence, then
\[
 \mathbb P\!\left(\bigcup_{m_0=1}^{\infty}\bigcap_{m\ge m_0}\mathcal A_m\right)=1;
\]
thus the convergence and likelihood conclusions hold almost surely for every sufficiently large $m$.
\end{theorem}

Theorem~\ref{thm:particles} holds for every fixed $\eta\in(0,1]$.
Appendix~B of the Supplement gives a common bound
$\eta_{\mathrm{adm}}>0$. For $0<\eta<\eta_{\mathrm{adm}}$, there is
an open neighborhood of $X^\triangle$ on which $\eta\le\eta_0$ for
every resampling initialization in \eqref{eq:particleinit}. On the
intersection of this neighborhood with $U_\eta$, both
Theorems~\ref{thm:convergence} and~\ref{thm:particles} apply to the
same update. The data neighborhood and $\varepsilon_\eta$ may depend
on $\eta$.

Under the conditions of Theorem~\ref{thm:particles}, the probability
of convergence to a suboptimal limit tends to one as $m\to\infty$.
Increasing the number of draws from the empirical distribution alone
therefore does not guarantee global optimality. For every $m$, the
initial measure is supported on the observed sites, so it falls
outside the full-support assumption. We obtain a lower bound on
the likelihood gap by comparing the limit with $\delta_0$. This
comparison requires no uniqueness assumption for the NPMLE.

\subsection{Removing reweighting}
The preceding theorems concern the same transport--reaction iteration.
We now start from a measure with full support and omit the reaction
step. In the notation of \citet[Definition~4]{YWR2024}, the split update
\eqref{eq:transport}--\eqref{eq:reaction} uses $\gamma=\eta>0$.
Pure transport sets $\gamma=0$ and updates the measure by
\begin{equation}\label{eq:pure-W}
 \rho_{n+1}=(\operatorname{id}+\eta\nabla g_{\rho_n})_\#\rho_n.
\end{equation}
Theorem~7 of \citet{YWR2024} assumes $\gamma\in(0,1)$. On p.~1793,
the authors ask whether it extends to $\gamma=0$. The next theorem
shows that the global optimality conclusion in part~(iii) can fail
at $\gamma=0$. Even from a centered Gaussian initialization, the
pure-transport iterates can converge to a strictly suboptimal limit.

\begin{theorem}[Pure transport from a smooth positive density]\label{thm:smooth-W}
Let
$d=2$, $N=3$, $R=\sqrt{\log7}$, $X_i=Re_i$ $(i=1,2,3)$, and $\rho_0=\mathcal N(0,s^2I_2)$ with $s>0$,
where $e_i$ are the vectors in \eqref{eq:triangle}. Initialize \eqref{eq:pure-W} at $\rho_0$, and put
\begin{equation}\label{eq:smooth-constants}
 \begin{gathered}
 M=(2\pi)^{-1},\qquad
 F(x)=\frac13\sum_{i=1}^3\phi(x-X_i),\qquad F_0=F(0)=M/\sqrt7,\\
 c_{\mathrm{init}}=\frac{M}{1+s^2}
       \exp\!\left\{-\frac{\log7}{2(1+s^2)}\right\},\qquad
 \kappa=F_0(1-\tfrac12\log7),\qquad r=\frac{\kappa}{8M},\\
 p_s=1-e^{-r^2/(2s^2)},\qquad
H=\frac{F_0}{3}\bigl(7^{3/8}+2\cdot 7^{-3/8}\bigr)>F_0.
\end{gathered}
\end{equation}
For every fixed $0<\eta\le c_{\mathrm{init}}/(2M)$,
\[
 d\rho_n(x)=f_n(x)\,dx,\qquad f_n\in C^\infty(\R^2),\qquad
 f_n(x)>0\quad(x\in\R^2,\ n\ge0).
\]
There exists $\rho_\infty\in\mathcal P(\R^2)$ such that
\[
 \begin{gathered}
 \#\operatorname{supp}(\rho_\infty)<\infty,\qquad \rho_n\Rightarrow\rho_\infty,\\
 W_p(\rho_n,\rho_\infty)\longrightarrow0\quad(1\le p<\infty),
 \end{gathered}
\]
and
\begin{equation}\label{eq:smooth-gap}
 \rho_\infty(\{0\})\ge p_s>0,\qquad
 L_X(\rho_\infty)-L_X^*
 \ge\log\!\left(1+\frac{p_s(H-F_0)}{M}\right)>0.
\end{equation}
In particular, $\rho_\infty$ is not an NPMLE. For this initialization, the threshold in \eqref{eq:threshold} satisfies
$(0,\eta_0]\subseteq(0,c_{\mathrm{init}}/(2M)]$.
\end{theorem}

This counterexample uses symmetric observations and a smooth, atom-free
Gaussian initialization. The iterates converge. Symmetry reduces
\eqref{eq:pure-W} to gradient ascent on a fixed Gaussian sum. The
initial measure assigns positive mass to the basin of a nonglobal
local maximum.

The exact one-dimensional comparison shows how a stationary atom at
zero can prevent optimality. Starting from $\mathcal N(0,s^2)$ with
$s>0$, the same orbit argument gives convergence to the unique NPMLE.
The step-size bound is computed for this Gaussian initialization.
\begin{proposition}[A full-support initialization with a persistent atom]\label{prop:atomic-W}
Let
\[
 \begin{gathered}
 d=1,\qquad N=2,\qquad X_1=-a,\qquad X_2=a,\qquad a>1,\\
 \rho_0=\alpha\delta_0+(1-\alpha)\mathcal N(0,s^2),
 \qquad 0<\alpha<1,\quad s>0.
 \end{gathered}
\]
Put
\[
 \begin{gathered}
 h(x)=\{\phi(a-x)+\phi(a+x)\}/2,\qquad M=(2\pi)^{-1/2},\\
 c_{\mathrm{init}}=\alpha\phi(a)
 +\frac{1-\alpha}{\sqrt{1+s^2}}\phi\!\left(\frac a{\sqrt{1+s^2}}\right).
 \end{gathered}
\]
The equation $b=a\tanh(ab)$ has a unique solution $b\in(0,a)$.
Fix $0<\eta\le c_{\mathrm{init}}/(2M)$. The iterates of
\eqref{eq:pure-W} satisfy $\operatorname{supp}(\rho_n)=\R$
for every $n\ge0$. For every $1\le p<\infty$, they converge in $W_p$ to the measure
\begin{equation}\label{eq:atomic-limit}
 \rho_\infty=\alpha\delta_0+\frac{1-\alpha}{2}(\delta_{-b}+\delta_b).
\end{equation}
Moreover,
\[
 \operatorname*{arg\,min}_{\rho\in\mathcal P(\R)}L_X(\rho)
 =\{\rho^*\},\qquad \rho^*=\tfrac12(\delta_{-b}+\delta_b),
\]
and
\begin{equation}\label{eq:atomic-gap}
 L_X(\rho_\infty)-L_X(\rho^*)
 =\log\frac{h(b)}{\alpha h(0)+(1-\alpha)h(b)}>0.
\end{equation}
When $\alpha=0$, let $\rho_0=\mathcal N(0,s^2)$ and evaluate
$c_{\mathrm{init}}$ at $\alpha=0$. For any fixed
$0<\eta\le c_{\mathrm{init}}/(2M)$, the iterates satisfy
$W_p(\rho_n,\rho^*)\to0$ for every $1\le p<\infty$.
\end{proposition}

\noindent The proof is given in Appendix~E of the Supplement.

In Proposition~\ref{prop:atomic-W}, the initial measure has full support,
but its atom at zero persists and the limit is suboptimal.
Theorem~\ref{thm:smooth-W} shows that pure transport can also converge
to a suboptimal limit from a smooth, atom-free initialization.

\section{A trajectory inequality by exact analytic compression}
\label{sec:compression}

Theorems~\ref{thm:convergence} and~\ref{thm:particles} concern the same explicit iteration, but
different properties of its limit.  We first prove convergence from an arbitrary
initial probability measure.  Sections~\ref{sec:compression}
and~\ref{sec:finite-length} yield $\rho_n\Rightarrow\widehat\rho$ and
\eqref{eq:stationary}, but not $L(\widehat\rho)=L^*$, which can fail by
Theorem~\ref{thm:particles}.  Full support is used only at the end of
Section~\ref{sec:finite-length}, to apply \citet[Theorem~2(iii)]{YWR2024}.

\subsection{Dissipation and the two substeps}
\label{sec:cv-estimates}
For the fixed initial measure, let $c_0$ and $\eta_0$ be the constants in
\eqref{eq:threshold}.  Write
\begin{equation}\label{cv:kernel-bounds}
 M=\phi(0),\qquad M_1=\sup_x|\nabla\phi(x)|,\qquad
 M_2=\sup_x\|\nabla^2\phi(x)\|_{\mathrm{op}}.
\end{equation}
Thus $0<M_1<\infty$ and $0<M\le M_2<\infty$.  For
$\rho\in\mathcal P(\mathbb R^d)$, define
\begin{equation}\label{cv:dissipation}
 \mathcal D(\rho)=\int\bigl\{|\nabla g_\rho(x)|^2+
                   (g_\rho(x)-1)^2\bigr\}\,\rho(dx).
\end{equation}
At step $n$, the quantities that occur directly in the two descent estimates are
\begin{equation}\label{cv:split-residuals}
 G_n=\int|\nabla g_{\rho_n}|^2\,d\rho_n,\qquad
 R_n=\int(g_{\widetilde\rho_n}-1)^2\,d\widetilde\rho_n,
 \qquad E_n=G_n+R_n.
\end{equation}
Note that the reaction residual is evaluated at the transported measure;
with a residual at $\rho_n$ instead, \eqref{cv:descent} would not follow
from the reaction estimate in Appendix~A of the Supplement.

\begin{lemma}\label{cv:sublevel}
For $\rho\in\mathcal P(\mathbb R^d)$ with $L(\rho)\le L(\rho_0)$,
\begin{equation}\label{cv:ai-bounds}
 c_0\le a_i(\rho)\le M\quad(1\le i\le N),
\end{equation}
and
\begin{equation}\label{cv:g-bounds}
 \|g_\rho\|_\infty\le M/c_0,\qquad
 \|\nabla g_\rho\|_\infty\le M_1/c_0,\qquad
 \|\nabla^2g_\rho\|_\infty\le M_2/c_0.
\end{equation}
Moreover,
\begin{equation}\label{cv:eta-bounds}
 \eta_0\le\tfrac12,\qquad \eta_0M/c_0<1,\qquad
 \eta_0M_2/c_0<1.
\end{equation}
\end{lemma}

\begin{proposition}
\label{cv:split-estimates}
For $0<\eta\le\eta_0$,
\[
 \rho_n,\widetilde\rho_n\in\mathcal P(\mathbb R^d),\qquad
 L(\rho_n),L(\widetilde\rho_n)\le L(\rho_0)
 \quad(n\ge0).
\]
There exist $c,C>0$ such that, for every $n\ge0$,
\begin{align}
 L(\rho_n)-L(\rho_{n+1})&\ge cE_n,
                                      \label{cv:descent}\\
 \mathcal D(\rho_n)&\le CE_n,
                                      \label{cv:residual-comparison}\\
 d_{\mathrm{BL}}(\rho_{n+1},\rho_n)&\le C\sqrt{E_n}.
                                      \label{cv:step-length}
\end{align}
\end{proposition}

The sublevel and descent estimates come from the analysis of
\citet{YWR2024}.  Appendix~A of the Supplement gives their
Gaussian derivation, the endpoint $\eta=\eta_0$, the comparison
\eqref{cv:residual-comparison} needed here, and the step bound
\eqref{cv:step-length}.  Since $a_i\le M$, we have $L\ge-\log M$.  Thus
\eqref{cv:descent} and \eqref{cv:residual-comparison} give
\begin{equation}\label{cv:loss-limit}
 L(\rho_n)\downarrow L_\infty\in\mathbb R,\qquad
 \sum_{n=0}^\infty E_n<\infty,\qquad
 \mathcal D(\rho_n)\longrightarrow0.
\end{equation}
The estimate $\sum_nE_n<\infty$ does not by itself give finite trajectory
length.  To obtain $\sum_n\sqrt{E_n}<\infty$, we establish a gradient
inequality with constants uniform over an auxiliary compact parameter set.
\subsection{A uniform gradient inequality near a level set}
We use the finite-dimensional {\L}ojasiewicz gradient inequality for
real-analytic functions \citep{Bolte2007}. Near a critical point $z_*$,
it gives $|H(z)-H(z_*)|^\theta\le C\|\operatorname{grad}H(z)\|$
for some $\theta\in[1/2,1)$. Here $\operatorname{grad}H$ denotes the
Riemannian gradient on the manifold. The compact set in the lemma
specifies where we apply the inequality.

\begin{lemma}[Compact-level uniformization]\label{cv:compact-level}
Let $\mathscr M$ be a finite-dimensional real-analytic manifold with an
analytic Riemannian metric.  Suppose that $\mathscr K\subset\mathscr M$ is compact, that $H$ is real
analytic on a neighborhood of $\mathscr K$, and that
$\mathscr K\cap H^{-1}(\{\lambda\})\ne\varnothing$.  Then there exist
$\theta\in[1/2,1)$, $C>0$ and $\delta>0$ such that
\begin{equation}\label{cv:compact-gradient}
 |H(z)-\lambda|^\theta\le C\|\operatorname{grad}H(z)\|
 \quad\text{for }z\in\mathscr K,\quad |H(z)-\lambda|<\delta.
\end{equation}
\end{lemma}

\noindent The complete proof of Lemma~\ref{cv:compact-level} is given in the Supplement.
\setcounter{equation}{12}\setcounter{theorem}{3}\setcounter{subsection}{2}

\subsection{Truncation with an error of fixed sign}
\label{sec:cv-truncation}
Fix the ball
\begin{equation}\label{cv:fixed-ball}
 R=\max_i|X_i|+\sqrt{2\log(2M/c_0)},\qquad
 B_R=\{x\in\mathbb R^d:|x|\le R\}.
\end{equation}
For $|x|>R$ and every $i$,
\[
 |x-X_i|\ge |x|-|X_i|>\sqrt{2\log(2M/c_0)},\qquad
 \psi_i(x)=Me^{-|x-X_i|^2/2}\le c_0/2.
\]
On the likelihood sublevel, the bound $a_i(\rho)\ge c_0$ gives
\begin{equation}\label{cv:tail-g}
 0<g_\rho(x)=\frac1N\sum_i\frac{\psi_i(x)}{a_i(\rho)}
 \le\frac12\qquad(x\notin B_R).
\end{equation}
Let
\begin{equation}\label{cv:truncation-definitions}
 m=\rho(B_R^c),\qquad a_i^R=\int_{B_R}\psi_i\,d\rho,
 \qquad L^R(\rho)=-\frac1N\sum_i\log a_i^R.
\end{equation}
We compute $a_i^R$ using the unnormalized restriction of $\rho$ to
$B_R$. If $\tau_i=a_i(\rho)-a_i^R$, then
\begin{equation}\label{cv:tail-moment}
 0\le\tau_i=\int_{B_R^c}\psi_i\,d\rho\le(c_0/2)m,
 \qquad
 a_i^R\ge c_0-(c_0/2)m\ge c_0/2.
\end{equation}
The reaction component controls the discarded mass.
\begin{equation}\label{cv:tail-control}
 \mathcal D(\rho)\ge\int_{B_R^c}(1-g_\rho)^2\,d\rho
 \ge\frac14\rho(B_R^c)=m/4.
\end{equation}
By $\log(1+t)\le t$ for $t\ge0$,
\begin{align}
 0\le L^R(\rho)-L(\rho)
 &=\frac1N\sum_i\log\left(1+\frac{\tau_i}{a_i^R}\right)
 \le\frac1N\sum_i\frac{\tau_i}{a_i^R}
 \le m\le4\mathcal D(\rho).
 \label{cv:truncated-loss-error}
\end{align}
Thus $L^R(\rho)\ge L(\rho)$, and the original dissipation controls
the difference.

\subsection{Exact representation of the moments and dissipation}
\label{sec:cv-compression}
For the current likelihood vector $a=a(\rho)$, put
\begin{equation}\label{cv:da}
 d_a(x)=|\nabla g_a(x)|^2+(g_a(x)-1)^2,
 \qquad F_a(x)=(\psi_1(x),\ldots,\psi_N(x),d_a(x)).
\end{equation}
The function $F_a:B_R\to\mathbb R^{N+1}$ is continuous. The retained
mass $t=1-m$ is positive, because
\begin{equation}\label{cv:inside-positive}
 c_0/2\le a_i^R=\int_{B_R}\psi_i\,d\rho\le Mt,
 \qquad t\ge c_0/(2M)>0.
\end{equation}
Let $\nu=\rho|_{B_R}/t$.

We give the finite-dimensional reduction explicitly. Consider
\[
 v=\sum_{j=1}^{q}\alpha_jF_a(x_j),\qquad
 x_j\in B_R,\quad \alpha_j>0,\quad \sum_{j=1}^{q}\alpha_j=1.
\]
If $q>N+2$, the vectors $(1,F_a(x_j))\in\mathbb R^{N+2}$ are linearly
dependent. Thus there is a nonzero $\beta\in\mathbb R^q$ such that
\[
 \sum_{j=1}^{q}\beta_j=0,\qquad
 \sum_{j=1}^{q}\beta_jF_a(x_j)=0.
\]
Hence $\max_j\beta_j>0$. Set
\[
 s=\min_{\beta_j>0}\frac{\alpha_j}{\beta_j}>0,
 \qquad \alpha'_j=\alpha_j-s\beta_j.
\]
Then
\begin{align*}
 \alpha'_j&\ge0\quad(1\le j\le q),\qquad
 \min_{1\le j\le q}\alpha'_j=0,\\
 \sum_{j=1}^{q}\alpha'_j
 &=\sum_{j=1}^{q}\alpha_j-s\sum_{j=1}^{q}\beta_j=1,\\
 \sum_{j=1}^{q}\alpha'_jF_a(x_j)
 &=v-s\sum_{j=1}^{q}\beta_jF_a(x_j)=v.
\end{align*}
Repeated elimination gives a representation with at most $N+2$ points.
In particular, with
$\Delta_{N+1}=\{\alpha\in[0,1]^{N+2}:\sum_j\alpha_j=1\}$,
\[
 \operatorname{conv}F_a(B_R)
 =\left\{\sum_{j=1}^{N+2}\alpha_jF_a(x_j):
     (\alpha,x)\in\Delta_{N+1}\times B_R^{N+2}\right\}.
\]
The right-hand side is a continuous image of a compact set and hence
is compact.

The integral $\int F_a\,d\nu$ lies in this convex hull. To see this,
fix $\epsilon>0$. By uniform continuity, we can choose a finite Borel
partition $(A_\ell)_\ell$ of $B_R$ and points $y_\ell\in A_\ell$ such
that $\sup_{x\in A_\ell}|F_a(x)-F_a(y_\ell)|\le\epsilon$ for every
$\ell$. Then
\[
 \left|\int F_a\,d\nu-\sum_\ell\nu(A_\ell)F_a(y_\ell)\right|
 \le\sum_\ell\int_{A_\ell}|F_a(x)-F_a(y_\ell)|\,\nu(dx)
 \le\epsilon.
\]
Writing $C_a=\operatorname{conv}F_a(B_R)$, the preceding estimate gives
\[
 \operatorname{dist}\left(\int F_a\,d\nu,C_a\right)\le\epsilon
 \quad(\epsilon>0),\qquad
 \int F_a\,d\nu\in\overline{C_a}=C_a.
\]
Take $K=N+2$, adding zero coefficients if necessary.  There exist
$x_j\in B_R$ and $\alpha_j\ge0$ with
\[
 \sum_{j=1}^K\alpha_j=1,\qquad
 \sum_{j=1}^K\alpha_jF_a(x_j)=\int F_a\,d\nu,
 \qquad w_j=t\alpha_j.
\]
Since $t\nu=\rho|_{B_R}$, these weights satisfy
\begin{align}
 \sum_{j=1}^K w_j&=1-m, \label{cv:compression-mass}\\
 \sum_{j=1}^Kw_j\psi_i(x_j)&=a_i^R\quad(1\le i\le N),
                                      \label{cv:compression-moments}\\
 \sum_{j=1}^Kw_jd_a(x_j)&=\int_{B_R}d_a\,d\rho.
                                      \label{cv:compression-dissipation}
\end{align}
We use the full fitted vector $a=a(\rho)$ in $d_a$.
Equation~\eqref{cv:compression-dissipation} therefore matches the
contribution of the retained mass to the original dissipation.

Use index $j=0$ for the discarded mass $w_0=m$. This index has no
associated location and contributes zero to each likelihood sum.
Write $w_j=s_j^2$ for $0\le j\le K$. Since the weights sum to one,
$s=(s_0,\ldots,s_K)$ lies on the unit sphere $\mathbb S^K$.
On $\mathscr M=\mathbb S^K\times(\mathbb R^d)^K$, set
\begin{equation}\label{cv:analytic-H}
 Q_i(s,x)=\sum_{j=1}^Ks_j^2\psi_i(x_j),\qquad
 H(s,x)=-\frac1N\sum_{i=1}^N\log Q_i(s,x),
\end{equation}
where $H$ is defined and real analytic on the open set
$\{(s,x)\in\mathscr M: Q_i(s,x)>0\text{ for every }i\}$.
All representatives constructed above lie in the compact set
\begin{equation}\label{cv:compact-K}
 \mathscr K=\left\{(s,x)\in\mathbb S^K\times B_R^K:
                       Q_i(s,x)\ge c_0/2\ (1\le i\le N)\right\}.
\end{equation}
The set $\{Q_i>c_0/4\text{ for every }i\}$ is an open neighborhood
of $\mathscr K$ on which $H$ is real analytic. For each representative
$z=z(\rho)$,
\begin{equation}\label{cv:represented-energy}
 Q_i(z)=a_i^R,\qquad H(z)=L^R(\rho).
\end{equation}
The representative and the moment used to match dissipation may
depend on $\rho$. The same function $H$ and the same compact set
$\mathscr K$ apply throughout the sublevel.

Equip $\mathscr M$ with the standard product metric. Set $V_0=0$ and
$V_j=g_Q(x_j)$ for $1\le j\le K$. Direct differentiation gives
\begin{equation}\label{cv:ambient-weight-gradient}
 \partial_{s_0}H=0,\qquad
 \partial_{s_j}H
 =-\frac1N\sum_i\frac{2s_j\psi_i(x_j)}{Q_i}
 =-2s_jV_j,
\end{equation}
\begin{equation}\label{cv:location-gradient}
 \nabla_{x_j}H
 =-\frac1N\sum_i\frac{s_j^2\nabla\psi_i(x_j)}{Q_i}
 =-w_j\nabla g_Q(x_j).
\end{equation}
Moreover,
\begin{equation}\label{cv:radial-identity}
 \sum_{j=0}^Ks_j^2V_j
 =\frac1N\sum_i\frac{\sum_{j=1}^Ks_j^2\psi_i(x_j)}{Q_i}=1,
 \qquad \langle\nabla_sH,s\rangle=-2.
\end{equation}
Projecting $\nabla_s H$ onto the tangent space of $\mathbb S^K$ gives
\[
 \operatorname{grad}_{\mathbb S^K}H
 =(I-ss^{\mathsf T})\nabla_sH
 =\nabla_sH-\langle\nabla_sH,s\rangle s
 =\nabla_sH+2s.
\]
Together with the location derivatives, this gives
\begin{equation}\label{cv:full-gradient}
 \begin{split}
 (\operatorname{grad}_{\mathbb S^K}H)_0&=2s_0,\\
 (\operatorname{grad}_{\mathbb S^K}H)_j&=-2s_j(g_Q(x_j)-1)
                              \quad(1\le j\le K),\\
 \nabla_{x_j}H&=-w_j\nabla g_Q(x_j)\quad(1\le j\le K).
 \end{split}
\end{equation}
The $j=0$ coordinate represents the discarded mass and contributes
$4s_0^2=4m$ to the squared gradient norm.

\begin{lemma}[Control of the lifted gradient]\label{cv:lifted-gradient}
For every $\rho\in\mathcal P(\mathbb R^d)$ with
$L(\rho)\le L(\rho_0)$, there exists $z\in\mathscr K$ satisfying
\eqref{cv:compression-mass}--\eqref{cv:compression-dissipation} and
\begin{equation}\label{cv:lifted-bound}
 H(z)=L^R(\rho)\ge L(\rho),\qquad
 \|\operatorname{grad}H(z)\|^2\le C\mathcal D(\rho).
\end{equation}
The constant is uniform over the sublevel and over all representatives
satisfying \eqref{cv:compression-mass}--\eqref{cv:compression-dissipation}.
\end{lemma}

\begin{proof}
Let $a=a(\rho)$, $Q=a^R$, $A=M/c_0$ and $B_1=M_1/c_0$.  By
\eqref{cv:tail-moment},
\begin{equation}\label{cv:inverse-error}
 0\le\frac1{Q_i}-\frac1{a_i}
 =\frac{a_i-Q_i}{a_iQ_i}
 \le\frac{(c_0/2)m}{c_0(c_0/2)}=m/c_0.
\end{equation}
Termwise comparison on the entire space gives
\begin{align}
 \|g_Q-g_a\|_\infty
 &\le\frac1N\sum_i M\left|\frac1{Q_i}-\frac1{a_i}\right|
 \le Am, \label{cv:uniform-g-error}\\
 \|\nabla g_Q-\nabla g_a\|_\infty
 &\le\frac1N\sum_i M_1\left|\frac1{Q_i}-\frac1{a_i}\right|
 \le B_1m. \label{cv:uniform-gradient-error}
\end{align}
In the sums below, $j$ ranges from $1$ to $K$. Since $0\le w_j\le1$
and $\sum_{j=1}^K w_j=1-m\le1$, the inequality
$|u+v|^2\le2|u|^2+2|v|^2$ yields
\begin{align}
 \sum_jw_j(g_Q(x_j)-1)^2
 &\le2\sum_jw_j(g_a(x_j)-1)^2
       +2\sum_jw_j(g_Q(x_j)-g_a(x_j))^2 \notag\\
 &\le2\sum_jw_j(g_a(x_j)-1)^2
       +2A^2m^2\sum_jw_j \notag\\
 &\le2\sum_jw_j(g_a(x_j)-1)^2+2A^2m^2,
                                      \label{cv:weighted-reaction}\\
 \sum_jw_j^2|\nabla g_Q(x_j)|^2
 &\le2\sum_jw_j^2|\nabla g_a(x_j)|^2+2B_1^2m^2\sum_jw_j^2
 \notag\\
 &\le2\sum_jw_j|\nabla g_a(x_j)|^2+2B_1^2m^2.
                                      \label{cv:weighted-location}
\end{align}
Using \eqref{cv:full-gradient} and the preserved dissipation moment, we obtain
\begin{align}
 \|\operatorname{grad}H(z)\|^2
 &=4m+4\sum_jw_j(g_Q(x_j)-1)^2
                   +\sum_jw_j^2|\nabla g_Q(x_j)|^2 \notag\\
 &\le4m+8\sum_jw_j(g_a(x_j)-1)^2
          +2\sum_jw_j|\nabla g_a(x_j)|^2+(8A^2+2B_1^2)m^2 \notag\\
 &\le4m+8\sum_jw_jd_a(x_j)+(8A^2+2B_1^2)m^2 \notag\\
 &=4m+8\int_{B_R}d_a\,d\rho+(8A^2+2B_1^2)m^2 \notag\\
 &\le\bigl\{16+8+4(8A^2+2B_1^2)\bigr\}\mathcal D(\rho) \notag\\
 &=(24+32A^2+8B_1^2)\mathcal D(\rho).
                                      \label{cv:expanded-lifted-bound}
\end{align}
Here $m^2\le m\le4\mathcal D(\rho)$ and
$\int_{B_R}d_a\,d\rho\le\mathcal D(\rho)$ give the last bound.  Finally,
\eqref{cv:represented-energy} and \eqref{cv:truncated-loss-error} give
\begin{equation*}
 H(z)=L^R(\rho)
 =L(\rho)+\{L^R(\rho)-L(\rho)\}\ge L(\rho).
\end{equation*}
\end{proof}

Writing $w_j=s_j^2$ keeps the formulas regular at $s_j=0$.
We take the location derivatives $\nabla_{x_j}H$ in
\eqref{cv:full-gradient} in the ambient space $(\mathbb R^d)^K$,
even when $x_j\in\partial B_R$. The compact set $\mathscr K$
gives uniform constants in the gradient inequality. Stationarity
is defined using the gradient on $\mathscr M$.

\begin{remark}[Why likelihood moments alone do not suffice]
\label{cv:moment-only-failure}
Take $N=2$, $d=1$, $X_1=-2$, $X_2=2$ and $\rho_0=\rho=\delta_0$.
Then
\[
 a_1=a_2=\phi(2),\qquad
 g_a(x)=e^{-x^2/2}\cosh(2x),\qquad \mathcal D(\delta_0)=0.
\]
The values
\[
 g_a(3)=\tfrac12(e^{3/2}+e^{-21/2})>1,\qquad
 g_a(4)=\tfrac12(1+e^{-16})<1
\]
give $r\in(3,4)$ such that $g_a(r)=1$.  These locations belong to the
fixed ball in \eqref{cv:fixed-ball}: here $b=Me^{-2}$, $c_0\le b/2$, and
\[
 R=2+\sqrt{2\log(2M/c_0)}
 \ge2+\sqrt{4+2\log4}>4.
\]
The measure $\sigma=(\delta_{-r}+\delta_r)/2$ has precisely the same
likelihood moments as $\delta_0$, since
\[
 a_i(\sigma)=\phi(2)e^{-r^2/2}\cosh(2r)=a_i(\delta_0)
 \quad(i=1,2).
\]
In \eqref{cv:analytic-H}, take
\[
 x_1=-r,\quad x_2=r,\quad x_j=0\ (3\le j\le K),
 \qquad
 s_1=s_2=2^{-1/2},\quad
 s_j=0\ (j\notin\{1,2\}).
\]
Then $Q=a$, $g_a(\pm r)=1$ and
$\operatorname{grad}_{\mathbb S^K}H=0$, whereas
\[
 \|\operatorname{grad}H\|^2
 =\tfrac12|g_a'(r)|^2
 =\tfrac12\{2\tanh(2r)-r\}^2>0,
\]
because $2\tanh(2r)-r<2-r<-1$.  Thus, for every $C<\infty$,
\[
 a(\sigma)=a(\rho),\qquad
 \|\operatorname{grad}H\|^2>0=C\mathcal D(\rho).
\]
Thus no finite constant $C$ can make
$\|\operatorname{grad}H\|^2\le C\mathcal D(\rho)$ hold uniformly
over all representatives that match the likelihood moments alone.
Equation~\eqref{cv:compression-dissipation} supplies the additional
control on dissipation needed for this bound.
\end{remark}

\begin{proposition}[Gradient inequality along the measure trajectory]
\label{cv:trajectory-gradient}
There exist $n_0\in\mathbb N$, $\theta\in[1/2,1)$ and $C_g>0$ such that
\begin{equation}\label{cv:trajectory-inequality}
 \{L(\rho_n)-L_\infty\}^\theta\le C_g\sqrt{E_n}
 \qquad(n\ge n_0).
\end{equation}
\end{proposition}

\begin{proof}
For each $n$, choose $z_n\in\mathscr K$ satisfying
\eqref{cv:compression-mass}--\eqref{cv:compression-dissipation}
with $\rho=\rho_n$.
By \eqref{cv:truncated-loss-error} and \eqref{cv:loss-limit},
\[
 0\le H(z_n)-L(\rho_n)\le4\mathcal D(\rho_n)\longrightarrow0,
 \qquad H(z_n)\longrightarrow L_\infty.
\]
Compactness and continuity give
\[
 z_{n_k}\longrightarrow z_*\in\mathscr K,\qquad
 H(z_*)=\lim_{k\to\infty}H(z_{n_k})=L_\infty,
 \qquad z_*\in\mathscr K\cap H^{-1}(\{L_\infty\}).
\]
Apply Lemma~\ref{cv:compact-level} with $\lambda=L_\infty$.
Since $H(z_n)\to L_\infty$, its constant $\delta>0$ gives $n_0$ such that
\[
 n\ge n_0\quad\Longrightarrow\quad
 |H(z_n)-L_\infty|<\delta,\qquad
 H(z_n)\ge L(\rho_n)\ge L_\infty.
\]
For every $n\ge n_0$, Lemmas~\ref{cv:compact-level} and
\ref{cv:lifted-gradient} and \eqref{cv:residual-comparison} therefore imply
\begin{align*}
 \{L(\rho_n)-L_\infty\}^\theta
 &\le\{H(z_n)-L_\infty\}^\theta\\
 &\le C\|\operatorname{grad}H(z_n)\|
 \le C_1\sqrt{\mathcal D(\rho_n)}
 \le C_g\sqrt{E_n}.
\end{align*}
The first comparison uses $H(z_n)\ge L(\rho_n)\ge L_\infty$.
We use the subsequence only to find a point in $\mathscr K$ at level
$L_\infty$. This step precedes any convergence result for the measures
or their likelihood vectors. The estimates hold for every such
representative, so we can choose $z_n$ separately for each $n$ without
requiring a continuous or measurable selection.
\end{proof}

\section{Finite length and convergence of the mixing measures}
\label{sec:finite-length}
We now prove the convergence assertions in Theorem~\ref{thm:convergence}.
By \eqref{cv:loss-limit}, $e_n=L(\rho_n)-L_\infty\ge0$ is nonincreasing.
With $\theta$ and $n_0$ from Proposition~\ref{cv:trajectory-gradient}, put
\begin{equation}\label{cv:desingularizing-function}
 \Phi(t)=\frac{t^{1-\theta}}{1-\theta}\quad(t\ge0),\qquad
 \Phi'(t)=t^{-\theta}\quad(t>0).
\end{equation}
Let $n\ge n_0$.  If $e_n=0$, then $e_{n+1}=0$ and \eqref{cv:descent} gives
$E_n=0$.  If $e_n>0$, then \eqref{cv:trajectory-inequality} gives
$0<e_n^\theta\le C_g\sqrt{E_n}$; using $t^{-\theta}\ge e_n^{-\theta}$ on
$[e_{n+1},e_n]$, \eqref{cv:descent} and this bound, we obtain
\begin{equation}\label{cv:telescoping}
 \Phi(e_n)-\Phi(e_{n+1})
 =\int_{e_{n+1}}^{e_n}t^{-\theta}\,dt
 \ge e_n^{-\theta}(e_n-e_{n+1})
 \ge c\,e_n^{-\theta}E_n
 \ge\frac{c}{C_g}\sqrt{E_n}.
\end{equation}
In either case $\Phi(e_n)-\Phi(e_{n+1})\ge(c/C_g)\sqrt{E_n}$, so for
$m\ge n\ge n_0$,
\begin{equation}\label{cv:finite-square-root-sum}
 \sum_{k=n}^m\sqrt{E_k}
 \le\frac{C_g}{c}\{\Phi(e_n)-\Phi(e_{m+1})\}
 \le\frac{C_g}{c(1-\theta)}e_n^{1-\theta}.
\end{equation}
Since each $E_n$ is finite by \eqref{cv:loss-limit},
$\sum_n\sqrt{E_n}<\infty$, and \eqref{cv:step-length} gives
\begin{equation}\label{cv:total-length}
 \sum_{n=0}^\infty d_{\mathrm{BL}}(\rho_{n+1},\rho_n)<\infty
\end{equation}
and, by the triangle inequality,
\begin{equation}\label{cv:Cauchy}
 \sup_{m>n}d_{\mathrm{BL}}(\rho_m,\rho_n)
 \le C\sum_{k=n}^\infty\sqrt{E_k}\longrightarrow0.
\end{equation}

We verify tightness rather than assuming that a
$d_{\mathrm{BL}}$-Cauchy sequence has a probability limit.  By
\eqref{cv:tail-control} and \eqref{cv:loss-limit}, the ball in
\eqref{cv:fixed-ball} satisfies
\begin{equation}\label{cv:tail-vanishing}
 \rho_n(B_R^c)\le4\mathcal D(\rho_n)\longrightarrow0.
\end{equation}
Given $\varepsilon>0$, choose $n_\varepsilon$ with
$\rho_n(B_R^c)<\varepsilon$ for $n\ge n_\varepsilon$, and $R_\varepsilon\ge R$
with $\rho_k(B_{R_\varepsilon}^c)<\varepsilon$ for the finitely many
$k<n_\varepsilon$.  Then $\sup_n\rho_n(B_{R_\varepsilon}^c)\le\varepsilon$,
so the sequence is uniformly tight, and Prokhorov's theorem gives a
subsequence $\rho_{n_j}\Rightarrow\widehat\rho\in\mathcal P(\mathbb R^d)$.
For $n_j>n$, the triangle inequality and \eqref{cv:Cauchy} give
\[
 d_{\mathrm{BL}}(\rho_n,\widehat\rho)
 \le C\sum_{k=n}^\infty\sqrt{E_k}
       +d_{\mathrm{BL}}(\rho_{n_j},\widehat\rho).
\]
Since $d_{\mathrm{BL}}$ metrizes weak convergence, the last term tends to
zero as $j\to\infty$.  Hence
$d_{\mathrm{BL}}(\rho_n,\widehat\rho)\le C\sum_{k\ge n}\sqrt{E_k}\to0$, and
$\rho_n\Rightarrow\widehat\rho$.

To pass dissipation to the limit, write $a_n=a(\rho_n)$ and
$\widehat a=a(\widehat\rho)$.  Each $\psi_i$ is bounded and continuous, so
\begin{equation}\label{cv:moment-limit}
 a_{n,i}\longrightarrow\widehat a_i=\int\psi_i\,d\widehat\rho,
 \qquad \widehat a\in[c_0,M]^N.
\end{equation}
The fitted values are bounded away from zero, so
\eqref{cv:moment-limit} gives
$a_{n,i}^{-1}\to\widehat a_i^{-1}$ for every $i$.
Since $\psi_i$ and $\nabla\psi_i$ are bounded, \eqref{eq:score}
gives uniform convergence of $g_{a_n}$ to $g_{\widehat a}$
and of $\nabla g_{a_n}$ to $\nabla g_{\widehat a}$.  With $A=M/c_0$ and
$B_1=M_1/c_0$, the identities for differences of squares give
\begin{equation}\label{cv:da-limit}
 \|d_{a_n}-d_{\widehat a}\|_\infty
 \le2B_1\|\nabla g_{a_n}-\nabla g_{\widehat a}\|_\infty
       +2(A+1)\|g_{a_n}-g_{\widehat a}\|_\infty
 \longrightarrow0.
\end{equation}
Since $d_{\widehat a}$ is bounded and continuous and
$\mathcal D(\rho_n)=\int d_{a_n}\,d\rho_n$,
\begin{equation}\label{cv:stationarity-limit}
 0\le\int d_{\widehat a}\,d\widehat\rho
 =\lim_{n\to\infty}\int d_{\widehat a}\,d\rho_n
 \le\lim_{n\to\infty}\bigl\{\mathcal D(\rho_n)
     +\|d_{\widehat a}-d_{a_n}\|_\infty\bigr\}=0.
\end{equation}
Therefore
\begin{equation}\label{cv:stationarity}
 g_{\widehat\rho}=1,\qquad\nabla g_{\widehat\rho}=0
 \qquad\widehat\rho\text{-almost everywhere},
\end{equation}
and, by \eqref{cv:moment-limit}, $L(\widehat\rho)=\lim_nL(\rho_n)=L_\infty$.

If $\operatorname{supp}(\rho_0)=\mathbb R^d$, the weak convergence
just proved and the step-size bound \eqref{eq:threshold} allow us
to apply \citet[Theorem~2(iii) and Remark~2]{YWR2024}.
Thus $L(\widehat\rho)=L^*$.

\begin{corollary}[Remaining distance and length]\label{cv:distance-to-limit}
For $\theta$ and $n_0$ as in Proposition~\ref{cv:trajectory-gradient},
there exists $C<\infty$ such that
\begin{equation}\label{cv:distance-rate}
 d_{\mathrm{BL}}(\rho_n,\widehat\rho)
 \le\sum_{k=n}^\infty d_{\mathrm{BL}}(\rho_{k+1},\rho_k)
 \le C\{L(\rho_n)-L_\infty\}^{1-\theta}
 \quad(n\ge n_0).
\end{equation}
\end{corollary}

\noindent The complete proof of Corollary~\ref{cv:distance-to-limit} is given in the Supplement.
\setcounter{equation}{11}\setcounter{theorem}{1}\setcounter{subsection}{0}

The exponent and constants may depend on the trajectory. The compact
representation serves only to prove \eqref{cv:trajectory-inequality}.
It does not approximate the iteration or yield a bound on the number
of support points of $\widehat\rho$.
\section{Finite particles and suboptimal attraction}\label{sec:particles}
We now prove Theorem~\ref{thm:particles}. We use the same transport--reaction update as in Theorem~\ref{thm:convergence}, but consider a different initial support. Throughout this section, we set $N=3$ and $d=2$, and write the dependence on the observations explicitly:
\[
 \begin{gathered}
 a_{X,i}(\rho)=\int\phi(X_i-x)\,d\rho(x),\qquad
 L_X(\rho)=-\frac13\sum_{i=1}^3\log a_{X,i}(\rho),\\
 g_{X,\rho}(x)=\frac13\sum_{i=1}^3\frac{\phi(X_i-x)}{a_{X,i}(\rho)}.
 \end{gathered}
\]
Let $\mathcal S_{X,\eta}$ denote one complete transport--reaction update, defined by
\begin{equation}\label{pt:split}
 T_{X,\rho}(x)=x+\eta\nabla g_{X,\rho}(x),\qquad
 \widetilde\rho=(T_{X,\rho})_\#\rho,\qquad
 \mathcal S_{X,\eta}(\rho)=(1-\eta+\eta g_{X,\widetilde\rho})\widetilde\rho.
\end{equation}
The reaction step uses the score $g_{X,\widetilde\rho}$ of the
transported measure $\widetilde\rho$.

\subsection{Exact aggregation and a symmetric orbit}
\begin{lemma}[Aggregation]\label{pt:aggregation}
For $0<\eta\le1$ and $\rho=\sum_{k=1}^Kp_k\delta_{z_k}$, where $K<\infty$, $p_k>0$ and $\sum_kp_k=1$,
\[
 \mathcal S_{X,\eta}^n(\rho)\in\mathcal P(\mathbb R^2),\qquad
 \#\operatorname{supp}\mathcal S_{X,\eta}^n(\rho)\le K
 \qquad(n\in\mathbb N_0).
\]
Coincident particles aggregate by summing their masses. In particular, the initialization in Theorem~\ref{thm:particles} can be represented exactly by at most three groups, initially located at $X_j$ with masses
\begin{equation}\label{pt:counts}
 q_j=C_j/m,\qquad C_j=\sum_{\ell=1}^m\mathbf1_{\{J_\ell=j\}},\qquad j=1,2,3.
\end{equation}
\end{lemma}
\noindent The complete proof of Lemma~\ref{pt:aggregation} is given in the Supplement.
\setcounter{equation}{2}\setcounter{theorem}{1}\setcounter{subsection}{1}

For clarity, recall the constants in Theorem~\ref{thm:particles}:
\begin{equation}\label{pt:triangle}
\begin{gathered}
 e_1=(1,0),\qquad e_2=(-1/2,\sqrt3/2),\qquad e_3=(-1/2,-\sqrt3/2),\\
 r_\triangle^2=\frac53\log3,\qquad X_i^\triangle=r_\triangle e_i,\qquad
 r_*=\frac25r_\triangle,\qquad
 \rho_*=\frac13\sum_{j=1}^3\delta_{r_*e_j},\\
 \delta_\triangle=\frac{22}{15}\log3-\log5.
\end{gathered}
\end{equation}
For $r>0$, define $\rho_r=3^{-1}\sum_{j=1}^3\delta_{re_j}$. At the equilateral configuration $X=X^\triangle$, symmetry implies that the three fitted values $a_{X^\triangle,i}(\rho_r)$, $i=1,2,3$, are equal. Denote their common value by $f(r)$, which is given by
\begin{equation}\label{pt:fitted}
 f(r)=\phi(0)\exp\!\left(-\frac{r_\triangle^2+r^2}{2}\right)
 \frac{\exp(r_\triangle r)+2\exp(-r_\triangle r/2)}3.
\end{equation}
Indeed, $e_i\cdot e_j=1$ when $i=j$ and $e_i\cdot e_j=-1/2$ when $i\ne j$. Substituting these identities into
\[
|r_\triangle e_i-re_j|^2
=r_\triangle^2+r^2-2r_\triangle r\,e_i\cdot e_j
\]
gives the stated formula for $f(r)$.
Let $F(x)=3^{-1}\sum_i\phi(r_\triangle e_i-x)$. Since $a_{X^\triangle,i}(\rho_r)=f(r)$ for every $i$, we have
\[
 g_{X^\triangle,\rho_r}(x)=F(x)/f(r),\qquad F(re_j)=f(r).
\]
Reflection across the line spanned by $e_j$ leaves $F$ invariant. Since $re_j$ lies on this line, the tangential derivative of $F$ at $re_j$ is zero. The radial derivative is $f'(r)$ because $F(re_j)=f(r)$. Hence the gradient is parallel to $e_j$, and
\begin{equation}\label{pt:radial}
 \nabla g_{X^\triangle,\rho_r}(re_j)=v(r)e_j,\qquad
 v(r)=\frac{f'(r)}{f(r)}
 =-r+r_\triangle\frac{q(r)-1}{q(r)+2},\qquad
 q(r)=\exp(3r_\triangle r/2).
\end{equation}
After transport, the three support points remain at a common radius. By symmetry, $g$ takes the same value at all three points. Since the weighted average of these values is one, their common value must also be one. The reaction factor is therefore equal to one at each support point, so the three masses remain equal.

\begin{lemma}[Entry from the empirical measure]\label{pt:radial-entry}
For every $0<\eta\le1$, the iterates starting from the empirical measure satisfy
\[
 \mathcal S_{X^\triangle,\eta}^n\!\left(\frac13\sum_j\delta_{X_j^\triangle}\right)=\rho_{r_n},\qquad n\in\mathbb N_0,
\]
where the radii satisfy
\[
 r_0=r_\triangle,\qquad r_{n+1}=r_n+\eta v(r_n),\qquad r_n\downarrow r_*.
\]
Moreover, the limiting measure is strictly suboptimal:
\begin{equation}\label{pt:exact-gap}
 L_{X^\triangle}(\rho_*)-L_{X^\triangle}(\delta_0)=\delta_\triangle>0.
\end{equation}
\end{lemma}
\noindent The complete proof of Lemma~\ref{pt:radial-entry} is given in the Supplement.
\setcounter{equation}{8}\setcounter{theorem}{2}\setcounter{subsection}{1}

\subsection{The complete position--weight Hessian}
We allow all three positions and all positive masses to vary. Set
\[
 E=\begin{pmatrix}
 1/\sqrt2&1/\sqrt6\\
 -1/\sqrt2&1/\sqrt6\\
 0&-2/\sqrt6
 \end{pmatrix},\qquad E^\top E=I_2,\qquad E^\top\mathbf1=0.
\]
Use the eight-dimensional chart
\begin{equation}\label{pt:chart}
 y=(z,u)\in\mathbb R^6\times\mathbb R^2,
 \qquad p=\mathbf1/3+Eu,\qquad
 \mathcal F_X(z,u)=L_X\!\left(\sum_{j=1}^3p_j\delta_{z_j}\right),
\end{equation}
on the open set $p_j>0$. Its base state is
\[
 y_*=(r_*e_1,r_*e_2,r_*e_3,0).
\]
The vectors $e_j$ describe the triangle; $u$ is the independent weight coordinate.

\begin{proposition}[Full reduced Hessian]\label{pt:hessian}
The matrix $H_*=D^2\mathcal F_{X^\triangle}(y_*)$ is positive definite, with blocks
\begin{equation}\label{pt:hessian-blocks}
 H_*\succ0,\qquad 0\prec(H_*)_{zz}\preceq I_6/3,
 \qquad (H_*)_{uu}=\frac{12}{25}I_2.
\end{equation}
\end{proposition}
\noindent The complete proof of Proposition~\ref{pt:hessian} is given in the Supplement.
\setcounter{equation}{10}\setcounter{theorem}{3}\setcounter{subsection}{2}

\subsection{Contraction of the exact split map}
At a positive-weight state, the gradients in the reduced chart are
\begin{equation}\label{pt:chart-gradients}
 \nabla_{z_j}\mathcal F_X=-p_j\nabla g_{X,\rho}(z_j),
 \qquad
 \nabla_u\mathcal F_X=-E^\top(g_{X,\rho}(z_j))_{j=1}^3.
\end{equation}
Define
\[
 A(p)=\operatorname{diag}(p_1^{-1}I_2,p_2^{-1}I_2,p_3^{-1}I_2),
 \qquad
 B(p)=E^\top\{\operatorname{diag}(p)-pp^\top\}E.
\]
In these coordinates, the transport--reaction map $\mathcal S_{X,\eta}$ is
\begin{equation}\label{pt:chart-split}
 z^+=z-\eta A(p)\nabla_z\mathcal F_X(z,u),
 \qquad
 u^+=u-\eta B(p)\nabla_u\mathcal F_X(z^+,u).
\end{equation}
Here the weight gradient is evaluated at the new positions and old weights. To verify the second identity, put $G=\operatorname{diag}(p)-pp^\top$ and let $g$ be the vector of support values computed at $(z^+,u)$. Since $p^\top g=1$,
\[
 Gg=\operatorname{diag}(p)(g-\mathbf1),\qquad
 G\mathbf1=0,\qquad GEE^\top=G.
\]
The reaction therefore satisfies
\[
 u^+-u=\eta E^\top\operatorname{diag}(p)(g-\mathbf1)
 =\eta E^\top GEE^\top g
 =-\eta B(p)\nabla_u\mathcal F_X(z^+,u).
\]
Both $A(p)$ and $B(p)$ are positive definite. For the latter, if $w\ne0$,
\[
 \begin{aligned}
 w^\top B(p)w
 &=\sum_jp_j(Ew)_j^2-\left(\sum_jp_j(Ew)_j\right)^2\\
 &=\frac12\sum_{j,k}p_jp_k\{(Ew)_j-(Ew)_k\}^2>0.
 \end{aligned}
\]
Indeed, equality would make the three values $(Ew)_j$ constant; their zero sum would then give $Ew=0$ and hence $w=0$.

Every interior critical point of $\mathcal F_X$ is a fixed point of \eqref{pt:chart-split}. Indeed,
\[
 \begin{gathered}
 \nabla_z\mathcal F_X=0\ \Longrightarrow\ \nabla g(z_j)=0\quad(j=1,2,3),\\
 \nabla_u\mathcal F_X=0
 \ \Longrightarrow\ E^\top g=0
 \ \Longrightarrow\ g=c\mathbf1,
 \qquad 1=p^\top g=c\sum_jp_j=c,\\
 z^+=z,\qquad u^+=u.
 \end{gathered}
\]

\begin{proposition}[Local attraction at every step size]\label{pt:contraction}
For every fixed $0<\eta\le1$, $y_*$ is an interior fixed point and
\begin{equation}\label{pt:strict-contraction}
 \kappa_\eta:=
 \|H_*^{1/2}D_y\mathcal S_{X^\triangle,\eta}(y_*)H_*^{-1/2}\|_2<1.
\end{equation}
\end{proposition}
\noindent The complete proof of Proposition~\ref{pt:contraction} is given in the Supplement.
Write $J_\eta=D_y\mathcal S_{X^\triangle,\eta}(y_*)$.
\setcounter{equation}{16}\setcounter{theorem}{4}\setcounter{subsection}{3}

The contraction holds in the eight-dimensional aggregated chart. When labeled particles coincide, redistributing their total mass leaves the measure unchanged and produces neutral directions in the labeled coordinates.

\subsection{An open basin for perturbed data}
Fix $\eta\in(0,1]$. The functions $\mathcal F_X$ and $\mathcal S_{X,\eta}$ are smooth on the positive-weight chart. Since $H_*\succ0$, the implicit function theorem applied to $\nabla_y\mathcal F_X(y)=0$ gives a smooth branch $y_*(X)$ for $X$ near $X^\triangle$, with $y_*(X^\triangle)=y_*$. Restrict the neighborhood so that $y_*(X)$ has positive weights and distinct support points. Each $y_*(X)$ is a fixed point of the exact split map, by the critical-point observation above.

Use the fixed norm $\|v\|_*=(v^\top H_*v)^{1/2}$, with the corresponding induced operator norm, and put $\bar\kappa_\eta=(1+\kappa_\eta)/2<1$. Continuity of the derivative permits a radius $r>0$ and a smaller data neighborhood such that the closed balls $\|y-y_*(X)\|_*\le r$ lie in the chart and
\begin{equation}\label{pt:uniform-derivative}
 \|y_*(X)-y_*\|_*<r/8,
 \qquad
 \sup_{\|y-y_*(X)\|_*\le r}
 \|D_y\mathcal S_{X,\eta}(y)-J_\eta\|_*
 \le\frac{1-\kappa_\eta}{2}.
\end{equation}
First choose a small fixed ball about $y_*$ on which the derivative is close to $J_\eta$, uniformly for $X$ near $X^\triangle$. Take $r$ smaller than its radius, then restrict the data neighborhood so that $\|y_*(X)-y_*\|_*<r/8$ and every radius-$r$ ball about $y_*(X)$ lies inside the fixed ball.

If $\|y-y_*(X)\|_*\le r$, the segment joining $y_*(X)$ to $y$ stays in the same ball. The integral mean-value formula gives
\begin{align}
 \mathcal S_{X,\eta}(y)-y_*(X)
 &=\int_0^1D_y\mathcal S_{X,\eta}\bigl(y_*(X)+t(y-y_*(X))\bigr)
       (y-y_*(X))\,dt,\notag\\
 \|\mathcal S_{X,\eta}(y)-y_*(X)\|_*
 &\le\left(\kappa_\eta+\frac{1-\kappa_\eta}{2}\right)
       \|y-y_*(X)\|_*
 =\bar\kappa_\eta\|y-y_*(X)\|_* .\label{pt:basin-contraction}
\end{align}
In particular these balls are invariant.

Lemma~\ref{pt:radial-entry} gives a finite integer $K$ such that
\begin{equation}\label{pt:base-entry}
 \|\mathcal S_{X^\triangle,\eta}^K(X^\triangle,0)-y_*\|_*<r/4.
\end{equation}
Here $(X,u)$ denotes the initial state with positions $z_j=X_j$ and weights $\mathbf1/3+Eu$. The finite composition
\[
 \mathcal H(X,u)=\mathcal S_{X,\eta}^K(X,u)
\]
is smooth near $(X^\triangle,0)$, since the weights along the base orbit up to time $K$ are positive and remain so for nearby $(X,u)$. On a smaller compact convex neighborhood, $D\mathcal H$ is bounded, so there is a constant $C\ge1$ such that
\begin{equation}\label{pt:pullback}
 \|\mathcal H(X,u)-\mathcal H(X^\triangle,0)\|_*
 \le C\bigl(|X-X^\triangle|+|u|\bigr).
\end{equation}
Choose
\begin{equation}\label{pt:epsilon}
 0<\varepsilon_\eta\le\min\left\{\frac1{12},\frac{r}{16\sqrt3 C}\right\},
\end{equation}
small enough to stay in this neighborhood, and restrict the data further so that $|X-X^\triangle|<r/(16C)$. If $q$ is any probability vector with $\max_j|q_j-1/3|\le\varepsilon_\eta$, then all $q_j>0$, and its chart coordinate satisfies
\[
 u=E^\top(q-\mathbf1/3),\qquad
 |u|=|q-\mathbf1/3|\le\sqrt3\varepsilon_\eta.
\]
Equations \eqref{pt:uniform-derivative}, \eqref{pt:base-entry} and \eqref{pt:pullback} imply
\begin{align*}
 \|\mathcal H(X,u)-y_*(X)\|_*
 &\le C\bigl(|X-X^\triangle|+|u|\bigr)
      +\|\mathcal H(X^\triangle,0)-y_*\|_*
      +\|y_*-y_*(X)\|_*\\
 &<r/8+r/4+r/8=r/2.
\end{align*}
Repeated application of \eqref{pt:basin-contraction} therefore gives, for every $n\ge K$,
\begin{equation}\label{pt:geometric}
 \|\mathcal S_{X,\eta}^n(X,u)-y_*(X)\|_*
 \le\frac r2\bar\kappa_\eta^{\,n-K}.
\end{equation}

For each fixed $X$, define a subset of the positive-weight chart by
\[
 \mathcal V_X=\bigl\{y:\ \|\mathcal S_{X,\eta}^K(y)-y_*(X)\|_*<r\bigr\}.
\]
This set is open by continuity of the finite-step map. Every orbit starting in $\mathcal V_X$ converges to $y_*(X)$ by \eqref{pt:basin-contraction}, and the initial states $(X,u)$ considered above belong to $\mathcal V_X$.

The constants $K$, $r$ and $\bar\kappa_\eta$ are uniform over the final data neighborhood and all initial weights satisfying $\max_j|q_j-1/3|\le\varepsilon_\eta$; in particular, they do not depend on $m$. Since $E^\top E=I_2$, norm equivalence gives the coordinate bound in Theorem~\ref{thm:particles} with $\theta=\bar\kappa_\eta$ and rate constant $(r/2)\sqrt{2/\lambda_{\min}(H_*)}$.

Equation~\eqref{pt:geometric} gives geometric convergence of the
positions and weights. The limiting measure is
\[
 \rho_{\mathrm{bad}}(X)=\sum_{j=1}^3p_{*,j}(X)\delta_{z_{*,j}(X)}.
\]
For every $h\in C_b(\mathbb R^2)$,
\[
 \sum_jp_{n,j}h(z_{n,j})
 \longrightarrow
 \sum_jp_{*,j}(X)h(z_{*,j}(X)),
\]
so $\rho_n\rightharpoonup\rho_{\mathrm{bad}}(X)$. Moreover,
\[
 a_{X,i}(\rho_n)
 \longrightarrow
 \sum_jp_{*,j}(X)\phi(X_i-z_{*,j}(X))>0.
\]
Since these limits are positive, continuity of the logarithm gives
$L_X(\rho_n)\to L_X(\rho_{\mathrm{bad}}(X))$.

Finally, shrink the open data neighborhood $U_\eta$ once more so that
\[
 |\mathcal F_X(y_*(X))-\mathcal F_{X^\triangle}(y_*)|<\delta_\triangle/4,
 \qquad
 |L_X(\delta_0)-L_{X^\triangle}(\delta_0)|<\delta_\triangle/4.
\]
Together with \eqref{pt:exact-gap}, this gives
\begin{align}
 L_X(\rho_{\mathrm{bad}}(X))-L_X^*
 &\ge L_X(\rho_{\mathrm{bad}}(X))-L_X(\delta_0)\notag\\
 &>L_{X^\triangle}(\rho_*)-L_{X^\triangle}(\delta_0)
       -\delta_\triangle/2
 =\delta_\triangle/2.\label{pt:persistent-gap}
\end{align}
The comparison uses only $\delta_0$, so the global optimizer need not be identified or unique. All constants and neighborhoods above are chosen for fixed $\eta$ and may depend on it.

\subsection{Random initialization}
By Lemma~\ref{pt:aggregation}, the particle iteration reduces to at most three groups with initial weights $q_j=C_j/m$. Appendix~F of the Supplement gives
\begin{equation}\label{pt:prob-summary}
 \mathbb P\{\max_j|C_j/m-1/3|>\varepsilon_\eta\}
 \le6e^{-2m\varepsilon_\eta^2}.
\end{equation}
On the complementary event, \eqref{pt:geometric} and \eqref{pt:persistent-gap} give convergence to $\rho_{\mathrm{bad}}(X)$ and the likelihood gap, proving \eqref{eq:particleprob}. No convergence result on the exceptional event is needed. Under the coupling in Theorem~\ref{thm:particles}, the bounds in \eqref{pt:prob-summary} are summable in $m$, so the first Borel--Cantelli lemma gives the same conclusions almost surely for all sufficiently large $m$.

\section{Failure of pure transport from a smooth full-support density}
\label{sec:smooth-W}

We remove the reaction step and start from a smooth, strictly positive
density. The proof treats $\gamma=0$ directly and is independent of
the convergence argument in Sections~3 and~4.

As in Theorem~\ref{thm:smooth-W}, take
$\rho_0=\mathcal N(0,s^2I_2)$ with $s>0$. Throughout this section, let
\begin{equation}\label{sw:data}
 R=\sqrt{\log 7},\qquad X_i=Re_i\quad(1\le i\le3),\qquad
 F(x)=\frac13\sum_{i=1}^3\phi(x-X_i),\qquad M=(2\pi)^{-1},
\end{equation}
where the vectors $e_i$ are defined in \eqref{eq:triangle} and $\phi$ is the
standard two-dimensional Gaussian density. Let $Q$ denote rotation through
$2\pi/3$, so that $QX_i=X_{i+1}$, with cyclic indices. Define
\begin{equation}\label{sw:constants}
 \begin{gathered}
 F_0=F(0)=M e^{-R^2/2}=\frac M{\sqrt7},\\
 \kappa=F_0(1-R^2/2)>0,\qquad r=\frac{\kappa}{8M},\qquad
 p_s=1-e^{-r^2/(2s^2)},\\
 H=F(X_1/2)=\frac{F_0}{3}(u+2/u),\qquad u=7^{3/8}.
 \end{gathered}
\end{equation}
For this choice of $R$, the origin is a strict local maximum of $F$.
Yet $F(X_1/2)>F(0)$.

\begin{lemma}[Geometry and a finite critical set]\label{sw:geometry}
For every orthogonal map $A$ on $\mathbb R^2$,
\[
 A\{X_1,X_2,X_3\}=\{X_1,X_2,X_3\}
 \quad\Longrightarrow\quad F(Ax)=F(x)\quad(x\in\mathbb R^2).
\]
Moreover,
\begin{equation}\label{sw:geometry-bounds}
 0<F\le M,\qquad \|\nabla^2F\|_{\mathrm{op}}\le M,
 \qquad \nabla F(0)=0,\qquad \nabla^2F(0)=-\kappa I,
\end{equation}
\begin{equation}\label{sw:local-Hessian}
 -MI\preceq\nabla^2F(x)\preceq-\frac\kappa2I
 \quad(|x|\le r),
\end{equation}
and
\[
 H>F_0,\qquad
 \mathcal C:=\{x\in\mathbb R^2:\nabla F(x)=0\}
 \subseteq\{x:|x|\le R\},\qquad \#\mathcal C<\infty.
\]
\end{lemma}

\noindent The complete proof of Lemma~\ref{sw:geometry} is given in the Supplement.
\setcounter{equation}{4}\setcounter{theorem}{1}\setcounter{subsection}{0}

Under the step-size condition of Theorem~\ref{thm:smooth-W},
symmetry gives a common fitted value $c_n$ and
\[
 a_i(\rho_n)=c_n,\qquad g_{\rho_n}=F/c_n,\qquad
 T_n=\operatorname{id}+(\eta/c_n)\nabla F.
\]
Thus every spatial trajectory ascends the same function $F$. In the central
ball the maps contract, and the Gaussian initialization gives that ball
positive mass:
\[
 |T_n(x)|\le\left(1-\frac{\eta\kappa}{2M}\right)|x|
 \quad(|x|\le r),\qquad
 \rho_\infty(\{0\})\ge p_s>0.
\]
After establishing convergence, write $c_\infty=a_i(\rho_\infty)$ and set
\[
 \pi=\frac13\sum_{i=1}^3\delta_{X_i/2},\qquad
 \rho^\dagger=\rho_\infty-p_s\delta_0+p_s\pi.
\]
The competitor improves all three fitted values simultaneously:
\[
 a_i(\rho^\dagger)=c_\infty+p_s(H-F_0)>c_\infty,
\]
\[
 L(\rho^\dagger)-L(\rho_\infty)
 =-\log\left(1+\frac{p_s(H-F_0)}{c_\infty}\right)
 \le-\log\left(1+\frac{p_s(H-F_0)}M\right)<0.
\]
The proof of Theorem~\ref{thm:smooth-W} in the Supplement also shows
that each finite iterate has a smooth, strictly positive density
and that the full sequence of measures converges.
\setcounter{equation}{23}\setcounter{theorem}{1}\setcounter{subsection}{0}

The limit also separates the two stationarity conditions:
\[
 \int|\nabla g_{\rho_\infty}|^2\,d\rho_\infty=0,
 \qquad \int(g_{\rho_\infty}-1)^2\,d\rho_\infty>0.
\]
A direct lower bound for the second integral is proved in
Appendix~D of the Supplement. Thus
$\rho_\infty$ is a fixed point of pure transport and has a strictly
positive reaction residual.

\section{Consequences and the role of initialization}
Theorem~\ref{thm:convergence} proves the weak convergence of the full
sequence assumed in the original Gaussian optimality result. The mixing
measures converge even for data sets with nonunique NPMLEs. We use the
auxiliary atomic representation only to establish the scalar gradient
inequality. It does not approximate the update or give a bound on the
number of support points of the limit.

For $0<\eta<\eta_{\mathrm{adm}}$ and $X\in U_\eta'$, as defined in
Appendix~B of the Supplement, Theorem~\ref{thm:convergence} gives
convergence from the resampling initialization. With probability
tending to one as $m\to\infty$, the limit is the suboptimal fixed point
identified in Theorem~\ref{thm:particles}. The resampling counts
determine the total weight at each of the three observed sites.
The two theorems have separate proofs. The full-support hypothesis
in Theorem~\ref{thm:convergence} applies only to the global optimality
conclusion.

Convergence for each full-support initialization leaves open whether
the number of iterations needed to reach a prescribed likelihood gap
can be bounded uniformly over a class of initializations.
Write $S_\eta\rho$ for one update of
\eqref{eq:transport}--\eqref{eq:reaction}.

\begin{corollary}[Nonuniformity over smooth full-support initializations]\label{cor:nonuniform}
Fix the observations and suppose that
\[
 \rho_{\mathrm b}=\sum_{j=1}^m w_j\delta_{z_j},\qquad
 w_j>0,\quad \sum_jw_j=1,\qquad
 S_\eta\rho_{\mathrm b}=\rho_{\mathrm b},\qquad
 \Delta=L(\rho_{\mathrm b})-L^*>0.
\]
Let $B>L(\rho_{\mathrm b})$, define $r_B,c_B,\eta_B$ by \eqref{eq:threshold} with $b=e^{-B}$, and assume $0<\eta\le\eta_B$. Put
\begin{equation}\label{nu:class}
 \mathcal A_B=\left\{p(x)\,dx:\ p\in C^\infty(\R^d),\quad
 p(x)>0\ (x\in\R^d),\quad \int p=1,\quad L(p\,dx)\le B\right\}.
\end{equation}
Then
\begin{equation}\label{nu:horizon}
 \begin{gathered}
 0<\varepsilon<\Delta,\quad T\in\mathbb N_0\quad\Longrightarrow\quad
 \exists\rho_0\in\mathcal A_B:\
 \min_{0\le n\le T}\{L(S_\eta^n\rho_0)-L^*\}>\varepsilon.
 \end{gathered}
\end{equation}
The measure $\rho_0$ in \eqref{nu:horizon} can be chosen as a finite Gaussian mixture, with all moments finite. For
\[
 \tau_\varepsilon(\rho_0)=\inf\{n\in\mathbb N_0:L(S_\eta^n\rho_0)-L^*\le\varepsilon\},
 \qquad \inf\varnothing=\infty,
\]
we have, for every $0<\varepsilon<\Delta$,
\begin{equation}\label{nu:hitting}
 \tau_\varepsilon(\rho_0)<\infty\quad(\rho_0\in\mathcal A_B),\qquad
 \sup_{\rho_0\in\mathcal A_B}\tau_\varepsilon(\rho_0)=\infty.
\end{equation}
\end{corollary}

The proof is in Appendix~G of the Supplement. The suboptimal fixed
points from Theorem~\ref{thm:particles} satisfy the fixed-point
hypothesis of the corollary. For these data, fix
$0<\eta\le\eta_B$ and $0<\varepsilon<\Delta$. The number of iterations
needed to reach a likelihood gap of at most $\varepsilon$ is
unbounded over $\mathcal A_B$.

In \eqref{nu:horizon}, the initial measure $\rho_0$ may depend on both
$T$ and $\varepsilon$. Under the same step-size condition, the
likelihood gap still tends to zero for each fixed initialization
in $\mathcal A_B$. The corollary leaves open iteration bounds that
use additional quantitative information about the initialization.

By \eqref{eq:stationary} and continuity,
\[
 \operatorname{supp}(\widehat\rho)
 \subseteq\{x:g_{\widehat\rho}(x)=1,\ \nabla g_{\widehat\rho}(x)=0\}
 \subseteq\operatorname{Crit}(g_{\widehat\rho}).
\]
The function
$g_{\widehat\rho}=N^{-1}\sum_i a_i(\widehat\rho)^{-1}\phi(\,\cdot-X_i)$
is a finite sum of Gaussian kernels with positive weights.
Dividing by the sum of its weights gives a finite mixture of isotropic
Gaussians with the same critical set. By
\citet[Theorem~1]{Wang2026Modes}, this critical set is finite.
So $\widehat\rho$ has finite support for every initialization
covered by Theorem~\ref{thm:convergence}. We use Wang's theorem here
only to deduce finite support of the limit.

The almost-everywhere uniqueness result of \citet{Wang2026Unique}
concerns the global NPMLE. It is compatible with the strictly
suboptimal attracting fixed points in Theorem~\ref{thm:particles}.

In the pure-transport example of Theorem~\ref{thm:smooth-W}, the
initial measure and every finite iterate have smooth, strictly
positive densities. Yet the limit is strictly suboptimal and assigns
positive mass to a nonglobal local maximum of the Gaussian sum.
Thus full support alone does not guarantee global optimality for
pure transport.

Proposition~\ref{prop:atomic-W} gives the exact one-dimensional limit
when the initial measure contains a stationary atom. When $\alpha=0$,
the initial measure is a centered Gaussian. The iterates then converge
to the unique NPMLE under the step-size bound computed for this
Gaussian initialization.

For each fixed step size covered by the particle theorem, the
counterexample holds on an open set of data triples and gives a
probability bound for the resampling initialization. The smooth
pure-transport counterexample uses a fixed symmetric data
configuration. For these data, it holds for every centered isotropic
Gaussian initialization and every step size in the stated range.

Under the stated step-size condition, the split iterates converge
from any initial probability measure. If the initial measure has full
support, the result of \citet{YWR2024} identifies the limit as an NPMLE.
The particle example shows that drawing more particles from the same
finite set can still lead to a suboptimal limit. The pure-transport
example shows that full support alone cannot guarantee optimality
once reweighting is removed.

\begin{acks}
Xingzhou Wang and Mingyan Liu are also affiliated with the
Department of Economics at North Carolina State University.
\end{acks}

\begin{supplement}
\stitle{Supplementary proofs}
\sdescription{The supplement contains proofs, split-step estimates,
and calculations for the particle and pure-transport examples.}
\end{supplement}

\section*{Statement on AI use}
We used ChatGPT (OpenAI) for language editing, reference and formatting
checks, and assistance in checking existing proofs in the article and
supplementary material. The authors take responsibility for the
submitted work.

\bibliographystyle{spr-ims-nameyear}
\bibliography{references}
\end{document}


\begin{frontmatter}
\title{Supplement to ``Convergence and global optimality of Wasserstein--Fisher--Rao descent for Gaussian mixtures''}
\runtitle{Supplementary proofs for WFR descent}
\runauthor{Wang et al.}

\begin{aug}
\author[A]{\fnms{Xingzhou}~\snm{Wang}\thanksref{authororder}%
  \ead[label=e1]{xwang268@ncsu.edu}}
\author[B]{\fnms{Ziyang}~\snm{Jiang}%
  \ead[label=e2]{zjiang29@ncsu.edu}}
\author[C]{\fnms{Zheng}~\snm{Li}%
  \ead[label=e3]{zli42@ncsu.edu}}
\author[A]{\fnms{Mingyan}~\snm{Liu}%
  \ead[label=e4]{mliu32@ncsu.edu}}
\author[D]{\fnms{Yiling}~\snm{Wang}%
  \ead[label=e5]{ywang327@scut.edu.cn}}
\author[B]{\fnms{Jingze}~\snm{Yang}%
  \ead[label=e6]{jyang76@ncsu.edu}}

\thankstext{authororder}{The first author led the research.
The remaining authors are listed alphabetically by surname.}

\address[A]{Department of Statistics, North Carolina State University%
  \printead[presep={,\ }]{e1,e4}}
\address[B]{Department of Mathematics, North Carolina State University%
  \printead[presep={,\ }]{e2,e6}}
\address[C]{Department of Agricultural and Resource Economics, North Carolina State University%
  \printead[presep={,\ }]{e3}}
\address[D]{School of Mathematics,
South China University of Technology
  \printead[presep={,\ }]{e5}}
\end{aug}

\begin{abstract}
This supplement contains the detailed estimates and calculations supporting the main paper. All notation and result numbers from the paper are retained through cross-references. 
\end{abstract}
\end{frontmatter}
This supplement collects the proofs and technical details deferred from
the main paper. Sections~3, 5 and~6 retain the notation, statements, explanatory text and equation numbers of the corresponding sections of the main paper; Section~4 contains only the proof of Corollary~4.1. The proof of Corollary~2.2 precedes those sections; Appendices~A--F retain the detailed auxiliary calculations. Appendix~G proves Corollary~7.1.
\subsection*{Proof of Corollary~2.2}
\begin{proof}
Let $1\le p<\infty$ and suppose that $\int |x|^p\,d\rho_0<\infty$.
Let $T_n$ and $h_n$ be as in
(1.4)--(1.5).
By Proposition~3.2 of the main paper, both $\rho_n$ and
$\widetilde\rho_n$ lie in the likelihood sublevel. Let $R$ be the radius
in equation~(3.13) of the main paper, and set
\[
 \delta=\eta M_1/c_0,\qquad \lambda=(1-\eta/2)^{1/p}<1.
\]
The gradient bound in Lemma~3.1 and the tail bound in equation~(3.14)
of the main paper give, respectively,
\[
 |T_n(x)-x|\le\delta\quad(x\in\mathbb R^d),\qquad
 h_n(y)\le\lambda^p\quad(|y|>R).
\]

For $K\ge R$ with $K>\delta$, put
$v_n(K)=\|(|x|-K)_+\|_{L^p(\rho_n)}$, and for $s>0$ define
$\tau(s)=\sup_{m\ge0}\rho_m(B_s^c)$. The displacement bound gives
\[
 (|T_n(x)|-K)_+
 \le (|x|-K)_+ + \delta\mathbf{1}_{\{|x|>K-\delta\}}.
\]
Since $(|y|-K)_+$ is nonzero only when $|y|>K\ge R$, the reaction
bound, the pushforward formula, and Minkowski's inequality yield
\[
\begin{aligned}
 v_{n+1}(K)
 &\le \lambda\left(\int (|T_n(x)|-K)_+^p\,d\rho_n(x)\right)^{1/p}\\
 &\le \lambda v_n(K)+\lambda\delta\,\tau(K-\delta)^{1/p}\\
 &\le \lambda v_n(K)+\delta\,\tau(K-\delta)^{1/p}.
\end{aligned}
\]
By induction,
\[
 \sup_{n\ge0}v_n(K)\le\max\left\{v_0(K),\
 \frac{\delta\,\tau(K-\delta)^{1/p}}{1-\lambda}\right\}
 \longrightarrow0\qquad(K\to\infty),
\]
by dominated convergence for $v_0$ and the uniform tightness proved in Section~4 of the main paper.  Since $|x|^p\le2^p(|x|-K)_+^p$ for
$|x|\ge2K$, the $p$-th moments of $(\rho_n)$ are uniformly integrable; in
particular $\int|x|^p\,d\widehat\rho<\infty$ by lower semicontinuity.
Together with $\rho_n\Rightarrow\widehat\rho$, this gives
$W_p(\rho_n,\widehat\rho)\to0$
\citep[Definition~6.8 and Theorem~6.9]{Villani2009}.  If
$\operatorname{supp}(\rho_0)=\R^d$, then $\widehat\rho$ is an NPMLE by
Theorem~2.1.  Neither full support nor $p>1$ is used in
the convergence argument.
\end{proof}
\setcounter{section}{2}
\section{A trajectory inequality by exact analytic compression}
\label{sec:compression}

Theorems~2.1 and~2.3 concern the same
explicit iteration, but different properties of its limit. We first prove
convergence from an arbitrary initial probability measure.
Sections~3 and~4 of the main paper yield
$\rho_n\Rightarrow\widehat\rho$ and (2.5).
The limit can fail to attain $L^*$, as Theorem~2.3 shows.
Full support is used only at the end of Section~4 of the main paper,
to apply \citet[Theorem~2(iii)]{YWR2024}.

\subsection{Dissipation and the two substeps}
\label{sec:cv-estimates}
For the fixed initial measure, let $c_0$ and $\eta_0$ be the constants in
(2.2).  Write
\begin{equation}\label{cv:kernel-bounds}
 M=\phi(0),\qquad M_1=\sup_x|\nabla\phi(x)|,\qquad
 M_2=\sup_x\|\nabla^2\phi(x)\|_{\mathrm{op}}.
\end{equation}
Thus $0<M_1<\infty$ and $0<M\le M_2<\infty$.  For
$\rho\in\mathcal P(\mathbb R^d)$, define
\begin{equation}\label{cv:dissipation}
 \mathcal D(\rho)=\int\bigl\{|\nabla g_\rho(x)|^2+
                   (g_\rho(x)-1)^2\bigr\}\,\rho(dx).
\end{equation}
At step $n$, the quantities that occur directly in the two descent estimates are
\begin{equation}\label{cv:split-residuals}
 G_n=\int|\nabla g_{\rho_n}|^2\,d\rho_n,\qquad
 R_n=\int(g_{\widetilde\rho_n}-1)^2\,d\widetilde\rho_n,
 \qquad E_n=G_n+R_n.
\end{equation}
Note that the reaction residual is evaluated at the transported measure;
with a residual at $\rho_n$ instead, \eqref{cv:descent} would not follow
from the reaction estimate in Appendix~\ref{app:explicit-estimates}.

\begin{lemma}\label{cv:sublevel}
For $\rho\in\mathcal P(\mathbb R^d)$ with $L(\rho)\le L(\rho_0)$,
\begin{equation}\label{cv:ai-bounds}
 c_0\le a_i(\rho)\le M\quad(1\le i\le N),
\end{equation}
and
\begin{equation}\label{cv:g-bounds}
 \|g_\rho\|_\infty\le M/c_0,\qquad
 \|\nabla g_\rho\|_\infty\le M_1/c_0,\qquad
 \|\nabla^2g_\rho\|_\infty\le M_2/c_0.
\end{equation}
Moreover,
\begin{equation}\label{cv:eta-bounds}
 \eta_0\le\tfrac12,\qquad \eta_0M/c_0<1,\qquad
 \eta_0M_2/c_0<1.
\end{equation}
\end{lemma}

\begin{proposition}
\label{cv:split-estimates}
For $0<\eta\le\eta_0$,
\[
 \rho_n,\widetilde\rho_n\in\mathcal P(\mathbb R^d),\qquad
 L(\rho_n),L(\widetilde\rho_n)\le L(\rho_0)
 \quad(n\ge0).
\]
There exist $c,C>0$ such that, for every $n\ge0$,
\begin{align}
 L(\rho_n)-L(\rho_{n+1})&\ge cE_n,
                                      \label{cv:descent}\\
 \mathcal D(\rho_n)&\le CE_n,
                                      \label{cv:residual-comparison}\\
 d_{\mathrm{BL}}(\rho_{n+1},\rho_n)&\le C\sqrt{E_n}.
                                      \label{cv:step-length}
\end{align}
\end{proposition}

The sublevel and descent estimates come from the analysis of
\citet{YWR2024}.  Appendix~\ref{app:explicit-estimates} gives their
Gaussian derivation, the endpoint $\eta=\eta_0$, the comparison
\eqref{cv:residual-comparison} needed here, and the step bound
\eqref{cv:step-length}.  Since $a_i\le M$, we have $L\ge-\log M$.  Thus
\eqref{cv:descent} and \eqref{cv:residual-comparison} give
\begin{equation}\label{cv:loss-limit}
 L(\rho_n)\downarrow L_\infty\in\mathbb R,\qquad
 \sum_{n=0}^\infty E_n<\infty,\qquad
 \mathcal D(\rho_n)\longrightarrow0.
\end{equation}
The estimate $\sum_nE_n<\infty$ does not by itself give finite trajectory
length.  To obtain $\sum_n\sqrt{E_n}<\infty$, we establish a gradient
inequality with constants uniform over an auxiliary compact parameter set.

\subsection{Uniformization on a compact level set}
We use the finite-dimensional real-analytic gradient inequality; see
\citet{Bolte2007}.  In an analytic chart near a critical point $z_*$, it gives
$|H(z)-H(z_*)|^\theta\le C\|\operatorname{grad}H(z)\|$ for some
$\theta\in[1/2,1)$.  The gradient below is the gradient on the surrounding
analytic manifold, rather than a gradient projected onto a compact constraint
set.

\begin{lemma}[Compact-level uniformization]\label{cv:compact-level}
Let $\mathscr M$ be a finite-dimensional real-analytic manifold with an
analytic Riemannian metric.  Suppose that $\mathscr K\subset\mathscr M$ is compact, that $H$ is real
analytic on a neighborhood of $\mathscr K$, and that
$\mathscr K\cap H^{-1}(\{\lambda\})\ne\varnothing$.  Then there exist
$\theta\in[1/2,1)$, $C>0$ and $\delta>0$ such that
\begin{equation}\label{cv:compact-gradient}
 |H(z)-\lambda|^\theta\le C\|\operatorname{grad}H(z)\|
 \quad\text{for }z\in\mathscr K,\quad |H(z)-\lambda|<\delta.
\end{equation}
\end{lemma}

\begin{proof}
Put $A=\mathscr K\cap H^{-1}(\{\lambda\})$.  For
$z_*\in A$ with $\operatorname{grad}H(z_*)=0$, choose a real-analytic chart
$z=\chi(v)$, and put $h=H\circ\chi$.  If $\Gamma(v)$ is the metric
matrix, restriction of the chart gives
\[
 0\prec\Gamma(v)\preceq b_*I,
 \qquad \Gamma(v)^{-1}\succeq b_*^{-1}I
 \quad(0<b_*<\infty).
\]
Consequently,
\begin{equation}\label{cv:chart-gradient}
 \|\operatorname{grad}H(\chi(v))\|^2
 =\nabla h(v)^{\mathsf T}\Gamma(v)^{-1}\nabla h(v)
 \ge b_*^{-1}|\nabla h(v)|^2.
\end{equation}
The analytic gradient inequality for $h$ therefore gives a neighborhood
$U_*$ of $z_*$ and constants $C_*>0$, $\theta_*\in[1/2,1)$ such that
\[
 |H(z)-\lambda|^{\theta_*}
 \le C_*|\nabla h(v)|
 \le C_*\sqrt{b_*}\,\|\operatorname{grad}H(z)\|,
 \qquad z\in U_*.
\]
For $z_*\in A$ with $\operatorname{grad}H(z_*)\ne0$, continuity gives
$U_*\ni z_*$ and $b_*>0$ such that
\[
 \inf_{z\in U_*}\|\operatorname{grad}H(z)\|\ge b_*>0.
\]
With $\theta_*=1/2$,
\[
 |H(z)-\lambda|^{1/2}\le1
 \le b_*^{-1}\|\operatorname{grad}H(z)\|
 \quad\text{when }z\in U_*,\quad |H(z)-\lambda|<1.
\]
Compactness gives a finite cover $A\subset\bigcup_{j=1}^JU_j$.  Absorb the
metric factors into $C_j$, and set
\[
 \theta=\max_{1\le j\le J}\theta_j<1,\qquad
 C=\max_{1\le j\le J}C_j,\qquad U=\bigcup_{j=1}^JU_j.
\]
For $z\in U_j$ and $t=|H(z)-\lambda|<1$,
$\theta\ge\theta_j$ implies $t^\theta\le t^{\theta_j}$; hence
\[
 |H(z)-\lambda|^\theta
 \le |H(z)-\lambda|^{\theta_j}
 \le C\|\operatorname{grad}H(z)\|.
\]
If $\mathscr K\setminus U\ne\varnothing$, compactness and
$(\mathscr K\setminus U)\cap H^{-1}(\{\lambda\})=\varnothing$ give
\[
 d_*:=\min_{z\in\mathscr K\setminus U}|H(z)-\lambda|>0,
 \qquad \delta=\min\{1,d_*/2\}.
\]
If $\mathscr K\setminus U=\varnothing$, set $\delta=1$.  Thus, in either case,
\[
 z\in\mathscr K,\quad |H(z)-\lambda|<\delta
 \quad\Longrightarrow\quad z\in U,
\]
and the preceding inequality proves \eqref{cv:compact-gradient}.
\end{proof}

\section{Finite length and convergence of the mixing measures}
\label{sec:finite-length}

\begin{proof}[Proof of Corollary~4.1]
Let $e_n=L(\rho_n)-L_\infty$ and let $C_s$ be the constant in
\eqref{cv:step-length}.  For $m>n\ge n_0$, \eqref{cv:step-length} and
(4.3) give
\[
 d_{\mathrm{BL}}(\rho_m,\rho_n)
 \le\sum_{k=n}^{m-1}d_{\mathrm{BL}}(\rho_{k+1},\rho_k)
 \le C_s\sum_{k=n}^{m-1}\sqrt{E_k}
 \le\frac{C_sC_g}{c(1-\theta)}\,e_n^{1-\theta}.
\]
Letting $m\to\infty$, with $d_{\mathrm{BL}}(\rho_m,\widehat\rho)\to0$,
yields (4.11).
\end{proof}

\section{Finite particles and suboptimal attraction}\label{sec:particles}
We now prove Theorem~2.3. The argument concerns the same transport--reaction update as Theorem~2.1, but with a different initial support. Throughout this section $N=3$, $d=2$, and the dependence on the observations is written explicitly:
\[
 \begin{gathered}
 a_{X,i}(\rho)=\int\phi(X_i-x)\,d\rho(x),\qquad
 L_X(\rho)=-\frac13\sum_{i=1}^3\log a_{X,i}(\rho),\\
 g_{X,\rho}(x)=\frac13\sum_{i=1}^3\frac{\phi(X_i-x)}{a_{X,i}(\rho)}.
 \end{gathered}
\]
Write $\mathcal S_{X,\eta}$ for one complete update. Thus
\begin{equation}\label{pt:split}
 T_{X,\rho}(x)=x+\eta\nabla g_{X,\rho}(x),\qquad
 \widetilde\rho=(T_{X,\rho})_\#\rho,\qquad
 \mathcal S_{X,\eta}(\rho)=(1-\eta+\eta g_{X,\widetilde\rho})\widetilde\rho.
\end{equation}
The reaction in this formula is evaluated at the transported measure.

\subsection{Exact aggregation and a symmetric orbit}
\begin{lemma}[Aggregation]\label{pt:aggregation}
For $0<\eta\le1$ and $\rho=\sum_{k=1}^Kp_k\delta_{z_k}$, where $K<\infty$, $p_k>0$ and $\sum_kp_k=1$,
\[
 \mathcal S_{X,\eta}^n(\rho)\in\mathcal P(\mathbb R^2),\qquad
 \#\operatorname{supp}\mathcal S_{X,\eta}^n(\rho)\le K
 \qquad(n\in\mathbb N_0).
\]
Coincident particles aggregate by summing their masses. In particular, the initialization in Theorem~2.3 is exactly equivalent to at most three groups, initially at $X_j$ with masses
\begin{equation}\label{pt:counts}
 q_j=C_j/m,\qquad C_j=\sum_{\ell=1}^m\mathbf1_{\{J_\ell=j\}},\qquad j=1,2,3.
\end{equation}
\end{lemma}
\begin{proof}
For $\rho=\sum_{k=1}^Kp_k\delta_{z_k}$, $p_k>0$, $\sum_kp_k=1$, write
\[
 a_i=\sum_{k=1}^Kp_k\phi(X_i-z_k)>0,\qquad
 z_k^+=z_k+\frac\eta3\sum_{i=1}^3
       \frac{\nabla\phi(z_k-X_i)}{a_i}.
\]
Every $z_k^+$ is finite. At $\widetilde\rho=\sum_kp_k\delta_{z_k^+}$,
\begin{align*}
 g_{X,\widetilde\rho}(z_k^+)&>0,
 &p_k^+&=p_k\{1-\eta+\eta g_{X,\widetilde\rho}(z_k^+)\}>0,\\
 \sum_kp_k^+
 &=1-\eta+\eta\int g_{X,\widetilde\rho}\,d\widetilde\rho\\
 &=1-\eta+\frac\eta3\sum_{i=1}^3
 \frac{a_{X,i}(\widetilde\rho)}{a_{X,i}(\widetilde\rho)}=1.
\end{align*}
The positivity of the updated weights follows from $0<\eta\le1$. Thus $\mathcal S_{X,\eta}^n(\rho)$ is a finitely supported probability measure for every finite $n$. For coincident particles,
\[
 z_k=z_\ell
 \ \Longrightarrow\ z_k^+=z_\ell^+
 \ \Longrightarrow\ p_k^++p_\ell^+
 =(p_k+p_\ell)\{1-\eta+\eta g_{X,\widetilde\rho}(z_k^+)\}.
\]
Thus aggregation is preserved at every iteration, and
\[
 \frac1m\sum_{\ell=1}^m\delta_{X_{J_\ell}}
 =\sum_{j=1}^3\frac{C_j}{m}\delta_{X_j},
\]
which is \eqref{pt:counts}.
\end{proof}

For clarity, recall the constants in Theorem~2.3:
\begin{equation}\label{pt:triangle}
\begin{gathered}
 e_1=(1,0),\qquad e_2=(-1/2,\sqrt3/2),\qquad e_3=(-1/2,-\sqrt3/2),\\
 r_\triangle^2=\frac53\log3,\qquad X_i^\triangle=r_\triangle e_i,\qquad
 r_*=\frac25r_\triangle,\qquad
 \rho_*=\frac13\sum_{j=1}^3\delta_{r_*e_j},\\
 \delta_\triangle=\frac{22}{15}\log3-\log5.
\end{gathered}
\end{equation}
For $r>0$ put $\rho_r=3^{-1}\sum_j\delta_{re_j}$. At $X=X^\triangle$, symmetry makes all three fitted values equal to
\begin{equation}\label{pt:fitted}
 f(r)=\phi(0)\exp\!\left(-\frac{r_\triangle^2+r^2}{2}\right)
 \frac{\exp(r_\triangle r)+2\exp(-r_\triangle r/2)}3.
\end{equation}
Indeed, $e_i\cdot e_j=1$ for $i=j$ and $-1/2$ otherwise, so this follows by expanding $|r_\triangle e_i-re_j|^2$.
Let $F(x)=3^{-1}\sum_i\phi(r_\triangle e_i-x)$. Then
\[
 g_{X^\triangle,\rho_r}(x)=F(x)/f(r),\qquad F(re_j)=f(r).
\]
Reflection in the line spanned by $e_j$ leaves $F$ invariant. The tangential derivative of $F$ at $re_j$ therefore vanishes. Its radial derivative is $f'(r)$, and hence
\begin{equation}\label{pt:radial}
 \nabla g_{X^\triangle,\rho_r}(re_j)=v(r)e_j,\qquad
 v(r)=\frac{f'(r)}{f(r)}
 =-r+r_\triangle\frac{q(r)-1}{q(r)+2},\qquad
 q(r)=\exp(3r_\triangle r/2).
\end{equation}
After transport the three support points still have a common radius. At that radius $g$ equals one at every support point, so the reaction leaves all three masses equal.

\begin{lemma}[Entry from the empirical measure]\label{pt:radial-entry}
For every $0<\eta\le1$,
\[
 \mathcal S_{X^\triangle,\eta}^n\!\left(\frac13\sum_j\delta_{X_j^\triangle}\right)=\rho_{r_n},\qquad n\in\mathbb N_0,
\]
where
\[
 r_0=r_\triangle,\qquad r_{n+1}=r_n+\eta v(r_n),\qquad r_n\downarrow r_*.
\]
Furthermore,
\begin{equation}\label{pt:exact-gap}
 L_{X^\triangle}(\rho_*)-L_{X^\triangle}(\delta_0)=\delta_\triangle>0.
\end{equation}
\end{lemma}
\begin{proof}
By \eqref{pt:triangle} and \eqref{pt:radial},
\[
 q(r_*)=\exp\!\left(\frac32r_\triangle\frac25r_\triangle\right)
 =\exp(\log3)=3,\qquad
 v(r_*)=-\frac25r_\triangle+r_\triangle\frac{3-1}{3+2}=0.
\]
Differentiating \eqref{pt:radial} gives
\begin{equation}\label{pt:radial-derivative}
 v'(r)=-1+\frac{9r_\triangle^2}{2}\frac{q(r)}{(q(r)+2)^2}.
\end{equation}
For $r\ge r_*$,
\[
 q(r)\ge3,\qquad
 \frac{d}{dq}\frac{q}{(q+2)^2}=\frac{2-q}{(q+2)^3}<0,
 \qquad \frac{q(r)}{(q(r)+2)^2}\le\frac3{25}.
\]
Consequently,
\begin{equation}\label{pt:radial-bounds}
 -1<v'(r)\le-1+\frac{27r_\triangle^2}{50}
 =-1+\frac9{10}\log3<0.
\end{equation}
The last strict inequality follows from
\[
 e^{10/9}>\sum_{k=0}^4\frac{(10/9)^k}{k!}
 =\frac{59453}{19683}>3
 \quad\Longrightarrow\quad \log3<10/9.
\]
Put $R_\eta(r)=r+\eta v(r)$. For $0<\eta\le1$,
\[
 R_\eta(r_*)=r_*,\qquad
 R_\eta'(r)=1+\eta v'(r)>0,\qquad
 r>r_*\ \Longrightarrow\ v(r)=\int_{r_*}^rv'(t)\,dt<0.
\]
Hence
\[
 r>r_*\ \Longrightarrow\ r_*=R_\eta(r_*)<R_\eta(r)<r,
 \qquad r_0=r_\triangle>r_*\ \Longrightarrow\ r_n\downarrow r_\infty\ge r_*.
\]
Passing to the limit in the recursion,
\[
 0=\lim_n\frac{r_{n+1}-r_n}{\eta}=v(r_\infty),\qquad
 v(r_\infty)=v(r_*)=0,\quad v'<0
 \ \Longrightarrow\ r_\infty=r_*.
\]
For the likelihood comparison,
\[
 r_\triangle r_*=\frac23\log3,\qquad \frac{r_*^2}{2}=\frac2{15}\log3,
\]
and \eqref{pt:fitted} yields
\begin{align*}
 \frac{f(r_*)}{f(0)}
 &=e^{-r_*^2/2}\frac{e^{r_\triangle r_*}+2e^{-r_\triangle r_*/2}}3\\
 &=3^{-2/15}\frac{3^{2/3}+2\,3^{-1/3}}3
 =\frac5{3^{22/15}},\\
 L_{X^\triangle}(\rho_*)-L_{X^\triangle}(\delta_0)
 &=-\log f(r_*)+\log f(0)
 =\frac{22}{15}\log3-\log5=\delta_\triangle>0.
\end{align*}
The last inequality is exact:
\[
 3^{22}=31381059609>30517578125=5^{15}.
\]
The argument uses $\delta_0$ only as a competitor and makes no claim that it is an NPMLE.
\end{proof}

\subsection{The complete position--weight Hessian}
We allow all three positions and all positive masses to vary. Set
\[
 E=\begin{pmatrix}
 1/\sqrt2&1/\sqrt6\\
 -1/\sqrt2&1/\sqrt6\\
 0&-2/\sqrt6
 \end{pmatrix},\qquad E^\top E=I_2,\qquad E^\top\mathbf1=0.
\]
Use the eight-dimensional chart
\begin{equation}\label{pt:chart}
 y=(z,u)\in\mathbb R^6\times\mathbb R^2,
 \qquad p=\mathbf1/3+Eu,\qquad
 \mathcal F_X(z,u)=L_X\!\left(\sum_{j=1}^3p_j\delta_{z_j}\right),
\end{equation}
on the open set $p_j>0$. Its base state is
\[
 y_*=(r_*e_1,r_*e_2,r_*e_3,0).
\]
The vectors $e_j$ describe the triangle; $u$ is the independent weight coordinate.

\begin{proposition}[Full reduced Hessian]\label{pt:hessian}
The matrix $H_*=D^2\mathcal F_{X^\triangle}(y_*)$ is positive definite, with blocks
\begin{equation}\label{pt:hessian-blocks}
 H_*\succ0,\qquad 0\prec(H_*)_{zz}\preceq I_6/3,
 \qquad (H_*)_{uu}=\frac{12}{25}I_2.
\end{equation}
\end{proposition}
\begin{proof}
We use the results from Appendix~C of the Supplement and write
\[
 \begin{gathered}
 t_{ij}=r_\triangle e_i-r_*e_j,\qquad
 D_{ij}=\phi(t_{ij})/f(r_*),\\
 S_j=\frac13\sum_iD_{ij}t_{ij}t_{ij}^\top,\qquad h=Ew.
 \end{gathered}
\]
Equations \textup{(C.3)}, \textup{(C.4)} and \textup{(C.7)} give
\begin{align*}
 (\xi,w)^\top H_*(\xi,w)
 &=\frac13\sum_i\left(\sum_jD_{ij}h_j+
        \frac13\sum_jD_{ij}t_{ij}\cdot\xi_j\right)^2
   +\frac13\sum_j\xi_j^\top(I_2-S_j)\xi_j,\\
 I_2-S_j&\succeq\left(1-\frac9{10}\log3\right)I_2\succ0,\\
 \xi=0\quad&\Longrightarrow\quad
 (0,w)^\top H_*(0,w)=\frac{12}{25}|w|^2.
\end{align*}
Thus $(\xi,w)\ne0$ implies $(\xi,w)^\top H_*(\xi,w)>0$. For $w=0$, the Jensen bound in Appendix~C yields
\[
 0<\xi^\top(H_*)_{zz}\xi
 \le\frac13\sum_j\xi_j^\top\{S_j+(I_2-S_j)\}\xi_j
 =\frac13|\xi|^2\qquad(\xi\ne0).
\]
These identities give \eqref{pt:hessian-blocks}, with all position--weight terms retained in the squares.
\end{proof}

\subsection{Contraction of the exact split map}
At a positive-weight state, the gradients in the reduced chart are
\begin{equation}\label{pt:chart-gradients}
 \nabla_{z_j}\mathcal F_X=-p_j\nabla g_{X,\rho}(z_j),
 \qquad
 \nabla_u\mathcal F_X=-E^\top(g_{X,\rho}(z_j))_{j=1}^3.
\end{equation}
Define
\[
 A(p)=\operatorname{diag}(p_1^{-1}I_2,p_2^{-1}I_2,p_3^{-1}I_2),
 \qquad
 B(p)=E^\top\{\operatorname{diag}(p)-pp^\top\}E.
\]
In these coordinates, the transport--reaction map $\mathcal S_{X,\eta}$ is
\begin{equation}\label{pt:chart-split}
 z^+=z-\eta A(p)\nabla_z\mathcal F_X(z,u),
 \qquad
 u^+=u-\eta B(p)\nabla_u\mathcal F_X(z^+,u).
\end{equation}
Here the weight gradient is evaluated at the new positions and old weights. To verify the second identity, put $G=\operatorname{diag}(p)-pp^\top$ and let $g$ be the vector of support values computed at $(z^+,u)$. Since $p^\top g=1$,
\[
 Gg=\operatorname{diag}(p)(g-\mathbf1),\qquad
 G\mathbf1=0,\qquad GEE^\top=G.
\]
The reaction therefore satisfies
\[
 u^+-u=\eta E^\top\operatorname{diag}(p)(g-\mathbf1)
 =\eta E^\top GEE^\top g
 =-\eta B(p)\nabla_u\mathcal F_X(z^+,u).
\]
Both $A(p)$ and $B(p)$ are positive definite. For the latter, if $w\ne0$,
\[
 \begin{aligned}
 w^\top B(p)w
 &=\sum_jp_j(Ew)_j^2-\left(\sum_jp_j(Ew)_j\right)^2\\
 &=\frac12\sum_{j,k}p_jp_k\{(Ew)_j-(Ew)_k\}^2>0.
 \end{aligned}
\]
Indeed, equality would make the three values $(Ew)_j$ constant; their zero sum would then give $Ew=0$ and hence $w=0$.

Every interior critical point of $\mathcal F_X$ is a fixed point of \eqref{pt:chart-split}. Indeed,
\[
 \begin{gathered}
 \nabla_z\mathcal F_X=0\ \Longrightarrow\ \nabla g(z_j)=0\quad(j=1,2,3),\\
 \nabla_u\mathcal F_X=0
 \ \Longrightarrow\ E^\top g=0
 \ \Longrightarrow\ g=c\mathbf1,
 \qquad 1=p^\top g=c\sum_jp_j=c,\\
 z^+=z,\qquad u^+=u.
 \end{gathered}
\]

\begin{proposition}[Local attraction at every step size]\label{pt:contraction}
For every fixed $0<\eta\le1$, $y_*$ is an interior fixed point and
\begin{equation}\label{pt:strict-contraction}
 \kappa_\eta:=
 \|H_*^{1/2}D_y\mathcal S_{X^\triangle,\eta}(y_*)H_*^{-1/2}\|_2<1.
\end{equation}
\end{proposition}
\begin{proof}
By \textup{(C.2)} and \textup{(C.1)}, at $y_*$,
\[
 \nabla g(z_j)=0,\quad g(z_j)=1
 \quad\Longrightarrow\quad \nabla_y\mathcal F_{X^\triangle}(y_*)=0,
 \qquad A=3I_6,\quad B=I_2/3.
\]
Set
\[
 P=\operatorname{diag}(3I_6,0_2),\qquad
 Q=\operatorname{diag}(0_6,I_2/3).
\]
Since $(D_yA)\nabla_z\mathcal F=(D_yB)\nabla_u\mathcal F=0$ at $y_*$, differentiation in transport--reaction order gives
\begin{equation}\label{pt:jacobian}
 J_\eta:=D_y\mathcal S_{X^\triangle,\eta}(y_*)
 =(I_8-\eta QH_*)(I_8-\eta PH_*).
\end{equation}
The right factor is the transport Jacobian, and the left factor is the reaction Jacobian.

Put $C=H_*^{1/2}PH_*^{1/2}$ and $D_0=H_*^{1/2}QH_*^{1/2}$. With $V=H_*^{1/2}P^{1/2}$,
\[
 C=VV^\top,\quad P^{1/2}H_*P^{1/2}=V^\top V
 =\operatorname{diag}(3(H_*)_{zz},0_2),
\]
so $C$ and $3(H_*)_{zz}$ have the same nonzero eigenvalues. The same argument with $Q$ identifies the nonzero eigenvalues of $D_0$ with those of $(H_*)_{uu}/3$. Proposition~\ref{pt:hessian} therefore gives
\begin{equation}\label{pt:two-factors}
 0\preceq C\preceq I_8,\qquad
 0\preceq D_0\preceq\frac4{25}I_8.
\end{equation}
For $M=M^\top$, $0\preceq M\preceq I_8$ and $0<\eta\le1$,
\begin{equation}\label{pt:norm-loss}
 |v|^2-|(I_8-\eta M)v|^2
 =\eta v^\top M(2I_8-\eta M)v\ge\eta v^\top Mv\ge0.
\end{equation}
Hence $|(I_8-\eta M)v|=|v|$ if and only if $Mv=0$. Now
\begin{align*}
 K_\eta&:=H_*^{1/2}J_\eta H_*^{-1/2}
 =(I_8-\eta D_0)(I_8-\eta C),\\
 |K_\eta v|&\le|(I_8-\eta C)v|\le|v|,\\
 |K_\eta v|=|v|
 &\Longrightarrow Cv=0\ \text{and}\ D_0(I_8-\eta C)v=0\\
 &\Longrightarrow Cv=D_0v=0
 \Longrightarrow v^\top(C+D_0)v=0
 \Longrightarrow v=0,
\end{align*}
because $C+D_0=H_*^{1/2}(P+Q)H_*^{1/2}\succ0$. Compactness of $\{v:|v|=1\}$ then yields
\[
 \kappa_\eta=\|K_\eta\|_2
 =\max_{|v|=1}|K_\eta v|<1,
\]
which is \eqref{pt:strict-contraction}.
\end{proof}

The contraction holds in the eight-dimensional aggregated chart. When labeled particles coincide, redistributing their total mass leaves the measure unchanged and produces neutral directions in the labeled coordinates.

\subsection{An open basin for perturbed data}
Fix $\eta\in(0,1]$. The functions $\mathcal F_X$ and $\mathcal S_{X,\eta}$ are smooth on the positive-weight chart. Since $H_*\succ0$, the implicit function theorem applied to $\nabla_y\mathcal F_X(y)=0$ gives a smooth branch $y_*(X)$ for $X$ near $X^\triangle$, with $y_*(X^\triangle)=y_*$. Restrict the neighborhood so that $y_*(X)$ has positive weights and distinct support points. Each $y_*(X)$ is a fixed point of the exact split map, by the critical-point observation above.

Use the fixed norm $\|v\|_*=(v^\top H_*v)^{1/2}$, with the corresponding induced operator norm, and put $\bar\kappa_\eta=(1+\kappa_\eta)/2<1$. Continuity of the derivative permits a radius $r>0$ and a smaller data neighborhood such that the closed balls $\|y-y_*(X)\|_*\le r$ lie in the chart and
\begin{equation}\label{pt:uniform-derivative}
 \|y_*(X)-y_*\|_*<r/8,
 \qquad
 \sup_{\|y-y_*(X)\|_*\le r}
 \|D_y\mathcal S_{X,\eta}(y)-J_\eta\|_*
 \le\frac{1-\kappa_\eta}{2}.
\end{equation}
First choose a small fixed ball about $y_*$ on which the derivative is close to $J_\eta$, uniformly for $X$ near $X^\triangle$. Take $r$ smaller than its radius, then restrict the data neighborhood so that $\|y_*(X)-y_*\|_*<r/8$ and every radius-$r$ ball about $y_*(X)$ lies inside the fixed ball.

If $\|y-y_*(X)\|_*\le r$, the segment joining $y_*(X)$ to $y$ stays in the same ball. The integral mean-value formula gives
\begin{align}
 \mathcal S_{X,\eta}(y)-y_*(X)
 &=\int_0^1D_y\mathcal S_{X,\eta}\bigl(y_*(X)+t(y-y_*(X))\bigr)
       (y-y_*(X))\,dt,\notag\\
 \|\mathcal S_{X,\eta}(y)-y_*(X)\|_*
 &\le\left(\kappa_\eta+\frac{1-\kappa_\eta}{2}\right)
       \|y-y_*(X)\|_*
 =\bar\kappa_\eta\|y-y_*(X)\|_* .\label{pt:basin-contraction}
\end{align}
In particular these balls are invariant.

Lemma~\ref{pt:radial-entry} gives a finite integer $K$ such that
\begin{equation}\label{pt:base-entry}
 \|\mathcal S_{X^\triangle,\eta}^K(X^\triangle,0)-y_*\|_*<r/4.
\end{equation}
Here $(X,u)$ denotes the initial state with positions $z_j=X_j$ and weights $\mathbf1/3+Eu$. The finite composition
\[
 \mathcal H(X,u)=\mathcal S_{X,\eta}^K(X,u)
\]
is smooth near $(X^\triangle,0)$, since the weights along the base orbit up to time $K$ are positive and remain so for nearby $(X,u)$. On a smaller compact convex neighborhood, $D\mathcal H$ is bounded, so there is a constant $C\ge1$ such that
\begin{equation}\label{pt:pullback}
 \|\mathcal H(X,u)-\mathcal H(X^\triangle,0)\|_*
 \le C\bigl(|X-X^\triangle|+|u|\bigr).
\end{equation}
Choose
\begin{equation}\label{pt:epsilon}
 0<\varepsilon_\eta\le\min\left\{\frac1{12},\frac{r}{16\sqrt3 C}\right\},
\end{equation}
small enough to stay in this neighborhood, and restrict the data further so that $|X-X^\triangle|<r/(16C)$. If $q$ is any probability vector with $\max_j|q_j-1/3|\le\varepsilon_\eta$, then all $q_j>0$, and its chart coordinate satisfies
\[
 u=E^\top(q-\mathbf1/3),\qquad
 |u|=|q-\mathbf1/3|\le\sqrt3\varepsilon_\eta.
\]
Equations \eqref{pt:uniform-derivative}, \eqref{pt:base-entry} and \eqref{pt:pullback} imply
\begin{align*}
 \|\mathcal H(X,u)-y_*(X)\|_*
 &\le C\bigl(|X-X^\triangle|+|u|\bigr)
      +\|\mathcal H(X^\triangle,0)-y_*\|_*
      +\|y_*-y_*(X)\|_*\\
 &<r/8+r/4+r/8=r/2.
\end{align*}
Repeated application of \eqref{pt:basin-contraction} therefore gives, for every $n\ge K$,
\begin{equation}\label{pt:geometric}
 \|\mathcal S_{X,\eta}^n(X,u)-y_*(X)\|_*
 \le\frac r2\bar\kappa_\eta^{\,n-K}.
\end{equation}

For each fixed $X$, define a subset of the positive-weight chart by
\[
 \mathcal V_X=\bigl\{y:\ \|\mathcal S_{X,\eta}^K(y)-y_*(X)\|_*<r\bigr\}.
\]
This set is open by continuity of the finite-step map. Every orbit starting in $\mathcal V_X$ converges to $y_*(X)$ by \eqref{pt:basin-contraction}, and the initial states $(X,u)$ considered
above belong to $\mathcal V_X$.

The constants $K$, $r$ and $\bar\kappa_\eta$ are uniform over the final data neighborhood and all initial weights satisfying $\max_j|q_j-1/3|\le\varepsilon_\eta$; in particular, they do not depend on $m$. Since $E^\top E=I_2$, norm equivalence gives the coordinate bound in Theorem~2.3 with $\theta=\bar\kappa_\eta$ and rate constant $(r/2)\sqrt{2/\lambda_{\min}(H_*)}$.

Equation~\eqref{pt:geometric} gives geometric convergence of the positions and weights. The limiting measure is
\[
 \rho_{\mathrm{bad}}(X)=\sum_{j=1}^3p_{*,j}(X)\delta_{z_{*,j}(X)}.
\]
For every $h\in C_b(\mathbb R^2)$,
\[
 \sum_jp_{n,j}h(z_{n,j})
 \longrightarrow
 \sum_jp_{*,j}(X)h(z_{*,j}(X)),
\]
so $\rho_n\rightharpoonup\rho_{\mathrm{bad}}(X)$. Moreover,
\[
 a_{X,i}(\rho_n)
 \longrightarrow
 \sum_jp_{*,j}(X)\phi(X_i-z_{*,j}(X))>0.
\]
Since these limits are positive, continuity of the logarithm gives $L_X(\rho_n)\to L_X(\rho_{\mathrm{bad}}(X))$.

Finally, shrink the open data neighborhood $U_\eta$ once more so that
\[
 |\mathcal F_X(y_*(X))-\mathcal F_{X^\triangle}(y_*)|<\delta_\triangle/4,
 \qquad
 |L_X(\delta_0)-L_{X^\triangle}(\delta_0)|<\delta_\triangle/4.
\]
Together with \eqref{pt:exact-gap}, this gives
\begin{align}
 L_X(\rho_{\mathrm{bad}}(X))-L_X^*
 &\ge L_X(\rho_{\mathrm{bad}}(X))-L_X(\delta_0)\notag\\
 &>L_{X^\triangle}(\rho_*)-L_{X^\triangle}(\delta_0)
       -\delta_\triangle/2
 =\delta_\triangle/2.\label{pt:persistent-gap}
\end{align}
The comparison uses only $\delta_0$, so the global optimizer need not be identified or unique. All constants and neighborhoods above are chosen for fixed $\eta$ and may depend on it.

\subsection{Random initialization}
By Lemma~\ref{pt:aggregation}, the particle iteration reduces to at most three groups with initial weights $q_j=C_j/m$. Appendix~F gives
\begin{equation}\label{pt:prob-summary}
 \mathbb P\{\max_j|C_j/m-1/3|>\varepsilon_\eta\}
 \le6e^{-2m\varepsilon_\eta^2}.
\end{equation}
On the complementary event, \eqref{pt:geometric} and \eqref{pt:persistent-gap} give convergence to $\rho_{\mathrm{bad}}(X)$ and the likelihood gap, proving (2.8). No convergence result on the exceptional event is needed. Under the coupling in Theorem~2.3, the bounds in \eqref{pt:prob-summary} are summable in $m$, so the first Borel--Cantelli lemma gives the same conclusions almost surely for all sufficiently large $m$.

\section{Failure of pure transport from a smooth full-support density}
\label{sec:smooth-W}

We remove the reaction step and start from a smooth, strictly positive
density. The proof treats $\gamma=0$ directly and is independent of
the convergence argument in Sections~3 and~4 of the main paper.

Fix $s>0$ as in Theorem~2.4, and set
\begin{equation}\label{sw:data}
 R=\sqrt{\log 7},\qquad X_i=Re_i\quad(1\le i\le3),\qquad
 F(x)=\frac13\sum_{i=1}^3\phi(x-X_i),\qquad M=(2\pi)^{-1},
\end{equation}
where the vectors $e_i$ are defined in (2.6) and $\phi$ is the
standard two-dimensional Gaussian density.  Let $Q$ denote rotation through
$2\pi/3$, so that $QX_i=X_{i+1}$, with cyclic indices.  Define
\begin{equation}\label{sw:constants}
 \begin{gathered}
 F_0=F(0)=M e^{-R^2/2}=\frac M{\sqrt7},\\
 \kappa=F_0(1-R^2/2)>0,\qquad r=\frac{\kappa}{8M},\qquad
 p_s=1-e^{-r^2/(2s^2)},\\
 H=F(X_1/2)=\frac{F_0}{3}(u+2/u),\qquad u=7^{3/8}.
 \end{gathered}
\end{equation}
The radius $R$ here differs from $r_\triangle$ in the particle theorem.
For this choice of $R$, the origin is a strict local maximum of $F$.
Yet $F(X_1/2)>F(0)$.

\begin{lemma}[Geometry and a finite critical set]\label{sw:geometry}
For every orthogonal symmetry $A$ of $\{X_1,X_2,X_3\}$,
\[
 A\{X_1,X_2,X_3\}=\{X_1,X_2,X_3\}
 \quad\Longrightarrow\quad F(Ax)=F(x)\quad(x\in\mathbb R^2).
\]
Moreover,
\begin{equation}\label{sw:geometry-bounds}
 0<F\le M,\qquad \|\nabla^2F\|_{\mathrm{op}}\le M,
 \qquad \nabla F(0)=0,\qquad \nabla^2F(0)=-\kappa I,
\end{equation}
\begin{equation}\label{sw:local-Hessian}
 -MI\preceq\nabla^2F(x)\preceq-\frac\kappa2I
 \quad(|x|\le r),
\end{equation}
and
\[
 H>F_0,\qquad
 \mathcal C:=\{x\in\mathbb R^2:\nabla F(x)=0\}
 \subseteq\{x:|x|\le R\},\qquad \#\mathcal C<\infty.
\]
\end{lemma}

\noindent The proof of Lemma~\ref{sw:geometry} is given in
Appendix~\ref{app:smooth-geometry}.

\begin{proof}[Proof of Theorem~2.4]
For the initialization $\rho_0=\mathcal N(0,s^2I_2)$,
Gaussian convolution gives
\begin{equation}\label{sw:c-init}
 c_{\mathrm{init}}=\int F\,d\rho_0
 =\frac{1}{2\pi(1+s^2)}
   \exp\left\{-\frac{R^2}{2(1+s^2)}\right\}>0.
\end{equation}
For $0<\eta\le c_{\mathrm{init}}/(2M)$, the pure transport iteration is
\begin{equation}\label{sw:pure-update}
 \rho_{n+1}=(T_n)_\#\rho_n,\qquad
 T_n=\operatorname{id}+\eta\nabla g_{\rho_n}.
\end{equation}
There is no reaction update in \eqref{sw:pure-update}.

\paragraph*{Equal fitted values and a closed ascent argument}
Since $Q_\#\rho_0=\rho_0$, the induction starts at $n=0$.
For $Q_\#\rho=\rho$, change of variables gives, with cyclic indices,
\[
 \begin{aligned}
 a_{i+1}(\rho)&=\int\phi(x-QX_i)\,\rho(dx)
 =\int\phi(Qy-QX_i)\,\rho(dy)=a_i(\rho),\\
 a_i(\rho)&=\int F\,d\rho=:c,\qquad g_\rho=F/c.
 \end{aligned}
\]
We prove $Q_\#\rho_n=\rho_n$ and
\begin{equation}\label{sw:c-monotone}
 c_n:=\int F\,d\rho_n=a_i(\rho_n),\qquad
 c_{\mathrm{init}}\le c_n\le c_{n+1}\le M.
\end{equation}
Assume $Q_\#\rho_n=\rho_n$ and $c_{\mathrm{init}}\le c_n\le M$.
Set $\beta_n=\eta/c_n$.  Then
\begin{equation}\label{sw:beta}
 T_n(x)=x+\beta_n\nabla F(x),\qquad
 \frac\eta M\le\beta_n\le\frac\eta{c_{\mathrm{init}}}
 \le\frac1{2M}.
\end{equation}
Rotation invariance gives
\[
 T_n\circ Q=Q\circ T_n
 \quad\Longrightarrow\quad
 Q_\#\rho_{n+1}
 =(Q\circ T_n)_\#\rho_n
 =(T_n\circ Q)_\#\rho_n
 =(T_n)_\#\rho_n=\rho_{n+1}.
\]
Taylor's formula with integral remainder and
$\nabla^2F\succeq-MI$ give
\begin{align}
 F(T_nx)-F(x)
 &=\beta_n|\nabla F(x)|^2
 +\beta_n^2\int_0^1(1-t)
   \nabla F(x)^{\mathsf T}
   \nabla^2F(x+t\beta_n\nabla F(x))\nabla F(x)\,dt\notag\\
 &\ge\beta_n\left(1-\frac{M\beta_n}{2}\right)|\nabla F(x)|^2
 \ge\frac{3\beta_n}{4}|\nabla F(x)|^2.\label{sw:point-ascent}
\end{align}
Consequently,
\[
 c_n\le c_n+\frac{3\beta_n}{4}\int|\nabla F|^2\,d\rho_n
 \le\int F(T_nx)\,\rho_n(dx)=c_{n+1}\le M.
\]
The last bound uses $F\le M$; the induction proves
\eqref{sw:c-monotone}--\eqref{sw:point-ascent} for every $n\ge0$.

\paragraph*{Positive smooth densities at every finite iteration}
For every $n\ge0$ and $x\in\mathbb R^2$,
\begin{equation}\label{sw:Jacobian}
 DT_n(x)=I+\beta_n\nabla^2F(x)\succeq(1-\beta_nM)I\succeq\tfrac12I.
\end{equation}
For fixed $y\in\mathbb R^2$, define
\[
 J_y(x)=\tfrac12|x|^2+\beta_nF(x)-y\cdot x.
\]
Then
\[
 J_y(x)\ge\tfrac12|x|^2-|y||x|\longrightarrow\infty
 \quad(|x|\to\infty),\qquad
 \nabla^2J_y=DT_n\succeq\tfrac12I.
\]
Coercivity and strict convexity give a unique minimizer $x_y$ of $J_y$.
Thus
\[
 T_n(x)=y\ \Longleftrightarrow\ \nabla J_y(x)=0
 \ \Longleftrightarrow\ x=x_y.
\]
So $T_n$ is bijective.  Since $T_n\in C^\infty$ and
$\det DT_n\ge1/4>0$, the inverse function theorem gives
$T_n^{-1}\in C^\infty$.  For the density $q_n$ of $\rho_n$,
\begin{equation}\label{sw:density-change}
 q_{n+1}(y)=
 \frac{q_n(T_n^{-1}y)}{\det DT_n(T_n^{-1}y)}.
\end{equation}
The change-of-variables formula yields the induction
\[
 \begin{gathered}
 q_0(y)=\frac{e^{-|y|^2/(2s^2)}}{2\pi s^2}>0,\qquad q_0\in C^\infty,\\
 \Longrightarrow\quad q_n>0,\quad q_n\in C^\infty,\quad
 \operatorname{supp}(\rho_n)=\mathbb R^2\quad(n\ge0).
 \end{gathered}
\]

\paragraph*{Bounded trajectories and convergence to a single critical point}
Set $\Psi_0=\operatorname{id}$ and
$\Psi_n=T_{n-1}\circ\cdots\circ T_0$ for $n\ge1$.
Fix $x\in\mathbb R^2$ and write $x_n=\Psi_n(x)$.
Using the weights $p_i$ from \eqref{sw:softmax}, we obtain from
\eqref{sw:gradient-barycenter}
\begin{equation}\label{sw:convex-step}
 x_{n+1}=(1-\lambda_n(x_n))x_n
       +\lambda_n(x_n)\sum_i p_i(x_n)X_i,
 \qquad \lambda_n(y)=\beta_nF(y)\in(0,1/2].
\end{equation}
Because $p_i>0$, $\sum_i p_i=1$ and $|X_i|=R$,
\[
 |x_{n+1}|\le(1-\lambda_n(x_n))|x_n|+\lambda_n(x_n)R
 \le\max\{|x_n|,R\}.
\]
Induction gives
\begin{equation}\label{sw:orbit-bound}
 |x_{n+1}|\le\max\{|x_n|,R\},\qquad
 |x_n|\le\max\{|x|,R\}.
\end{equation}
Summing \eqref{sw:point-ascent} and using $\beta_n\ge\eta/M$ gives
\begin{equation}\label{sw:gradient-sum}
 \frac{3\eta}{4M}\sum_{n=0}^{K}|\nabla F(x_n)|^2
 \le F(x_{K+1})-F(x)\le M-F(x).
\end{equation}
So $\sum_{n\ge0}|\nabla F(x_n)|^2<\infty$, and hence
\begin{equation}\label{sw:gradient-vanishing}
 \nabla F(x_n)\longrightarrow0,\qquad
 |x_{n+1}-x_n|\le\frac{\eta}{c_{\mathrm{init}}}
                         |\nabla F(x_n)|\longrightarrow0.
\end{equation}
To verify $\operatorname{dist}(x_n,\mathcal C)\to0$, suppose otherwise.
Boundedness then gives $\varepsilon_0>0$, a subsequence and a point $z$
with
\[
 x_{n_j}\to z,\qquad
 \operatorname{dist}(x_{n_j},\mathcal C)\ge\varepsilon_0,
 \qquad
 \nabla F(z)=\lim_j\nabla F(x_{n_j})=0.
\]
But $z\in\mathcal C$ implies
$\operatorname{dist}(x_{n_j},\mathcal C)\le|x_{n_j}-z|\to0$,
a contradiction.

By Lemma~\ref{sw:geometry}, $1\le\#\mathcal C<\infty$.
If $\mathcal C=\{z\}$, then $|x_n-z|\to0$.
Otherwise put
\[
 \varepsilon=\tfrac14\min_{z\ne z',\ z,z'\in\mathcal C}|z-z'|>0.
\]
Choose $n_*$ so that, for $n\ge n_*$,
\[
 \operatorname{dist}(x_n,\mathcal C)<\varepsilon,
 \qquad |x_{n+1}-x_n|<\varepsilon.
\]
The balls around distinct $z,z'\in\mathcal C$ satisfy
\[
 \operatorname{dist}(B(z,\varepsilon),B(z',\varepsilon))
 \ge|z-z'|-2\varepsilon\ge2\varepsilon.
\]
Thus $x_{n_*}\in B(z,\varepsilon)$ forces $x_n\in B(z,\varepsilon)$
for all $n\ge n_*$, and
\[
 |x_n-z|=\operatorname{dist}(x_n,\mathcal C)\longrightarrow0.
\]
We have proved
\begin{equation}\label{sw:S}
 S(x)=\lim_{n\to\infty}\Psi_n(x)\in\mathcal C
 \quad\text{for every }x\in\mathbb R^2.
\end{equation}
Each $\Psi_n$ is continuous; its pointwise limit $S$ is Borel measurable.

\paragraph*{Weak and Wasserstein convergence of the measures}
Since $\rho_n=(\Psi_n)_\#\rho_0$, for every bounded continuous $f$,
\[
 \begin{gathered}
 |f(\Psi_n(x))|\le\|f\|_\infty,\\
 \int f\,d\rho_n=\int f(\Psi_n(x))\,\rho_0(dx)
 \longrightarrow\int f(S(x))\,\rho_0(dx)
 =\int f\,d(S_\#\rho_0).
 \end{gathered}
\]
The limit follows from \eqref{sw:S} and dominated convergence.  Therefore
\begin{equation}\label{sw:weak-limit}
 \rho_n\Rightarrow\rho_\infty:=S_\#\rho_0,
 \qquad\operatorname{supp}(\rho_\infty)\subseteq\mathcal C.
\end{equation}
In particular, $\#\operatorname{supp}(\rho_\infty)<\infty$.
For $1\le p<\infty$, the coupling
$(\Psi_n,S)_\#\rho_0\in\Pi(\rho_n,\rho_\infty)$ satisfies
\[
 |\Psi_n(x)-S(x)|^p
 \le(\max\{|x|,R\}+R)^p\le(|x|+2R)^p,
 \qquad \int(|x|+2R)^p\,\rho_0(dx)<\infty.
\]
By \eqref{sw:S} and dominated convergence,
\begin{equation}\label{sw:Wp}
 W_p(\rho_n,\rho_\infty)^p
 \le\int|\Psi_n(x)-S(x)|^p\,\rho_0(dx)\longrightarrow0.
\end{equation}

\paragraph*{A positive mass converges to the suboptimal center}
For $|x|\le r$, $\nabla F(0)=0$ gives
\[
 T_n(x)=A_n(x)x,\qquad
 A_n(x):=I+\beta_n\int_0^1\nabla^2F(tx)\,dt.
\]
Equations~\eqref{sw:local-Hessian} and \eqref{sw:beta} imply
\[
 \tfrac12I\preceq(1-\beta_nM)I
 \preceq A_n(x)\preceq(1-\beta_n\kappa/2)I.
\]
Consequently,
\begin{equation}\label{sw:center-contraction}
 |T_n(x)|\le(1-\beta_n\kappa/2)|x|
 \le\left(1-\frac{\eta\kappa}{2M}\right)|x|.
\end{equation}
With $\vartheta:=1-\eta\kappa/(2M)\in(0,1)$,
\[
 |x|<r\quad\Longrightarrow\quad
 |\Psi_n(x)|\le\vartheta^n|x|<r,
 \qquad S(x)=0.
\]
Using the radial distribution of $\mathcal N(0,s^2I_2)$,
\begin{equation}\label{sw:center-mass}
 \rho_\infty(\{0\})\ge\rho_0(B(0,r))
 =1-e^{-r^2/(2s^2)}=p_s>0.
\end{equation}

\paragraph*{An explicit competitor and a strict likelihood gap}
Rotation equivariance passes to the limit:
\[
 \begin{gathered}
 \Psi_n\circ Q=Q\circ\Psi_n
 \quad\Longrightarrow\quad S\circ Q=Q\circ S,\\
 Q_\#\rho_\infty=(Q\circ S)_\#\rho_0
 =(S\circ Q)_\#\rho_0=\rho_\infty.
 \end{gathered}
\]
Since each $\psi_i$ is bounded and continuous, \eqref{sw:weak-limit} gives
\[
 c_n\longrightarrow c_\infty:=a_i(\rho_\infty)
 =\int F\,d\rho_\infty\in[c_{\mathrm{init}},M].
\]
Consider the symmetric measures
\begin{equation}\label{sw:competitor}
 \pi=\tfrac13\sum_{i=1}^3\delta_{X_i/2},\qquad
 \rho^\dagger=\rho_\infty-p_s\delta_0+p_s\pi.
\end{equation}
By \eqref{sw:center-mass},
\[
 \rho_\infty-p_s\delta_0\ge0,\quad
 \rho^\dagger(\mathbb R^2)=1-p_s+p_s=1,
 \quad\text{so }\rho^\dagger\in\mathcal P(\mathbb R^2).
\]
Symmetry and \eqref{sw:constants} yield, for $i=1,2,3$,
\[
 a_i(\pi)=\frac13\sum_{j=1}^3a_j(\pi)
 =\int F\,d\pi=H,\qquad a_i(\delta_0)=F_0,
\]
\[
 a_i(\rho^\dagger)=a_i(\rho_\infty)-p_sF_0+p_sH
 =c_\infty+p_s(H-F_0)>c_\infty.
\]
Since the fitted values are constant across $i$ for each of
$\rho_\infty$ and $\rho^\dagger$, we obtain
\begin{align}\label{sw:strict-gap}
 L(\rho_\infty)-L^*
 &\ge L(\rho_\infty)-L(\rho^\dagger)\notag\\
 &=\log\left(1+\frac{p_s(H-F_0)}{c_\infty}\right)
 \ge\log\left(1+\frac{p_s(H-F_0)}{M}\right)>0.
\end{align}
Here $L^*\le L(\rho^\dagger)$, $c_\infty\le M$ and $p_s(H-F_0)>0$.
Thus $\rho_\infty$ is not an NPMLE, and
\[
 L(\rho_n)=-\log c_n\longrightarrow-\log c_\infty=L(\rho_\infty).
\]

Finally, for \eqref{sw:c-init}, the constants in (2.2)
satisfy
\[
 b=e^{-L(\rho_0)}=c_{\mathrm{init}},\qquad
 c_0=\frac{c_{\mathrm{init}}}{2}
      e^{-(r_0+D_X)^2/2}\le\frac{c_{\mathrm{init}}}{2},
 \qquad M_2=M,\quad M_1>0.
\]
Consequently,
\begin{equation}\label{sw:original-range}
 \eta_0=\frac{c_0}{M+M_1^2/c_0}
 <\frac{c_0}{M}\le\frac{c_{\mathrm{init}}}{2M}.
\end{equation}
Thus the counterexample includes the whole original small-step range.
\end{proof}

See Remark~\ref{sw:dissipation} in Appendix~\ref{app:smooth-geometry}
for the two stationarity conditions at the limit.

\begin{appendix}

\section{Estimates for the explicit splitting}
\label{app:explicit-estimates}

We prove the estimates used in Section~\ref{sec:cv-estimates},
including the comparison of the two reaction residuals.
The bounds hold up to and including $\eta=\eta_0$. The Gaussian descent argument follows \citet{YWR2024}; the calculations
below also specify all constants needed for the compression argument.

\subsection{Likelihood sublevel bounds}
\label{app:sublevel-proof}
Recall the constants $b$, $r_0$, $c_0$ and $\eta_0$ in (2.2).
Since $0<a_i(\rho_0)\le M$,
\[
 0<b=\left(\prod_{i=1}^Na_i(\rho_0)\right)^{1/N}\le M.
\]
If $L(\rho)\le L(\rho_0)$, then
\[
 \max_{1\le j\le N}a_j(\rho)
 \ge\left(\prod_{i=1}^Na_i(\rho)\right)^{1/N}
 =e^{-L(\rho)}\ge b.
\]
Choose $j$ with $a_j(\rho)\ge b$.  Then
\[
 x\notin B(X_j,r_0)\implies
 \psi_j(x)\le Me^{-r_0^2/2}=b/2.
\]
For $p=\rho(B(X_j,r_0))$,
\[
 b\le a_j(\rho)\le Mp+(b/2)(1-p)\le Mp+b/2,
 \qquad p\ge b/(2M).
\]
For $x\in B(X_j,r_0)$ and $1\le i\le N$,
\[
 |x-X_i|\le|x-X_j|+|X_j-X_i|\le r_0+D_X.
\]
Therefore
\begin{equation}\label{app:sublevel-lower-bound}
 a_i(\rho)\ge\int_{B(X_j,r_0)}Me^{-|x-X_i|^2/2}\,\rho(dx)
 \ge pMe^{-(r_0+D_X)^2/2}\ge c_0.
\end{equation}
The upper bound is $a_i(\rho)\le\int M\,d\rho=M$.  Differentiation of
the finite sum defining $g_\rho$ gives
\[
 \|g_\rho\|_\infty\le\frac1N\sum_i\frac M{a_i(\rho)}\le M/c_0,
 \quad
 \|\nabla g_\rho\|_\infty\le M_1/c_0,
 \quad
 \|\nabla^2g_\rho\|_\infty\le M_2/c_0.
\]
Since $\nabla^2\phi(0)=-MI$, we have $M_2\ge M$, and
$c_0\le b/2\le M/2$.  Thus
\begin{equation}\label{app:closed-step-range}
 \eta_0\le\frac{c_0}{M_2}\le\frac12,\qquad
 \frac{\eta_0M}{c_0}=\frac M{M_2+M_1^2/c_0}<1,\qquad
 \frac{\eta_0M_2}{c_0}=\frac {M_2}{M_2+M_1^2/c_0}<1.
\end{equation}
The last inequalities are strict even at $\eta=\eta_0$, because $M_1>0$.
This proves Lemma~\ref{cv:sublevel}.

\subsection{Descent at the transport and reaction steps}
\label{app:descent-proof}
Fix $0<\eta\le\eta_0$.  Let $\rho$ lie in the likelihood sublevel, let
$a=a(\rho)$, and put $v(x)=\eta\nabla g_a(x)$, $T(x)=x+v(x)$.  For each
$i$, $\pi_i(dx)=\psi_i(x)\rho(dx)/a_i$ is a probability measure, and the
Gaussian identity
\begin{equation}\label{app:gaussian-ratio}
 \frac{\psi_i(x+v(x))}{\psi_i(x)}
 =\exp\{(X_i-x)\cdot v(x)-|v(x)|^2/2\}
\end{equation}
implies
\[
 \frac{a_i(T_\#\rho)}{a_i(\rho)}
 =\int e^{(X_i-x)\cdot v(x)-|v(x)|^2/2}\,\pi_i(dx)\in(0,M/c_0].
\]
Since $|v|\le\eta M_1/c_0$, $|x-X_i|\psi_i(x)\le M_1$ and $a_i\ge c_0$, the
exponent lies in $L^1(\pi_i)$; no moment assumption on $\rho$ is needed.
Jensen's inequality for the logarithm and
$\nabla\psi_i(x)=(X_i-x)\psi_i(x)$ now yield
\begin{align}
 L(T_\#\rho)-L(\rho)
 &=-\frac1N\sum_i\log\int
       e^{(X_i-x)\cdot v(x)-|v(x)|^2/2}\,\pi_i(dx)\notag\\
 &\le-\frac1N\sum_i\int
       \{(X_i-x)\cdot v(x)-|v(x)|^2/2\}
                               \frac{\psi_i(x)}{a_i}\,\rho(dx)\notag\\
 &=-\int\nabla g_a\cdot v\,d\rho
                         +\frac12\int g_a|v|^2\,d\rho\notag\\
 &=-\eta\int|\nabla g_a|^2\,d\rho
                    +\frac{\eta^2}{2}\int g_a|\nabla g_a|^2\,d\rho\notag\\
 &\le-\eta\left(1-\frac{\eta M}{2c_0}\right)
                     \int|\nabla g_a|^2\,d\rho
 \le-\frac\eta2\int|\nabla g_a|^2\,d\rho.
                                      \label{app:transport-descent}
\end{align}
In particular, the transported measure remains in the same sublevel.

For a reaction step, let $\sigma$ be a probability measure in that sublevel,
put $g=g_\sigma$, $h=1-\eta+\eta g$, and $\sigma^+=h\sigma$.
Since $\int g\,d\sigma=1$ and $0<g\le M/c_0$,
\begin{equation}\label{app:reaction-probability}
 \tfrac12\le1-\eta\le h\le B:=1-\eta+\eta M/c_0,\qquad
 \int d\sigma^+=\int h\,d\sigma=1.
\end{equation}
Define the posterior measures
$\pi_i^\sigma(dx)=\psi_i(x)\sigma(dx)/a_i(\sigma)$.  Jensen's
inequality gives
\begin{align}
 L(\sigma^+)-L(\sigma)
 &=-\frac1N\sum_i\log\int h\,d\pi_i^\sigma
 \le-\frac1N\sum_i\int\log h\,d\pi_i^\sigma
 =-\int g\log h\,d\sigma.\label{app:reaction-jensen}
\end{align}
Because $g-h=(1-\eta)(g-1)$ and
$(g-1)\log(1+\eta(g-1))\ge0$ pointwise,
\[
 \int g\log h\,d\sigma-\int h\log h\,d\sigma
 =(1-\eta)\int(g-1)\log h\,d\sigma\ge0.
\]
Thus
\begin{equation}\label{app:reaction-KL-descent}
 L(\sigma^+)-L(\sigma)\le-\operatorname{KL}(\sigma^+\Vert\sigma).
\end{equation}
For $f(t)=t\log t-t+1$, we have $f(1)=f'(1)=0$ and $f''(t)=1/t$.  Since
$1+t(h-1)\le\max\{1,h\}\le B$, where $B\ge1$ because $M\ge c_0$, Taylor's
integral formula gives
\begin{equation}\label{app:KL-quadratic}
 f(h)=(h-1)^2\int_0^1\frac{1-t}{1+t(h-1)}\,dt
 \ge\frac{(h-1)^2}{2B}.
\end{equation}
As $\int(h-1)\,d\sigma=0$, it follows that
\begin{equation}\label{app:reaction-descent}
 \operatorname{KL}(h\sigma\Vert\sigma)
 =\int f(h)\,d\sigma
 \ge\frac{\eta^2}{2B}\int(g-1)^2\,d\sigma.
\end{equation}
Induction using \eqref{app:transport-descent} and
\eqref{app:reaction-KL-descent} gives, for every $n\ge0$,
\[
 \rho_n,\widetilde\rho_n\in\mathcal P(\mathbb R^d),\qquad
 L(\rho_{n+1})\le L(\widetilde\rho_n)\le L(\rho_n)\le L(\rho_0).
\]
Combining the two substeps proves \eqref{cv:descent}:
\[
 L(\rho_n)-L(\rho_{n+1})
 \ge\frac\eta2G_n+\frac{\eta^2}{2B}R_n\ge c(G_n+R_n)=cE_n,
\]
with
\begin{equation}\label{app:descent-constant}
 c=\min\{\eta/2,\eta^2/(2B)\}>0.
\end{equation}

\subsection{Residual comparison and bounded-Lipschitz length}
Let $a_n=a(\rho_n)$ and $\widetilde a_n=a(\widetilde\rho_n)$.  Kernel
Lipschitz continuity and Cauchy--Schwarz imply
\begin{align}
 |\widetilde a_{n,i}-a_{n,i}|
 &\le\int|\psi_i(T_n(x))-\psi_i(x)|\,\rho_n(dx)\notag\\
 &\le M_1\eta\int|\nabla g_{a_n}|\,d\rho_n
 \le M_1\eta\sqrt{G_n},\qquad
 |\widetilde a_n-a_n|\le\sqrt N M_1\eta\sqrt{G_n}.
                                      \label{app:moment-transport}
\end{align}
For any $a,b\in[c_0,M]^N$, direct coefficient comparison gives
\begin{align}
 \|g_b-g_a\|_\infty
 &\le\frac MN\sum_i\frac{|b_i-a_i|}{a_ib_i}
 \le\frac M{\sqrt N c_0^2}|b-a|,
                                      \label{app:g-coefficient-comparison}\\
 \|\nabla g_b-\nabla g_a\|_\infty
 &\le\frac {M_1}N\sum_i\frac{|b_i-a_i|}{a_ib_i}
 \le\frac {M_1}{\sqrt N c_0^2}|b-a|.
                                      \label{app:gradient-coefficient-comparison}
\end{align}
Set $K_0=M_1(M/c_0^2+1/c_0)$.  The triangle inequality, first for
the coefficient change and then for the spatial change, yields
\begin{align}
 \|g_{\widetilde a_n}\circ T_n-g_{a_n}\|_{L^2(\rho_n)}
 &\le\|g_{\widetilde a_n}-g_{a_n}\|_\infty
             +\|g_{a_n}\circ T_n-g_{a_n}\|_{L^2(\rho_n)}\notag\\
 &\le\frac M{\sqrt N c_0^2}|\widetilde a_n-a_n|
             +\frac {M_1}{c_0}\|T_n-\mathrm{id}\|_{L^2(\rho_n)}\notag\\
 &\le\left(\frac{MM_1}{c_0^2}+\frac{M_1}{c_0}\right)
                       \eta\sqrt{G_n}
 =K_0\eta\sqrt{G_n}.\label{app:residual-pullback-error}
\end{align}
Here $g_{a_n}$ is globally $M_1/c_0$-Lipschitz by
Lemma~\ref{cv:sublevel}.  Pulling back the post-transport reaction residual
gives
\[
 \|g_{\widetilde a_n}\circ T_n-1\|_{L^2(\rho_n)}^2
 =\int(g_{\widetilde\rho_n}-1)^2\,d\widetilde\rho_n=R_n.
\]
Consequently,
\begin{align}
 \int(g_{a_n}-1)^2\,d\rho_n
 &\le2\int(g_{\widetilde a_n}(T_n(x))-1)^2\,\rho_n(dx)
       +2\|g_{\widetilde a_n}\circ T_n-g_{a_n}\|_{L^2(\rho_n)}^2\notag\\
 &\le2R_n+2K_0^2\eta^2G_n.
                                      \label{app:reaction-residual-comparison}
\end{align}
Adding $G_n$ proves
\begin{equation}\label{app:dissipation-E}
 \mathcal D(\rho_n)\le(1+2K_0^2\eta^2)G_n+2R_n
 \le\max\{1+2K_0^2\eta^2,2\}E_n,
\end{equation}
which is \eqref{cv:residual-comparison}.

Finally, let $f$ satisfy $\|f\|_\infty\le1$ and
$\operatorname{Lip}(f)\le1$.  For the transport step,
\begin{equation}\label{app:BL-transport}
 \begin{aligned}
 \left|\int f\,d\widetilde\rho_n-\int f\,d\rho_n\right|
 &=\left|\int\{f(T_n(x))-f(x)\}\,\rho_n(dx)\right|\\
 &\le\int|T_n(x)-x|\,\rho_n(dx)
 =\eta\int|\nabla g_{a_n}|\,d\rho_n\le\eta\sqrt{G_n}.
 \end{aligned}
\end{equation}
For the reaction step,
\begin{equation}\label{app:BL-reaction}
 \begin{aligned}
 \left|\int f\,d\rho_{n+1}-\int f\,d\widetilde\rho_n\right|
 &=\eta\left|\int f(g_{\widetilde\rho_n}-1)\,d\widetilde\rho_n\right|\\
 &\le\eta\|f\|_{L^2(\widetilde\rho_n)}
   \|g_{\widetilde\rho_n}-1\|_{L^2(\widetilde\rho_n)}
 \le\eta\sqrt{R_n}.
 \end{aligned}
\end{equation}
Taking the supremum and using
$\sqrt{G_n}+\sqrt{R_n}\le\sqrt{2E_n}$ gives
\begin{equation}\label{app:BL-step}
 d_{\mathrm{BL}}(\rho_{n+1},\rho_n)
 \le\eta(\sqrt{G_n}+\sqrt{R_n})\le\eta\sqrt{2E_n}.
\end{equation}
Only the bounded displacement and the bounded test function are integrated;
no first moment of $\rho_n$ is required.  This completes the proof of
Proposition~\ref{cv:split-estimates}.

\section{Compatibility of the counterexample with the conservative step bound}\label{app:particle-step}
The counterexample holds for every fixed $\eta\in(0,1]$. We give a positive lower bound for the threshold in \textup{(2.2)}, uniform over all empirical initial weight vectors.

\begin{proposition}\label{pt:admissible}
Let the observation diameter be at most $D<\infty$. Write
\[
 M=\phi(0),\qquad M_1=\sup_x|\nabla\phi(x)|,\qquad
 M_2=\sup_x\|\nabla^2\phi(x)\|_{\mathrm{op}},
\]
and define
\begin{equation}\label{pt:uniform-step}
\begin{gathered}
 b_{\min}=M e^{-D^2/2},\qquad \bar r=\sqrt{2\log2+D^2},\\
 c_{\min}=\frac{b_{\min}}2e^{-(\bar r+D)^2/2},\qquad
 \eta_{\mathrm{adm}}=\frac{c_{\min}}{M_2+M_1^2/c_{\min}}>0.
\end{gathered}
\end{equation}
For every probability measure initially supported on the observations, the threshold in \textup{(2.2)} is at least $\eta_{\mathrm{adm}}$. Thus Theorem~2.3 holds, in particular, for fixed steps strictly smaller than this common admissible threshold.
\end{proposition}
\begin{proof}
Let $\rho_0=\sum_jq_j\delta_{X_j}$, $q_j\ge0$, $\sum_jq_j=1$; missing groups are permitted. For $D_X=\max_{i,j}|X_i-X_j|\le D$,
\begin{align*}
 b_{\min}=Me^{-D^2/2}
 &\le\sum_jq_jMe^{-|X_i-X_j|^2/2}
 =a_{X,i}(\rho_0)\le M,\\
 b_{\min}&\le b:=e^{-L_X(\rho_0)}
 =\left\{\prod_{i=1}^Na_{X,i}(\rho_0)\right\}^{1/N}\le M.
\end{align*}
For the Gaussian kernel, the quantities in \textup{(2.2)} satisfy
\begin{align*}
 r_0&=\sqrt{2\log(2M/b)}
 \le\sqrt{2\log(2M/b_{\min})}=\bar r,\\
 c_0&=\frac b2e^{-(r_0+D_X)^2/2}
 \ge\frac{b_{\min}}2e^{-(\bar r+D)^2/2}=c_{\min},\\
 \eta_0&=\frac{c_0}{M_2+M_1^2/c_0}
 =\frac{c_0^2}{M_2c_0+M_1^2}
 \ge\frac{c_{\min}^2}{M_2c_{\min}+M_1^2}
 =\eta_{\mathrm{adm}}.
\end{align*}
The last inequality follows from
\[
 \frac{d}{dc}\frac{c^2}{M_2c+M_1^2}
 =\frac{M_2c^2+2M_1^2c}{(M_2c+M_1^2)^2}>0\qquad(c>0).
\]
For the Gaussian kernel, $M_2\ge\|\nabla^2\phi(0)\|_{\mathrm{op}}=M$ and $M_1>0$. Moreover,
\[
 \begin{gathered}
 c_{\min}\le\frac M2e^{-D^2/2}e^{-\bar r^2/2}
 =\frac M4e^{-D^2}\le\frac M4,\\
 0<\eta_{\mathrm{adm}}<\frac{c_{\min}}{M_2}\le\frac14<1.
 \end{gathered}
\]
To apply the bound to Theorem~2.3, take
\[
 D=4r_\triangle,\quad 0<\eta<\eta_{\mathrm{adm}},\quad
 U_\eta'=U_\eta\cap\{X:D_X<D\}.
\]
Then $U_\eta'$ is open and
\[
 \begin{gathered}
 D_{X^\triangle}=\sqrt3r_\triangle<4r_\triangle=D
 \ \Longrightarrow\ X^\triangle\in U_\eta',\\
 X\in U_\eta'\ \Longrightarrow\
 \inf_{q\in\{q\ge0:\,\sum_jq_j=1\}}\eta_0(X,q)
 \ge\eta_{\mathrm{adm}}>\eta.
 \end{gathered}
\]
This includes every realized vector $q_j=C_j/m$.
\end{proof}

\section{The complete position--weight Hessian calculation}\label{app:particle-hessian}
We prove Proposition~5.3 of the main paper using the reduced chart \textup{(5.9)}. All variations include both positions and independent weights.
\begin{proof}
Write $f=f(r_*)$, $t_{ij}=r_\triangle e_i-r_*e_j$, and
\begin{equation}\label{pt:kernel-matrix}
 D_{ij}=\phi(t_{ij})/f,
 \qquad D=\frac35(2I_3+\mathbf1\mathbf1^\top).
\end{equation}
For $i\ne j$,
\begin{align*}
 |t_{ij}|^2-|t_{jj}|^2&=3r_\triangle r_*=2\log3,
 &\frac{\phi(t_{jj})}{\phi(t_{ij})}&=e^{\log3}=3,\\
 f&=\frac13\left\{\phi(t_{jj})+2\phi(t_{ij})\right\},
 &D_{jj}&=\frac95,\qquad D_{ij}=\frac35,\\
 D\mathbf1&=3\mathbf1,
 &D^\top\mathbf1&=3\mathbf1.
\end{align*}
Let $e_j^\perp$ be a unit vector perpendicular to $e_j$. We compute
\begin{align}
 \frac13\sum_{i=1}^3D_{ij}t_{ij}&=0,\label{pt:mean-zero}\\
 S_j:=\frac13\sum_{i=1}^3D_{ij}t_{ij}t_{ij}^\top
 &=\frac{27r_\triangle^2}{50}e_je_j^\top
   +\frac{3r_\triangle^2}{10}e_j^\perp(e_j^\perp)^\top.\label{pt:second-moment}
\end{align}
For $j=1$, in the basis $(e_1,e_1^\perp)$,
\[
 (t_{11},t_{21},t_{31})
 =r_\triangle\begin{pmatrix}3/5&-9/10&-9/10\\0&\sqrt3/2&-\sqrt3/2\end{pmatrix},
 \qquad (D_{11},D_{21},D_{31})/3=(3/5,1/5,1/5).
\]
Consequently,
\begin{align*}
 \frac35\frac{3r_\triangle}{5}
 +\frac25\left(-\frac{9r_\triangle}{10}\right)&=0,\\
 \frac15\frac{\sqrt3r_\triangle}{2}
 +\frac15\left(-\frac{\sqrt3r_\triangle}{2}\right)&=0,\\
 \frac35\left(\frac{3r_\triangle}{5}\right)^2
 +\frac25\left(\frac{9r_\triangle}{10}\right)^2
 &=\frac{27r_\triangle^2}{50},\\
 \frac25\left(\frac{\sqrt3r_\triangle}{2}\right)^2
 &=\frac{3r_\triangle^2}{10},\\
 \frac15\left(-\frac{9r_\triangle}{10}\right)
 \frac{\sqrt3r_\triangle}{2}
 +\frac15\left(-\frac{9r_\triangle}{10}\right)
 \left(-\frac{\sqrt3r_\triangle}{2}\right)&=0.
\end{align*}
Rotation by $2\pi(j-1)/3$ gives \eqref{pt:mean-zero}--\eqref{pt:second-moment} for every $j$. The eigenvalues of $S_j$ are
\begin{equation}\label{pt:moment-eigenvalues}
 \frac{27r_\triangle^2}{50}=\frac9{10}\log3<1,
 \qquad \frac{3r_\triangle^2}{10}=\frac12\log3<1.
\end{equation}

For $(\xi,w)\in\mathbb R^6\times\mathbb R^2$, set $h=Ew$ and
\[
 z_j(s)=r_*e_j+s\xi_j,\qquad p_j(s)=1/3+sh_j,
 \qquad A_i(s)=\sum_{j=1}^3p_j(s)\phi(r_\triangle e_i-z_j(s)).
\]
The Gaussian derivative identities are
\begin{align*}
 \left.\frac{d}{ds}\phi(t_{ij}-s\xi_j)\right|_{s=0}
 &=\phi(t_{ij})t_{ij}\cdot\xi_j,\\
 \left.\frac{d^2}{ds^2}\phi(t_{ij}-s\xi_j)\right|_{s=0}
 &=\phi(t_{ij})\{(t_{ij}\cdot\xi_j)^2-|\xi_j|^2\}.
\end{align*}
Thus
\begin{align}
 \frac{A_i'(0)}f
 &=\sum_jD_{ij}h_j+\frac13\sum_jD_{ij}t_{ij}\cdot\xi_j,\label{pt:first-variation}\\
 \frac{A_i''(0)}f
 &=\sum_jD_{ij}\left[
 \frac13\{(t_{ij}\cdot\xi_j)^2-|\xi_j|^2\}
 +2h_jt_{ij}\cdot\xi_j\right].\label{pt:second-variation}
\end{align}
The column sums of $D$ and \eqref{pt:mean-zero} imply
\begin{align*}
 \frac23\sum_jh_j\left(\sum_iD_{ij}t_{ij}\right)\cdot\xi_j&=0,\\
 -\frac13\sum_i\frac{A_i''(0)}f
 &=\frac19\sum_{i,j}D_{ij}|\xi_j|^2
   -\frac19\sum_{i,j}D_{ij}(t_{ij}\cdot\xi_j)^2\\
 &=\frac13\sum_j\xi_j^\top(I_2-S_j)\xi_j.
\end{align*}
Twice differentiating $-3^{-1}\sum_i\log A_i(s)$ now gives
\begin{align}
 (\xi,w)^\top H_*(\xi,w)
 &=\frac13\sum_i\left\{\left(\frac{A_i'(0)}f\right)^2-
 \frac{A_i''(0)}f\right\}\notag\\
 &=\frac13\sum_i\left(\sum_jD_{ij}h_j+
        \frac13\sum_jD_{ij}t_{ij}\cdot\xi_j\right)^2
   +\frac13\sum_j\xi_j^\top(I_2-S_j)\xi_j.\label{pt:full-hessian}
\end{align}
All remaining position--weight mixed terms are retained inside these squares. By \eqref{pt:moment-eigenvalues},
\[
 \xi\ne0\ \Longrightarrow\
 \frac13\sum_j\xi_j^\top(I_2-S_j)\xi_j
 \ge\frac13\left(1-\frac9{10}\log3\right)|\xi|^2>0.
\]
For $\xi=0$, $h=Ew$ gives $\mathbf1^\top h=0$, $|h|=|w|$, and
\[
 Dh=\frac65h,\qquad
 (0,w)^\top H_*(0,w)
 =\frac13|Dh|^2=\frac{12}{25}|h|^2
 =\frac{12}{25}|w|^2.
\]
Therefore $H_*\succ0$ and $(H_*)_{uu}=(12/25)I_2$.

For the $zz$ block, put $b_{ij}=D_{ij}/3$. Since $b_{ij}>0$ and $\sum_jb_{ij}=1$, Jensen's inequality yields
\begin{align*}
 \frac13\sum_i\left(\sum_jb_{ij}t_{ij}\cdot\xi_j\right)^2
 &\le\frac13\sum_{i,j}b_{ij}(t_{ij}\cdot\xi_j)^2\\
 &=\frac13\sum_j\xi_j^\top S_j\xi_j.
\end{align*}
Substituting into \eqref{pt:full-hessian} with $w=0$,
\begin{align*}
 \xi^\top(H_*)_{zz}\xi
 &\le\frac13\sum_j\xi_j^\top S_j\xi_j
     +\frac13\sum_j\xi_j^\top(I_2-S_j)\xi_j
 =\frac13|\xi|^2,\\
 H_*\succ0&\ \Longrightarrow\ (H_*)_{zz}\succ0.
\end{align*}
This proves both block bounds.
\end{proof}
\section{Geometry for the smooth pure-transport counterexample}\label{app:smooth-geometry}
We prove Lemma~\ref{sw:geometry} using the data and constants in
\eqref{sw:data} and \eqref{sw:constants}. 
\begin{proof}
For every orthogonal symmetry $A$ of $\{X_1,X_2,X_3\}$,
\[
 \begin{gathered}
 F(Ax)=\frac13\sum_i\phi(Ax-X_i)
 =\frac13\sum_i\phi(x-A^{-1}X_i)=F(x),\\
 0<F(x)=\frac13\sum_i\phi(x-X_i)\le M.
 \end{gathered}
\]
For $z\in\mathbb R^2$, direct differentiation gives
\begin{equation}\label{sw:Gaussian-derivatives}
 \nabla\phi(z)=-z\phi(z),\qquad
 \nabla^2\phi(z)=\phi(z)(zz^{\mathsf T}-I).
\end{equation}
The Hessian eigenvalues give
\begin{align*}
 \|\nabla^2\phi(z)\|_{\mathrm{op}}
 &=M e^{-|z|^2/2}\max\{1,\,||z|^2-1|\}\\
 &\le M\max\{1,\sup_{a\ge1}(a-1)e^{-a/2}\}
 =M\max\{1,2e^{-3/2}\}=M.
\end{align*}
For unit vectors $v,w,t$,
\begin{align*}
 D^3\phi(z)[v,w,t]
 =\phi(z)\{&-(z\cdot v)(z\cdot w)(z\cdot t)
 +(v\cdot w)(z\cdot t)\\
 &+(v\cdot t)(z\cdot w)+(w\cdot t)(z\cdot v)\}.
\end{align*}
Consequently,
\begin{equation}\label{sw:third-bound}
 \begin{split}
 \sup_z\|D^3\phi(z)\|_{\mathrm{op}}
 &\le M\sup_{a\ge0}(a^3+3a)e^{-a^2/2}\\
 &\le M(3\sqrt3e^{-3/2}+3e^{-1/2})<4M.
 \end{split}
\end{equation}
The last bound uses
$\sup_{a\ge0}a^3e^{-a^2/2}=3\sqrt3e^{-3/2}$ and
$\sup_{a\ge0}ae^{-a^2/2}=e^{-1/2}$.
Averaging yields, for $k=2,3$,
\[
 \|D^kF(x)\|_{\mathrm{op}}
 \le\frac13\sum_i\|D^k\phi(x-X_i)\|_{\mathrm{op}},
 \qquad \|\nabla^2F\|_{\mathrm{op}}\le M,\quad
 \|D^3F\|_{\mathrm{op}}\le4M.
\]
Using $\sum_iX_i=0$, $\frac13\sum_iX_iX_i^{\mathsf T}=\frac{R^2}{2}I$
and $\phi(-X_i)=F_0$, we obtain
\begin{align*}
 \nabla F(0)&=\frac{F_0}{3}\sum_iX_i=0,\\
 \nabla^2F(0)
 &=F_0\left(\frac13\sum_iX_iX_i^{\mathsf T}-I\right)
 =F_0\left(\frac{R^2}{2}-1\right)I=-\kappa I\prec0.
\end{align*}
Here $R^2=\log7<2$.  For $|x|\le r$, the fundamental theorem of calculus
gives
\begin{align*}
 \|\nabla^2F(x)-\nabla^2F(0)\|_{\mathrm{op}}
 &=\left\|\int_0^1D(\nabla^2F)(tx)[x]\,dt\right\|_{\mathrm{op}}\\
 &\le\int_0^1\|D^3F(tx)\|_{\mathrm{op}}|x|\,dt
 \le4M|x|\le4Mr=\kappa/2.
\end{align*}
Together with $\nabla^2F\succeq-MI$, this proves
\[
 -MI\preceq\nabla^2F(x)
 \preceq\nabla^2F(0)+(\kappa/2)I
 =-(\kappa/2)I,
\]
which proves \eqref{sw:local-Hessian}.

The squared distances satisfy
\[
 |X_1/2-X_1|^2=R^2/4,\qquad
 |X_1/2-X_2|^2=|X_1/2-X_3|^2=7R^2/4.
\]
Hence
\[
 \frac{F(X_1/2)}{F_0}
 =\frac{e^{-R^2/8}+2e^{-7R^2/8}}{3e^{-R^2/2}}
 =\frac13\left(e^{3R^2/8}+2e^{-3R^2/8}\right)
 =\frac13(u+2/u).
\]
Since $u^8=7^3=343>256=2^8$, we have $u>2$ and
\[
 \frac{H}{F_0}-1
 =\frac{u^2-3u+2}{3u}
 =\frac{(u-1)(u-2)}{3u}>0.
\]

To prove finiteness of $\mathcal C$, put
\begin{equation}\label{sw:softmax}
 p_i(x)=\frac{e^{X_i\cdot x}}{\sum_j e^{X_j\cdot x}}.
\end{equation}
Since $|X_i|=R$,
\[
 \phi(x-X_i)=M e^{-(|x|^2+R^2)/2}e^{X_i\cdot x},\qquad
 \frac{\phi(x-X_i)}{\sum_j\phi(x-X_j)}=p_i(x).
\]
Therefore
\begin{equation}\label{sw:gradient-barycenter}
 \nabla F(x)=F(x)\left(\sum_i p_i(x)X_i-x\right).
\end{equation}
For $x\in\mathcal C$, $F(x)>0$ gives
\[
 x=\sum_i p_i(x)X_i,\qquad
 |x|\le\sum_i p_i(x)|X_i|=R.
\]
As $X_i\cdot X_j=-R^2/2$ for $i\ne j$,
\[
 X_i\cdot x=R^2p_i(x)-\frac{R^2}{2}\sum_{j\ne i}p_j(x)
 =\frac{3R^2}{2}p_i(x)-\frac{R^2}{2}.
\]
Taking logarithms in \eqref{sw:softmax} now gives
\begin{equation}\label{sw:critical-p}
 \log p_i(x)-\frac{3R^2}{2}p_i(x)
 =-\frac{R^2}{2}-\log\sum_j e^{X_j\cdot x},
 \qquad i=1,2,3.
\end{equation}
For $\ell(t)=\log t-(3R^2/2)t$,
\[
 \begin{gathered}
 \ell'(t)=t^{-1}-3R^2/2,\qquad \ell''(t)=-t^{-2}<0
 \quad(0<t<1),\\
 \Longrightarrow\quad
 \#\{t\in(0,1):\ell(t)=v\}\le2\quad(v\in\mathbb R).
 \end{gathered}
\]
If $\ell(t)-v$ had three distinct roots, Rolle's theorem would give
two distinct zeros of $\ell'$. But $\ell'$ is strictly decreasing,
so it has at most one zero. Equation~\eqref{sw:critical-p} therefore
implies
\[
 \#\{p_1(x),p_2(x),p_3(x)\}\le2
 \quad\Longrightarrow\quad
 p_j(x)=p_k(x)
 \quad\text{for some }\{i,j,k\}=\{1,2,3\}.
\]
For this choice,
\[
 x=p_iX_i+p_j(X_j+X_k)=(p_i-p_j)X_i\in\mathbb R e_i.
\]
Define $h_i(t)=\frac{d}{dt}F(te_i)$.  Then
\[
 h_i\in C^\omega(\mathbb R),\qquad
 h_i'(0)=e_i^{\mathsf T}\nabla^2F(0)e_i=-\kappa\ne0.
\]
So $h_i\not\equiv0$.  Since the function is real analytic and not identically zero,
its zeros are isolated. Compactness of $[-R,R]$ then gives
\[
 Z_i:=\{t\in[-R,R]:h_i(t)=0\},\qquad \#Z_i<\infty.
\]
Finally,
\[
 x\in\mathcal C\cap\mathbb R e_i
 \quad\Longrightarrow\quad
 x=te_i,\quad |t|\le R,\quad h_i(t)=e_i\cdot\nabla F(x)=0,
\]
\[
 \mathcal C\subseteq\bigcup_{i=1}^3\{te_i:t\in Z_i\},
 \qquad \#\mathcal C\le\sum_{i=1}^3\#Z_i<\infty.
\]
\end{proof}
\begin{remark}[The two stationarity conditions]\label{sw:dissipation}
By \eqref{sw:weak-limit} and $g_{\rho_\infty}=F/c_\infty$,
\[
 \int|\nabla g_{\rho_\infty}|^2\,d\rho_\infty
 =\frac1{c_\infty^2}\int_{\mathcal C}|\nabla F|^2\,d\rho_\infty=0.
\]
In contrast,
\begin{equation}\label{sw:reaction-positive}
 \int(g_{\rho_\infty}-1)^2\,d\rho_\infty>0.
\end{equation}
Set $\delta=(H-F_0)/2>0$. Since $F(X_1/2)=H$, continuity gives
$\varepsilon>0$ such that
\[
 U:=B(X_1/2,\varepsilon)\subseteq\{F\ge F_0+\delta\}.
\]
The initial density $q_0$ is strictly positive, so
$q:=\rho_0(U)>0$.
The positivity follows from $q_0>0$.  Pointwise ascent gives
\[
 F(S(x))=\lim_nF(\Psi_n(x))\ge F(x),
\]
\[
 \rho_\infty\{F\ge F_0+\delta\}
 =\rho_0\{x:F(S(x))\ge F_0+\delta\}\ge\rho_0(U)=q,
 \qquad \rho_\infty\{0\}\ge p_s.
\]
Writing $A=\{0\}$ and $B=\{F\ge F_0+\delta\}$, we have
$A\cap B=\varnothing$ and $|F(x)-F(y)|\ge\delta$ on
$(A\times B)\cup(B\times A)$.  Since $\int F\,d\rho_\infty=c_\infty$,
the variance identity yields
\begin{align*}
 \int(g_{\rho_\infty}-1)^2\,d\rho_\infty
 &=\frac{1}{2c_\infty^2}
   \iint(F(x)-F(y))^2\,\rho_\infty(dx)\rho_\infty(dy)\\
 &\ge\frac{2\rho_\infty(A)\rho_\infty(B)\delta^2}{2c_\infty^2}
 \ge\frac{p_s q\delta^2}{c_\infty^2}
 \ge\frac{p_s q\delta^2}{M^2}>0.
\end{align*}
Thus $\rho_\infty$ is a fixed point of pure transport and has a
strictly positive reaction residual. Its atom at the origin forms
in the limit, since every finite iterate has a smooth density.
\end{remark}
\section{A one-dimensional obstruction with a persistent atom}\label{aw:sec}

We prove Proposition~2.5 by following the transport
trajectory from each initial point. We then identify the unique NPMLE
and compute the limiting likelihood gap. Throughout this section, $d=1$ and $N=2$. The observations and initial measure are given by
\begin{equation}\label{aw:data-initial}
 X_1=-a,\qquad X_2=a,\qquad a>1,\qquad
 \rho_0=\alpha\delta_0+(1-\alpha)\mathcal N(0,s^2),
 \quad 0<\alpha<1,\quad s>0.
\end{equation}
Write $M=\phi(0)=(2\pi)^{-1/2}$ and define
\begin{align}
 h(x)&=\frac{\phi(a-x)+\phi(a+x)}2
      =M e^{-(x^2+a^2)/2}\cosh(ax),\label{aw:h}\\
 c_{\mathrm{init}}&=\int h\,d\rho_0
 =\alpha\phi(a)+\frac{1-\alpha}{\sqrt{1+s^2}}
           \phi\!\left(\frac{a}{\sqrt{1+s^2}}\right)>0.
 \label{aw:cinit}
\end{align}
The Gaussian term follows by convolving the standard Gaussian density
with the density of $\mathcal N(0,s^2)$. Fix a step size
\begin{equation}\label{aw:step}
 0<\eta\le \frac{c_{\mathrm{init}}}{2M}.
\end{equation}

\subsection{The two maxima of the symmetric kernel average}
Logarithmic differentiation of \eqref{aw:h} gives
\begin{equation}\label{aw:F}
 h'(x)=h(x)F(x),\qquad F(x)=a\tanh(ax)-x.
\end{equation}
We have
\begin{align}
 F(0)&=0,\qquad F'(0)=a^2-1>0,\qquad
 F(a)=a\{\tanh(a^2)-1\}<0,\label{aw:F-signs}\\
 F'(x)&=a^2\operatorname{sech}^2(ax)-1,\qquad
 F''(x)=-2a^3\operatorname{sech}^2(ax)\tanh(ax)<0
 \quad(x>0).
 \label{aw:F-concavity}
\end{align}
Equations \eqref{aw:F-signs}--\eqref{aw:F-concavity} give
\[
 \begin{gathered}
 F''(x)<0\ (x>0),\qquad F'(0)>0,\qquad
 \lim_{x\to\infty}F'(x)=-1,\\
 \exists!t>0:\quad F'(t)=0,\quad
 F'(x)>0\ (0<x<t),\quad F'(x)<0\ (x>t),\\
 F(0)=0<F(t),\quad F(a)<0
 \quad\Longrightarrow\quad \exists!b\in(0,a):\ F(b)=0.
 \end{gathered}
\]
So,
\begin{equation}\label{aw:b}
 b=a\tanh(ab),\qquad
 F(x)>0\ (0<x<b),\qquad F(x)<0\ (x>b).
\end{equation}
Since $h$ is positive and even, \eqref{aw:F}--\eqref{aw:b} imply
\begin{equation}\label{aw:h-maxima}
 \begin{gathered}
 h'(x)>0\quad(0<x<b),\qquad h'(x)<0\quad(x>b),\\
 \operatorname*{arg\,max}_{x\in\mathbb R}h(x)=\{-b,b\},
 \qquad h(b)>h(0).
 \end{gathered}
\end{equation}
In particular,
\[
 \{x\in\R:h'(x)=0\}=\{-b,0,b\}.
\]

\subsection{Monotone maps and a uniform lower bound for the fitted values}
For a symmetric probability measure $\rho$, the two fitted values coincide:
\begin{equation}\label{aw:equal-fits}
 \int\phi(a-x)\,d\rho(x)
 =\int\phi(a+x)\,d\rho(x)
 =\int h\,d\rho.
\end{equation}
Hence, whenever $\rho_n$ is symmetric, pure Wasserstein descent takes the form
\begin{equation}\label{aw:iteration}
 c_n=\int h\,d\rho_n,\qquad
 g_{\rho_n}(x)=\frac{h(x)}{c_n},\qquad
 T_n(x)=x+\frac{\eta}{c_n}h'(x),\qquad
 \rho_{n+1}=(T_n)_\#\rho_n.
\end{equation}
For $S(x)=-x$, oddness of $T_n$ gives
\[
 S\circ T_n=T_n\circ S,\qquad
 S_\#\rho_n=\rho_n
 \ \Longrightarrow\ 
 S_\#(T_n)_\#\rho_n=(T_n)_\#S_\#\rho_n=(T_n)_\#\rho_n.
\]
Gaussian differentiation gives
\begin{equation}\label{aw:gaussian-derivatives}
 \begin{gathered}
 \phi'(x)=-x\phi(x),\qquad \phi''(x)=(x^2-1)\phi(x),\\
 \sup_x|\phi'(x)|=Me^{-1/2},\qquad \sup_x|\phi''(x)|=M.
 \end{gathered}
\end{equation}
Indeed,
\[
 \sup_{0\le u\le1}(1-u)e^{-u/2}=1,
 \qquad \sup_{u\ge1}(u-1)e^{-u/2}=2e^{-3/2}<1.
\]
Taking averages of the shifted derivatives yields
\begin{equation}\label{aw:h-derivatives}
 \sup_x|h'(x)|\le Me^{-1/2},\qquad
 \sup_x|h''(x)|\le M,\qquad 0<h(x)\le M.
\end{equation}

For the induction, the initial conditions are
\[
 S_\#\rho_0=\rho_0,\qquad \int h\,d\rho_0=c_{\mathrm{init}}.
\]
Assume $S_\#\rho_n=\rho_n$ and $c_n\ge c_{\mathrm{init}}$. By \eqref{aw:step} and \eqref{aw:h-derivatives}, for every $x\in\R$,
\begin{equation}\label{aw:T-derivative}
 T_n'(x)=1+\frac{\eta}{c_n}h''(x)
 \ge1-\frac{\eta M}{c_n}
 \ge1-\frac{\eta M}{c_{\mathrm{init}}}
 \ge\frac12.
\end{equation}
Moreover,
\begin{equation}\label{aw:T-proper}
 |T_n(x)-x|\le\frac{\eta Me^{-1/2}}{c_{\mathrm{init}}},
 \qquad\lim_{x\to\pm\infty}T_n(x)=\pm\infty.
\end{equation}
Thus $T_n:\R\to\R$ is an increasing homeomorphism and
\[
 T_n(0)=0,\qquad T_n(\pm b)=\pm b.
\]
Using $T_n'>0$ and \eqref{aw:h-maxima},
\begin{equation}\label{aw:orbit-intervals}
 \begin{cases}
 x<T_n(x)<b,&0<x<b,\\
 T_n(b)=b,\\
 b<T_n(x)<x,&x>b.
 \end{cases}
\end{equation}
Since $T_n(-x)=-T_n(x)$ and $h(-x)=h(x)$, \eqref{aw:orbit-intervals} and \eqref{aw:h-maxima} give
\begin{equation}\label{aw:h-increase}
 h(T_n(x))\ge h(x)\quad(x\in\mathbb R),\qquad
 c_{n+1}=\int h(T_n(x))\,d\rho_n(x)\ge c_n.
\end{equation}
The symmetry calculation above gives $S_\#\rho_{n+1}=\rho_{n+1}$. Thus
\[
 S_\#\rho_n=\rho_n,\ c_n\ge c_{\mathrm{init}}
 \quad\Longrightarrow\quad
 S_\#\rho_{n+1}=\rho_{n+1},\ c_{n+1}\ge c_n\ge c_{\mathrm{init}},
\]
which proves the induction and yields
\begin{equation}\label{aw:c-bounds}
 c_{\mathrm{init}}\le c_n\le M\qquad(n\ge0).
\end{equation}
The required lower bound follows directly from the scalar dynamics.

For the initialization in \eqref{aw:data-initial}, symmetry gives
$e^{-L(\rho_0)}=c_{\mathrm{init}}$, and the data diameter is $D_X=2a$.
The constants in (2.2) of the main paper therefore satisfy
\begin{equation}\label{aw:original-step}
 \begin{gathered}
 r_0=\sqrt{2\log(2M/c_{\mathrm{init}})},\qquad
 c_0=\frac{c_{\mathrm{init}}}{2}e^{-(r_0+2a)^2/2}
 \le\frac{c_{\mathrm{init}}}{2},\\
 \eta_0=\frac{c_0}{M+M^2e^{-1}/c_0}
 <\frac{c_0}{M}\le\frac{c_{\mathrm{init}}}{2M}.
 \end{gathered}
\end{equation}
Thus the counterexample covers the full step-size range
$0<\eta\le\eta_0$.

\subsection{Full support at finite times and convergence in finite-order Wasserstein distances}
Put $P_0(x)=x$ and $P_{n+1}=T_n\circ P_n$. Then
\begin{equation}\label{aw:P}
 \rho_n=(P_n)_\#\rho_0.
\end{equation}
Each $P_n$ is an increasing homeomorphism. For any nonempty open
interval $I\subset\R$, its preimage $P_n^{-1}(I)$ is nonempty and open.
Since $\rho_0$ has full support,
\begin{equation}\label{aw:full-support}
 \rho_n(I)=\rho_0(P_n^{-1}(I))>0.
\end{equation}
In particular, $\operatorname{supp}\rho_n=\mathbb R$ for every finite $n$.

Fix $x>0$ and put $x_n=P_n(x)$. Equation \eqref{aw:orbit-intervals} gives
\[
 \begin{cases}
 x\le x_n<x_{n+1}<b,&0<x<b,\\
 x_n=b,&x=b,\\
 b<x_{n+1}<x_n\le x,&x>b,
 \end{cases}
 \qquad
 x_n\longrightarrow\ell_x\in[\min\{x,b\},\max\{x,b\}]\subset(0,\infty).
\]
By \eqref{aw:iteration} and \eqref{aw:c-bounds},
\begin{equation}\label{aw:limit-critical}
 0\leftarrow |x_{n+1}-x_n|
 =\frac{\eta}{c_n}|h'(x_n)|
 \ge\frac{\eta}{M}|h'(x_n)|.
\end{equation}
Continuity and \eqref{aw:h-maxima} give
\[
 h'(\ell_x)=\lim_{n\to\infty}h'(x_n)=0,\qquad
 \ell_x\in\{y>0:h'(y)=0\}=\{b\}.
\]
Together with $P_n(-x)=-P_n(x)$ and $P_n(0)=0$, this yields
\begin{equation}\label{aw:pointwise-limit}
 P_n(x)\longrightarrow P_\infty(x):=
 \begin{cases}
 -b,&x<0,\\
 0,&x=0,\\
 b,&x>0.
 \end{cases}
\end{equation}
Since
\[
 \rho_0((0,\infty))=\rho_0((-\infty,0))=\frac{1-\alpha}{2},
 \qquad \rho_0(\{0\})=\alpha,
\]
we obtain
\begin{equation}\label{aw:limit-measure}
 (P_\infty)_\#\rho_0
 =\rho_\infty
 :=\alpha\delta_0+\frac{1-\alpha}{2}(\delta_{-b}+\delta_b).
\end{equation}
For $f\in C_b(\R)$, dominated convergence gives
\[
 \int f\,d\rho_n=\int f(P_n(x))\,d\rho_0(x)
 \longrightarrow\int f(P_\infty(x))\,d\rho_0(x)
 =\int f\,d\rho_\infty,
\]
and hence $\rho_n\Rightarrow\rho_\infty$.

The interval bounds also give
\begin{equation}\label{aw:moment-domination}
 |P_n(x)|\le\max\{|x|,b\},\qquad
 |P_n(x)-P_\infty(x)|^p\le2^p\max\{|x|^p,b^p\}
 \quad(p\ge1).
\end{equation}
For $1\le p<\infty$,
\[
 \int 2^p\max\{|x|^p,b^p\}\,d\rho_0(x)
 \le2^p\left(\int|x|^p\,d\rho_0(x)+b^p\right)<\infty.
\]
Thus the coupling $(P_n,P_\infty)_\#\rho_0$ and dominated convergence give
\begin{equation}\label{aw:Wp}
 W_p(\rho_n,\rho_\infty)^p
 \le\int|P_n(x)-P_\infty(x)|^p\,d\rho_0(x)
 \longrightarrow0,\qquad 1\le p<\infty.
\end{equation}

\subsection{The unique NPMLE and the exact likelihood gap}
The global optimum can be determined without restricting the competing measure to be symmetric. For any $\rho\in\mathcal P(\mathbb R)$, write
\begin{equation}\label{aw:Apm}
 A_+(\rho)=\int\phi(a-x)\,d\rho(x),\qquad
 A_-(\rho)=\int\phi(a+x)\,d\rho(x).
\end{equation}
Since $A_\pm(\rho)>0$, the arithmetic--geometric mean inequality and \eqref{aw:h-maxima} give
\begin{equation}\label{aw:global-bound}
 e^{-L(\rho)}
 =\sqrt{A_+(\rho)A_-(\rho)}
 \le\frac{A_+(\rho)+A_-(\rho)}2
 =\int h\,d\rho
 \le h(b).
\end{equation}
The measure
\begin{equation}\label{aw:optimal-measure}
 \rho^\star=\frac12(\delta_{-b}+\delta_b)
\end{equation}
satisfies
\[
 A_+(\rho^\star)=A_-(\rho^\star)=h(b),\qquad
 L(\rho^\star)=-\log h(b)=\inf_{\rho\in\mathcal P(\R)}L(\rho).
\]
For uniqueness, $h(b)-h\ge0$ and $\{x:h(x)=h(b)\}=\{-b,b\}$ imply that equality in the last step of \eqref{aw:global-bound} is equivalent to
\begin{equation}\label{aw:support-optimum}
 0=\int\{h(b)-h(x)\}\,d\rho(x)
 \quad\Longleftrightarrow\quad \rho(\{-b,b\})=1.
\end{equation}
For $\rho=q\delta_b+(1-q)\delta_{-b}$, $0\le q\le1$, equality in the first inequality of \eqref{aw:global-bound} is equivalent to
\begin{equation}\label{aw:unique-weight}
 0=A_+(\rho)-A_-(\rho)
 =(2q-1)\{\phi(a-b)-\phi(a+b)\}.
\end{equation}
Since
\[
 \phi(a-b)-\phi(a+b)
 =Me^{-(a-b)^2/2}(1-e^{-2ab})>0,
\]
the two equality conditions force $q=1/2$. Hence $\rho^\star$ is the
unique NPMLE over $\mathcal P(\R)$.

At the algorithmic limit \eqref{aw:limit-measure}, the common fitted value equals
\begin{equation}\label{aw:c-limit}
 h(0)<c_\infty=\int h\,d\rho_\infty
 =\alpha h(0)+(1-\alpha)h(b)<h(b).
\end{equation}
Thus the limiting loss exceeds the optimum by the explicit amount
\begin{equation}\label{aw:exact-gap}
 L(\rho_\infty)-L(\rho^\star)
 =\log\frac{h(b)}{\alpha h(0)+(1-\alpha)h(b)}>0.
\end{equation}
At the limit,
\[
 h'(x)=0\quad(x\in\operatorname{supp}\rho_\infty=\{-b,0,b\}),\qquad
 (\operatorname{id}+\eta g_{\rho_\infty}')_\#\rho_\infty=\rho_\infty.
\]
Nevertheless, its score values satisfy
\begin{equation}\label{aw:score-limit}
 g_{\rho_\infty}(0)=\frac{h(0)}{c_\infty}<1,
 \qquad
 g_{\rho_\infty}(-b)=g_{\rho_\infty}(b)
 =\frac{h(b)}{c_\infty}>1,
 \qquad
 g_{\rho_\infty}'=0\quad\rho_\infty\text{-a.e.}
\end{equation}
Thus the limit is spatially stationary, and the reaction stationarity
condition fails at every point of its support.

When $\alpha=0$, the initial measure is $\rho_0=\mathcal N(0,s^2)$.
The same trajectory argument applies under \eqref{aw:step}, with
$c_{\mathrm{init}}$ evaluated at $\alpha=0$. It gives
\[
 W_p(\rho_n,\rho^\star)\longrightarrow0
 \qquad(1\le p<\infty).
\]
This proves Proposition~2.5.

The two-dimensional example in Theorem~2.4 also starts
from a Gaussian. In that example, an atom forms at the origin in the
limit.

\section{The particle initialization probability}\label{app:particle-counts}

For $B\sim\operatorname{Bernoulli}(p)$, let
\[
 \psi(t)=\log\mathbb E e^{t(B-p)}
 =\log(1-p+pe^t)-pt,\qquad
 p_t=\frac{pe^t}{1-p+pe^t}.
\]
Then
\[
 \begin{gathered}
 \psi(0)=\psi'(0)=0,\qquad \psi'(t)=p_t-p,\qquad
 0\le\psi''(t)=p_t(1-p_t)\le\frac14,\\
 \psi(t)=t^2\int_0^1(1-s)\psi''(st)\,ds
 \le\frac{t^2}{8}\qquad(t\in\mathbb R).
 \end{gathered}
\]
For independent copies $B_1,\ldots,B_m$, $\varepsilon>0$ and $t>0$, Markov's inequality and independence give
\begin{align*}
 \mathbb P\left\{\sum_{\ell=1}^m(B_\ell-p)>m\varepsilon\right\}
 &\le e^{-mt\varepsilon}\mathbb E
       e^{t\sum_\ell(B_\ell-p)}
 =e^{-mt\varepsilon+m\psi(t)}\\
 &\le\exp\!\left(-mt\varepsilon+\frac{mt^2}{8}\right),\\
 \mathbb P\left\{\sum_{\ell=1}^m(B_\ell-p)<-m\varepsilon\right\}
 &\le e^{-mt\varepsilon+m\psi(-t)}
 \le\exp\!\left(-mt\varepsilon+\frac{mt^2}{8}\right).
\end{align*}
Taking $t=4\varepsilon$ in each tail,
\[
 \mathbb P\left\{\left|m^{-1}\sum_{\ell=1}^mB_\ell-p\right|>\varepsilon\right\}
 \le2e^{-2m\varepsilon^2}.
\]
The draws in equation~(2.7) of the main paper are independent
and uniform. Hence each count $C_j$ in \eqref{pt:counts} has
distribution $\operatorname{Binomial}(m,1/3)$. The union bound gives
\begin{equation}\label{pt:count-concentration}
 \begin{aligned}
 \mathbb P\left\{\max_j|C_j/m-1/3|>\varepsilon_\eta\right\}
 &\le\sum_{j=1}^3\mathbb P\{|C_j/m-1/3|>\varepsilon_\eta\}\\
 &\le6e^{-2m\varepsilon_\eta^2}.
 \end{aligned}
\end{equation}
No independence across $C_1,C_2,C_3$ is needed. Let
\[
 \mathcal A_m=\{\max_j|C_j/m-1/3|\le\varepsilon_\eta\}.
\]
By Lemma~5.1 and \textup{(5.22)}--\textup{(5.23)},
\[
 \mathcal A_m\ \Longrightarrow\
 \rho_n^{(m)}\rightharpoonup\rho_{\mathrm{bad}}(X),\qquad
 L_X(\rho_n^{(m)})\longrightarrow L_X(\rho_{\mathrm{bad}}(X))
 >L_X^*+\delta_\triangle/2.
\]
Together with $\mathbb P(\mathcal A_m)\ge1-6e^{-2m\varepsilon_\eta^2}$, this proves the probability assertion in Theorem~2.3. This probability argument does not require convergence on the exceptional count event.

Under the nested coupling stated in Theorem~2.3,
\[
 \sum_{m=1}^\infty\mathbb P(\mathcal A_m^c)
 \le6\sum_{m=1}^\infty e^{-2m\varepsilon_\eta^2}
 =\frac{6e^{-2\varepsilon_\eta^2}}
        {1-e^{-2\varepsilon_\eta^2}}
 <\infty.
\]
The first Borel--Cantelli lemma therefore implies that, almost surely,
$\mathcal A_m$ holds for all sufficiently large $m$. For every such $m$,
the particle orbit converges to $\rho_{\mathrm{bad}}(X)$ and satisfies
the strict likelihood gap above. This completes the proof of
Theorem~2.3.

\section{Nonuniformity over smooth full-support initializations}\label{app:nonuniform}

We prove Corollary~7.1 for the transport--reaction
update (1.4)--(1.5). The only dynamical
assumption on $\rho_{\mathrm b}$ is the fixed-point identity
$S_\eta\rho_{\mathrm b}=\rho_{\mathrm b}$.

\subsection{A common admissible step size}
For $0<b\le M$, set
\[
 r(b)=\sqrt{2\log(2M/b)},\qquad
 c(b)=\frac b2e^{-(r(b)+D_X)^2/2},\qquad
 H(c)=\frac{c^2}{M_2c+M_1^2}.
\]
Since $r'(b)=-1/(br(b))$,
\begin{equation}\label{nu:monotonicity}
 \begin{aligned}
 \frac{d\log c(b)}{d\log b}
 &=1-(r(b)+D_X)br'(b)
 =2+\frac{D_X}{r(b)}>0,\\
 H'(c)&=\frac{c(M_2c+2M_1^2)}{(M_2c+M_1^2)^2}>0.
 \end{aligned}
\end{equation}
For $\rho_0\in\mathcal A_B$, write $b_0=e^{-L(\rho_0)}$ and $b_B=e^{-B}$. Equations~(2.2) and~\eqref{nu:monotonicity} give
\begin{equation}\label{nu:common-step}
 \begin{gathered}
 0<b_B\le b_0\le M,
 \qquad c_B=c(b_B)\le c(b_0)=c_0,\\
 0<\eta\le\eta_B=H(c_B)\le H(c_0)=\eta_0(\rho_0),\\
 \eta_B\le\frac{c_B}{M_2}\le\frac{M}{2M_2}\le\frac12.
 \end{gathered}
\end{equation}
The last line uses $c_B\le b_B/2\le M/2$ and $M\le M_2$. Thus Theorem~2.1 applies to every $\rho_0\in\mathcal A_B$ with the same fixed $\eta$.

\subsection{Weak continuity of one complete update}
For $\mu\in\mathcal P(\R^d)$, put
\[
 T_\mu=\operatorname{id}+\eta\nabla g_\mu,\qquad
 \widetilde\mu=(T_\mu)_\#\mu,\qquad
 h_\mu=1-\eta+\eta g_{\widetilde\mu},\qquad
 S_\eta\mu=h_\mu\widetilde\mu.
\]
For $0<\eta\le1$, the update is a probability measure:
\begin{equation}\label{nu:probability}
 h_\mu\ge1-\eta\ge0,\qquad
 \int h_\mu\,d\widetilde\mu
 =1-\eta+\frac\eta N\sum_{i=1}^N
 \frac{a_i(\widetilde\mu)}{a_i(\widetilde\mu)}=1.
\end{equation}
Suppose $\mu_k\Rightarrow\mu$. The functions $\psi_i$ are bounded, continuous and strictly positive, so
\begin{equation}\label{nu:coefficients}
 a_i(\mu_k)\longrightarrow a_i(\mu)>0\qquad(1\le i\le N).
\end{equation}
For $j=0,1$, with $D^0g=g$ and $D^1g=\nabla g$,
\begin{equation}\label{nu:score-continuity}
 \begin{aligned}
 \|D^jg_{\mu_k}-D^jg_\mu\|_\infty
 &\le\frac1N\sum_{i=1}^N\|D^j\psi_i\|_\infty
 \left|\frac1{a_i(\mu_k)}-\frac1{a_i(\mu)}\right|
 \longrightarrow0,\\
 \|T_{\mu_k}-T_\mu\|_\infty
 &=\eta\|\nabla g_{\mu_k}-\nabla g_\mu\|_\infty\longrightarrow0.
 \end{aligned}
\end{equation}
For every bounded Lipschitz function $f$,
\begin{equation}\label{nu:transport-continuity}
 \begin{aligned}
 \left|\int f\,d(\widetilde\mu_k-\widetilde\mu)\right|
 &\le\int|f(T_{\mu_k}(x))-f(T_\mu(x))|\,d\mu_k(x)\\
 &\quad+\left|\int f\circ T_\mu\,d(\mu_k-\mu)\right|\\
 &\le\operatorname{Lip}(f)\|T_{\mu_k}-T_\mu\|_\infty
 +\left|\int f\circ T_\mu\,d(\mu_k-\mu)\right|\longrightarrow0.
 \end{aligned}
\end{equation}
Here $f\circ T_\mu\in C_b(\R^d)$, so the last integral tends to zero by $\mu_k\Rightarrow\mu$. Bounded Lipschitz tests determine weak convergence; so
\[
 \widetilde\mu_k\Rightarrow\widetilde\mu,
 \qquad
 \|h_{\mu_k}-h_\mu\|_\infty
 =\eta\|g_{\widetilde\mu_k}-g_{\widetilde\mu}\|_\infty\longrightarrow0,
\]
where the second assertion follows from \eqref{nu:score-continuity} applied to the transported measures. Since $h_\mu\in C_b(\R^d)$, every $f\in C_b(\R^d)$ satisfies
\begin{equation}\label{nu:reaction-continuity}
 \begin{aligned}
 \left|\int f\,d(S_\eta\mu_k-S_\eta\mu)\right|
 &\le\|f\|_\infty\|h_{\mu_k}-h_\mu\|_\infty
 +\left|\int fh_\mu\,d(\widetilde\mu_k-\widetilde\mu)\right|\\
 &\longrightarrow0.
 \end{aligned}
\end{equation}
Also, \eqref{nu:coefficients} and continuity of $\log$ on $(0,\infty)$ give
\begin{equation}\label{nu:loss-continuity}
 L(\mu_k)=-\frac1N\sum_i\log a_i(\mu_k)
 \longrightarrow-\frac1N\sum_i\log a_i(\mu)=L(\mu).
\end{equation}
Thus both $S_\eta$ and $L$ are weakly continuous.

\subsection{Proof of Corollary~7.1}
\begin{proof}
Let $Z\sim\mathcal N(0,I_d)$ and, for $\sigma>0$, define
\begin{equation}\label{nu:smoothing}
 \begin{gathered}
 \rho_\sigma=\rho_{\mathrm b}*\mathcal N(0,\sigma^2I_d),\\
 p_\sigma(x)=\sum_{j=1}^m
 \frac{w_j}{(2\pi\sigma^2)^{d/2}}
 e^{-|x-z_j|^2/(2\sigma^2)}\in C^\infty(\R^d),\\
 p_\sigma(x)>0\ (x\in\R^d),\qquad
 \int p_\sigma=\sum_jw_j=1,\qquad
 \operatorname{supp}(\rho_\sigma)=\R^d.
 \end{gathered}
\end{equation}
For every $p\ge1$,
\begin{equation}\label{nu:moments}
 \begin{aligned}
 \int|x|^p\,d\rho_\sigma(x)
 &=\sum_jw_j\mathbb E|z_j+\sigma Z|^p\\
 &\le2^{p-1}\left(\sum_jw_j|z_j|^p+\sigma^p\mathbb E|Z|^p\right)<\infty,\\
 W_2(\rho_\sigma,\rho_{\mathrm b})^2
 &\le\sum_jw_j\mathbb E|\sigma Z|^2=d\sigma^2\longrightarrow0.
 \end{aligned}
\end{equation}
The last inequality follows from the coupling $(z_J+\sigma Z,z_J)$,
where $\mathbb P(J=j)=w_j$ and $J$ is independent of $Z$. Thus
$\rho_\sigma\Rightarrow\rho_{\mathrm b}$ as $\sigma\downarrow0$.
Using \eqref{nu:reaction-continuity}--\eqref{nu:loss-continuity}
and $S_\eta\rho_{\mathrm b}=\rho_{\mathrm b}$, we obtain by induction
the following limits for each fixed $n$ as $\sigma\downarrow0$:
\begin{equation}\label{nu:finite-iterate}
 S_\eta^n\rho_\sigma\Rightarrow S_\eta^n\rho_{\mathrm b}=\rho_{\mathrm b},
 \qquad L(S_\eta^n\rho_\sigma)\longrightarrow L(\rho_{\mathrm b}).
\end{equation}
Fix $0<\varepsilon<\Delta$ and $T\in\mathbb N_0$. Then
\begin{equation}\label{nu:finite-maximum}
 E_T(\sigma):=\max_{0\le n\le T}
 |L(S_\eta^n\rho_\sigma)-L(\rho_{\mathrm b})|\longrightarrow0.
\end{equation}
Choose $\sigma>0$ so small that
\[
 E_T(\sigma)<\delta,
 \qquad \delta=\min\left\{\frac{B-L(\rho_{\mathrm b})}{2},
 \frac{\Delta-\varepsilon}{2}\right\}>0.
\]
For $\rho_0=\rho_\sigma$, we obtain
\begin{equation}\label{nu:gap-bound}
 \begin{aligned}
 L(\rho_0)&\le L(\rho_{\mathrm b})+E_T(\sigma)
 <L(\rho_{\mathrm b})+\frac{B-L(\rho_{\mathrm b})}{2}<B,\\
 \min_{0\le n\le T}\{L(S_\eta^n\rho_0)-L^*\}
 &\ge\Delta-E_T(\sigma)
 >\Delta-\delta
 \ge\frac{\Delta+\varepsilon}{2}>\varepsilon.
 \end{aligned}
\end{equation}
Equations~\eqref{nu:smoothing} and~\eqref{nu:gap-bound} give $\rho_0\in\mathcal A_B$ and (7.2).

For every fixed $\rho_0\in\mathcal A_B$, \eqref{nu:common-step} and Theorem~2.1 give
\[
 \rho_n=S_\eta^n\rho_0\Rightarrow\widehat\rho,\qquad
 L(\widehat\rho)=L^*,\qquad
 L(\rho_n)\longrightarrow L^*.
\]
The last limit follows from \eqref{nu:loss-continuity}.
This limit and the finite-horizon estimate \eqref{nu:gap-bound}
give, respectively,
\[
 \forall\rho_0\in\mathcal A_B:\ \tau_\varepsilon(\rho_0)<\infty,
 \qquad
 \forall T\in\mathbb N_0\ \exists\rho_0\in\mathcal A_B:\
 \tau_\varepsilon(\rho_0)>T,
\]
which proves (7.3).
\end{proof}

\end{appendix}
\bibliographystyle{spr-ims-nameyear}
\bibliography{references}